\documentstyle[twoside]{article}

\catcode`\@=11
\long\def\@makefntext#1{
\protect\noindent \hbox to 3.2pt {\hskip-.9pt
$^{{\eightrm\@thefnmark}}$\hfil}#1\hfill}		

\def\thefootnote{\fnsymbol{footnote}}
\def\@makefnmark{\hbox to 0pt{$^{\@thefnmark}$\hss}}	

\def\ps@myheadings{\let\@mkboth\@gobbletwo
\def\@oddhead{\hbox{}
\rightmark\hfil\eightrm\thepage}
\def\@oddfoot{}\def\@evenhead{\eightrm\thepage\hfil
\leftmark\hbox{}}\def\@evenfoot{}
\def\sectionmark##1{}\def\subsectionmark##1{}}

\oddsidemargin=\evensidemargin
\renewcommand{\thefootnote}{\fnsymbol{footnote}}

\newcounter{sectionc}
\newcounter{subsectionc}
\newcounter{subsubsectionc}
\renewcommand{\section}[1] {\vspace{12pt}\addtocounter{sectionc}{1}
\setcounter{subsectionc}{0}\setcounter{subsubsectionc}{0}\noindent
	{\tenbf\thesectionc. #1}\par\vspace{5pt}}
\renewcommand{\subsection}[1] {\vspace{12pt}
\addtocounter{subsectionc}{1}\setcounter{subsubsectionc}{0}\noindent
	{\bf\thesectionc.\thesubsectionc.
        {\kern1pt \bfit #1}}\par\vspace{5pt}}
\renewcommand{\subsubsection}[1] {\vspace{12pt}
\addtocounter{subsubsectionc}{1}\noindent
        {\tenrm\thesectionc.\thesubsectionc.\thesubsubsectionc.
	{\kern1pt \tenit #1}}\par\vspace{5pt}}

\newcounter{appendixc}
\newcounter{subappendixc}[appendixc]
\newcounter{subsubappendixc}[subappendixc]
\renewcommand{\thesubappendixc}{\Alph{appendixc}.
        \arabic{subappendixc}}
\renewcommand{\thesubsubappendixc}{\Alph{appendixc}.
        \arabic{subappendixc}.\arabic{subsubappendixc}}

\renewcommand{\appendix}[1] {\vspace{12pt}
        \refstepcounter{appendixc}
        \setcounter{figure}{0}
        \setcounter{table}{0}
        \setcounter{lemma}{0}
        \setcounter{theorem}{0}
        \setcounter{corollary}{0}
        \setcounter{definition}{0}
        \setcounter{equation}{0}
        \renewcommand{\thefigure}{\Alph{appendixc}.\arabic{figure}}
        \renewcommand{\thetable}{\Alph{appendixc}.\arabic{table}}
        \renewcommand{\theappendixc}{\Alph{appendixc}}
        \renewcommand{\thelemma}{\Alph{appendixc}.\arabic{lemma}}
        \renewcommand{\thetheorem}{\Alph{appendixc}.\arabic{theorem}}
        \renewcommand{\thedefinition}{\Alph{appendixc}.
         \arabic{definition}}
        \renewcommand{\thecorollary}{\Alph{appendixc}.
         \arabic{corollary}}
        \renewcommand{\theequation}{\Alph{appendixc}.
         \arabic{equation}}
        \noindent{\tenbf Appendix \theappendixc #1}\par\vspace{5pt}}
\newcommand{\subappendix}[1] {\vspace{12pt}
        \refstepcounter{subappendixc}
        \noindent{\bf Appendix \thesubappendixc. {\kern1pt \bfit #1}}
	\par\vspace{5pt}}
\newcommand{\subsubappendix}[1] {\vspace{12pt}
        \refstepcounter{subsubappendixc}
        \noindent{\rm Appendix \thesubsubappendixc.
        {\kern1pt \tenit #1}}\par\vspace{5pt}}

\newcommand{\textlineskip}{\baselineskip=13pt}
\newcommand{\smalllineskip}{\baselineskip=10pt}

\def\eightcirc{
\begin{picture}(0,0)
\put(4.4,1.8){\circle{6.5}}
\end{picture}}
\def\eightcopyright{\eightcirc\kern2.7pt\hbox{\eightrm c}}

\newcommand{\copyrightheading}[1]
  {\vspace*{-2.5cm}\smalllineskip{\flushleft
  {\footnotesize International Journal of Modern Physics A, #1}\\
  {\footnotesize $\eightcopyright$\, World Scientific Publishing
   Company}\\
  }}

\newcommand{\publisher}[2]{{\begin{center}\footnotesize\smalllineskip
	Received #1\\
	Revised #2
	\end{center}
	}}

\def\abstracts#1#2#3{{
	\centering{\begin{minipage}{4.5in}\baselineskip=10pt
        \footnotesize
	\parindent=0pt #1\par
	\parindent=15pt #2\par
	\parindent=15pt #3
	\end{minipage}}\par}}

\renewenvironment{thebibliography}[1]
	{\frenchspacing
	 \ninerm\baselineskip=11pt
	 \begin{list}{\arabic{enumi}.}
	{\usecounter{enumi}\setlength{\parsep}{0pt}
	 \setlength{\leftmargin 12.7pt}{\rightmargin 0pt}
	 \setlength{\itemsep}{0pt} \settowidth
	{\labelwidth}{#1.}\sloppy}}{\end{list}}

\newcounter{itemlistc}
\newcounter{romanlistc}
\newcounter{alphlistc}
\newcounter{arabiclistc}

\newcommand{\fcaption}[1]{
        \refstepcounter{figure}
        \setbox\@tempboxa = \hbox{\footnotesize Fig.~\thefigure. #1}
        \ifdim \wd\@tempboxa > 5in
           {\begin{center}
        \parbox{5in}{\footnotesize\smalllineskip Fig.~\thefigure. #1}
            \end{center}}
        \else
             {\begin{center}
             {\footnotesize Fig.~\thefigure. #1}
              \end{center}}
        \fi}

\newcommand{\tcaption}[1]{
        \refstepcounter{table}
        \setbox\@tempboxa = \hbox{\footnotesize Table~\thetable. #1}
        \ifdim \wd\@tempboxa > 5in
           {\begin{center}
        \parbox{5in}{\footnotesize\smalllineskip Table~\thetable. #1}
            \end{center}}
        \else
             {\begin{center}
             {\footnotesize Table~\thetable. #1}
              \end{center}}
        \fi}

\def\@citex[#1]#2{\if@filesw\immediate\write\@auxout
	{\string\citation{#2}}\fi
\def\@citea{}\@cite{\@for\@citeb:=#2\do
	{\@citea\def\@citea{,}\@ifundefined
	{b@\@citeb}{{\bf ?}\@warning
	{Citation `\@citeb' on page \thepage \space undefined}}
	{\csname b@\@citeb\endcsname}}}{#1}}

\newif\if@cghi
\def\cite{\@cghitrue\@ifnextchar [{\@tempswatrue
	\@citex}{\@tempswafalse\@citex[]}}
\def\citelow{\@cghifalse\@ifnextchar [{\@tempswatrue
	\@citex}{\@tempswafalse\@citex[]}}
\def\@cite#1#2{{$\null^{#1}$\if@tempswa\typeout
	{IJCGA warning: optional citation argument
	ignored: `#2'} \fi}}

\def\pmb#1{\setbox0=\hbox{#1}
	\kern-.025em\copy0\kern-\wd0
	\kern.05em\copy0\kern-\wd0
	\kern-.025em\raise.0433em\box0}

\def\fnt#1#2{\footnotetext{\kern-.3em
	{$^{\mbox{\scriptsize #1}}$}{#2}}}

\def\fpage#1{\begingroup
\voffset=.3in
\thispagestyle{empty}\begin{table}[b]\centerline{\footnotesize #1}
	\end{table}\endgroup}

\def\runninghead#1#2{\pagestyle{myheadings}
\markboth{{\protect\footnotesize\it{\quad #1}}\hfill}
{\hfill{\protect\footnotesize\it{#2\quad}}}}
\font\tenrm=cmr10
\font\tenit=cmti10
\font\tenbf=cmbx10
\font\bfit=cmbxti10 at 10pt
\font\ninerm=cmr9

\font\eightrm=cmr8

\def\qed{\hbox{${\vcenter{\vbox{		
   \hrule height 0.4pt\hbox{\vrule width 0.4pt height 6pt
   \kern5pt\vrule width 0.4pt}\hrule height 0.4pt}}}$}}

\renewcommand{\thefootnote}{\fnsymbol{footnote}}

\input epsf
\global\arraycolsep=2pt
\def\spose#1{\hbox to 0pt{#1\hss}}
\def\lsim{\mathrel{\spose{\lower 3pt\hbox{$\mathchar"218$}}
 \raise 2.0pt\hbox{$\mathchar"13C$}}}
\def\gsim{\mathrel{\spose{\lower 3pt\hbox{$\mathchar"218$}}
 \raise 2.0pt\hbox{$\mathchar"13E$}}}

\begin{document}

\begin{titlepage}

\begin{flushright}
CERN-TH/96-55\\
hep-ph/9604412
\end{flushright}

\vspace{2.5cm}

\begin{center}
\Large\bf B Decays and CP Violation
\end{center}

\vspace{2.0cm}

\begin{center}
Matthias Neubert\\
{\sl Theory Division, CERN, CH-1211 Geneva 23, Switzerland}
\end{center}

\vspace{2.0cm}

\begin{abstract}
We review the status of the theory and phenomenology of heavy-quark
symmetry, exclusive weak decays of $B$ mesons, inclusive decay rates
and lifetimes of $b$ hadrons, and CP violation in $B$-meson decays.
\end{abstract}

\vspace{1.5cm}

\centerline{(To appear in International Journal of Modern Physics A)}

\vspace{3.5cm}

\noindent
CERN-TH/96-55\\
April 1996
\vfil

\end{titlepage}

\thispagestyle{empty}
\vbox{}
\newpage

\normalsize\textlineskip
\thispagestyle{empty}
\setcounter{page}{1}

\copyrightheading{}			

\vspace*{0.88truein}

\fpage{1}

\centerline{\bf B DECAYS AND CP VIOLATION}
\vspace*{0.37truein}

\centerline{\footnotesize MATTHIAS NEUBERT}
\vspace*{0.015truein}
\centerline{\footnotesize\it Theory Division, CERN}
\baselineskip=10pt
\centerline{\footnotesize\it CH-1211 Geneva 23, Switzerland}

\vspace*{0.225truein}
\publisher{(received date)}{(revised date)}

\vspace*{0.21truein}
\abstracts{We review the status of the theory and phenomenology of
heavy-quark symmetry, exclusive weak decays of $B$ mesons, inclusive
decay rates and lifetimes of $b$ hadrons, and CP violation in
$B$-meson decays.}{}{}
\vspace*{1pt}\textlineskip

\textheight=7.8truein
\setcounter{footnote}{0}
\renewcommand{\thefootnote}{\alph{footnote}}

\runninghead{Introduction} {Introduction}
\section{Introduction}
\noindent
The rich phenomenology of weak decays has always been a source of
information about the nature of elementary particle interactions. A
long time ago, $\beta$- and $\mu$-decay experiments revealed the
structure of the effective flavour-changing interactions at low
momentum transfer. Today, weak decays of hadrons containing heavy
quarks are employed for tests of the Standard Model and measurements
of its parameters. In particular, they offer the most direct way to
determine the weak mixing angles, to test the unitarity of the
Cabibbo--Kobayashi--Maskawa (CKM) matrix, and to explore the physics
of CP violation. On the other hand, hadronic weak decays also serve
as a probe of that part of strong-interaction phenomenology which is
least understood: the confinement of quarks and gluons inside
hadrons.

The structure of weak interactions in the Standard Model is rather
simple. Flavour-changing decays are mediated by the coupling of the
charged current $J_{\rm CC}^\mu$ to the $W$-boson field:
\begin{equation}
   {\cal L}_{\rm CC} = - {g\over\sqrt{2}}\,J_{\rm CC}^\mu\,
   W_\mu^\dagger + \mbox{h.c.,}
\end{equation}
where
\begin{equation}
   J_{\rm CC}^\mu =
   (\bar\nu_e, \bar\nu_\mu, \bar\nu_\tau)\,\gamma^\mu
   \left( \begin{array}{c} e_{\rm L} \\ \mu_{\rm L} \\ \tau_{\rm L}
   \end{array} \right)
   + (\bar u_{\rm L}, \bar c_{\rm L}, \bar t_{\rm L})\,\gamma^\mu\,
   V_{\rm CKM} \left( \begin{array}{c} d_{\rm L} \\ s_{\rm L} \\
   b_{\rm L} \end{array} \right)
\end{equation}
contains the left-handed lepton and quark fields, and
\begin{equation}
   V_{\rm CKM} = \left( \begin{array}{ccc}
    V_{ud} & V_{us} & V_{ub} \\
    V_{cd} & V_{cs} & V_{cb} \\
    V_{td} & V_{ts} & V_{tb}
   \end{array} \right)
\end{equation}
is the CKM matrix. At low energies, the charged-current interaction
gives rise to local four-fermion couplings of the form
\begin{equation}\label{LFermi}
   {\cal L}_{\rm eff} = - 2\sqrt{2} G_F\,J_{\rm CC}^\mu
   J_{{\rm CC},\mu}^\dagger \,,
\end{equation}
where
\begin{equation}
   G_F = {g^2\over 8 M_W^2} = 1.16639(2)~\mbox{GeV}^{-2}
\end{equation}
is the Fermi constant.

\begin{figure}[htb]
   \epsfxsize=6cm
   \centerline{\epsffile{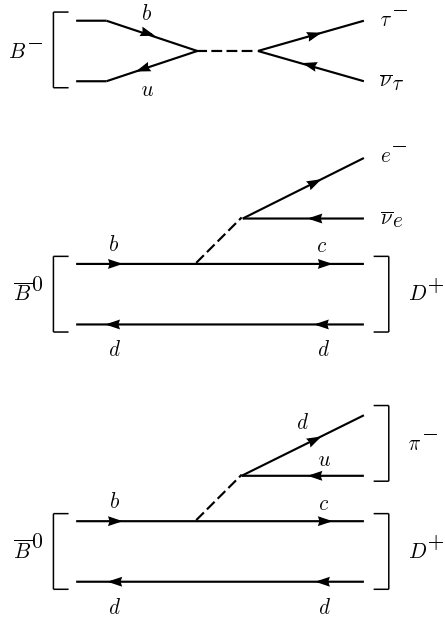}}
   \centerline{\parbox{11.5cm}{\caption{\label{fig:classes}
Examples of leptonic ($B^-\to\tau^-\bar\nu_\tau$), semileptonic
($\bar B^0\to D^+ e^-\bar\nu_e$), and non-leptonic ($\bar B^0\to
D^+\pi^-$) decays of $B$ mesons.
   }}}
\end{figure}

\begin{figure}[htb]
   \epsfxsize=8.5cm
   \centerline{\epsffile{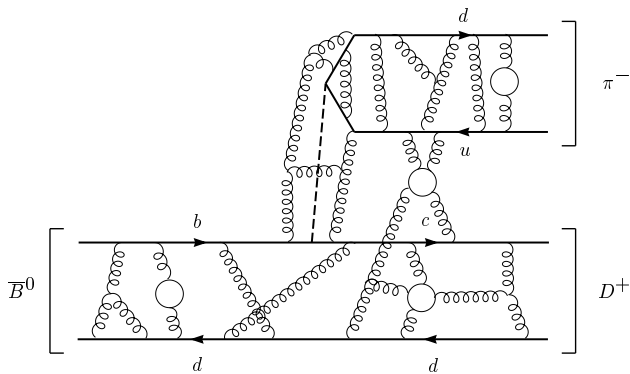}}
   \centerline{\parbox{11.5cm}{\caption{\label{fig:nonlep}
More realistic representation of a non-leptonic decay.
   }}}
\end{figure}

According to the structure of the charged-current interaction, weak
decays of had\-rons can be divided into three classes: leptonic
decays, in which the quarks of the decaying hadron annihilate each
other and only leptons appear in the final state; semileptonic
decays, in which both leptons and hadrons appear in the final state;
and non-leptonic decays, in which the final state consists of hadrons
only. Representative examples of these three types of decays are
shown in Fig.~\ref{fig:classes}. The simple quark-line graphs shown
in this figure are a gross oversimplification, however. In the real
world, quarks are confined inside hadrons, bound by the exchange of
soft gluons. The simplicity of the weak interactions is
over\-sha\-dowed by the complexity of the strong interactions. A
complicated interplay between the weak and strong forces
characterizes the phenomenology of hadronic weak decays. As an
example, a more realistic picture of a non-leptonic decay is shown in
Fig.~\ref{fig:nonlep}.

The complexity of strong-interaction effects increases with the
number of quarks appearing in the final state. Bound-state effects in
leptonic decays can be lumped into a single parameter (a ``decay
constant''), while those in semileptonic decays are described by
invariant form factors, depending on the momentum transfer $q^2$
between the hadrons. Approximate symmetries of the strong
interactions help to constrain the properties of these form factors.
For non-leptonic decays, on the other hand, we are still far from
having a quantitative understanding of strong-interaction effects
even in the simplest decay modes.

Over the last decade, a lot of information on heavy-quark decays has
been collected in experiments at $e^+ e^-$ storage rings operating at
the $\Upsilon(4s)$ resonance, and more recently at high-energy $e^+
e^-$ and hadron colliders. This has led to a rather detailed
knowledge of the flavour sector of the Standard Model and many of the
parameters associated with it. There have been several great
discoveries in this field, such as $B^0$--$\bar B^0$
mixing\cite{BBbar1,BBbar2}, $b\to u$
transitions\cite{btou1}$^-$\cite{Bpirho}, and rare decays induced by
penguin operators\cite{BKstar,btos}. Yet there is much more to come.
Upgrades of the existing facilities at Cornell and Fermilab, as well
as the $B$-factories to be operated at SLAC, KEK, HERA-B and LHC-B,
will provide a wealth of new results within the coming years. The
experimental progress in heavy-flavour physics has been accompanied
by a significant progress in theory, which was related to the
discovery of heavy-quark symmetry, the development of the heavy-quark
effective theory, and the establishment of the heavy-quark expansion
for inclusive decay rates. The excitement about these developments is
caused by the fact that they allow (some) model-independent
predictions in an area in which ``progress'' in theory often meant
nothing more than the construction of a new model, which could be
used to estimate some strong-interaction hadronic matrix elements. In
section~\ref{sec:2}, we explain the physical picture behind
heavy-quark symmetry and discuss the construction, as well as simple
applications, of the heavy-quark effective theory.
Section~\ref{sec:3} deals with applications of these concepts to
exclusive weak decays of $B$ mesons. Applications of the heavy-quark
expansion to the description of inclusive decay rates and lifetimes
of $b$ hadrons are the topic of section~\ref{sec:4}.
Section~\ref{sec:5} is devoted to a discussion of CP violation in
meson decays and the physics of the unitarity triangle.


\runninghead{Heavy-Quark Symmetry} {The Physical Picture}
\section{Heavy-Quark Symmetry}
\label{sec:2}
\noindent
This section provides an introduction to the ideas of heavy-quark
symmetry\cite{Shu1}$^-$\cite{Isgu} and the heavy-quark effective
theory\cite{EiFe}$^-$\cite{Mann}, which provide the modern
theoretical framework for the description of the properties and
decays of hadrons containing a heavy quark. For a more detailed
description of this subject, the reader is referred to the review
articles in Refs.\cite{review}$^-$\cite{ShifRev}.

\subsection{The Physical Picture}
\noindent
There are several reasons why the strong interactions of systems
containing heavy quarks are easier to understand than those of
systems containing only light quarks. The first is asymptotic
freedom, the fact that the effective coupling constant of QCD becomes
weak in processes with large momentum transfer, corresponding to
interactions at short-distance scales\cite{Gros,Poli}. At large
distances, on the other hand, the coupling becomes strong, leading to
non-perturbative phenomena such as the confinement of quarks and
gluons on a length scale $R_{\rm had}\sim 1/\Lambda_{\rm QCD}\sim
1$~fm, which determines the size of hadrons. Roughly speaking,
$\Lambda_{\rm QCD}\sim 0.2$ GeV is the energy scale that separates
the regions of large and small coupling constant. When the mass of a
quark $Q$ is much larger than this scale, $m_Q\gg\Lambda_{\rm QCD}$,
it is called a heavy quark. The quarks of the Standard Model fall
naturally into two classes: up, down and strange are light quarks,
whereas charm, bottom and top are heavy quarks\footnote{Ironically,
the top quark is of no relevance to my discussion here, since it is
too heavy to form hadronic bound states before it decays.}. For
heavy quarks, the effective coupling constant $\alpha_s(m_Q)$ is
small, implying that on length scales comparable to the Compton
wavelength $\lambda_Q\sim 1/m_Q$ the strong interactions are
perturbative and much like the electromagnetic interactions. In fact,
the quarkonium systems $(\bar QQ)$, whose size is of order
$\lambda_Q/\alpha_s(m_Q)\ll R_{\rm had}$, are very much
hydrogen-like. Since the discovery of asymptotic freedom, their
properties could be predicted\cite{Appe} before the observation of
charmonium, and later of bottomonium states.

Systems composed of a heavy quark and other light constituents are
more complicated. The size of such systems is determined by $R_{\rm
had}$, and the typical momenta exchanged between the heavy and light
constituents are of order $\Lambda_{\rm QCD}$. The heavy quark is
surrounded by a most complicated, strongly interacting cloud of light
quarks, antiquarks, and gluons. In this case it is the fact that
$\lambda_Q\ll R_{\rm had}$, i.e.\ that the Compton wavelength of the
heavy quark is much smaller than the size of the hadron, which leads
to simplifications. To resolve the quantum numbers of the heavy quark
would require a hard probe; the soft gluons exchanged between the
heavy quark and the light constituents can only resolve distances
much larger than $\lambda_Q$. Therefore, the light degrees of freedom
are blind to the flavour (mass) and spin orientation of the heavy
quark. They experience only its colour field, which extends over
large distances because of confinement. In the rest frame of the
heavy quark, it is in fact only the electric colour field that is
important; relativistic effects such as colour magnetism vanish as
$m_Q\to\infty$. Since the heavy-quark spin participates in
interactions only through such relativistic effects, it decouples.
That the heavy-quark mass becomes irrelevant can be seen as follows:
as $m_Q\to\infty$, the heavy quark and the hadron that contains it
have the same velocity. In the rest frame of the hadron, the heavy
quark is at rest, too. The wave function of the light constituents
follows from a solution of the field equations of QCD subject to the
boundary condition of a static triplet source of colour at the
location of the heavy quark. This boundary condition is independent
of $m_Q$, and so is the solution for the configuration of the light
constituents.

It follows that, in the limit $m_Q\to\infty$, hadronic systems which
differ only in the flavour or spin quantum numbers of the heavy quark
have the same configuration of their light degrees of
freedom\cite{Shu1}$^-$\cite{Isgu}. Although this observation still
does not allow us to calculate what this configuration is, it
provides relations between the properties of such particles as the
heavy mesons $B$, $D$, $B^*$ and $D^*$, or the heavy baryons
$\Lambda_b$ and $\Lambda_c$ (to the extent that corrections to the
infinite quark-mass limit are small in these systems). These
relations result from some approximate symmetries of the effective
strong interactions of heavy quarks at low energies. The
configuration of light degrees of freedom in a hadron containing a
single heavy quark with velocity $v$ does not change if this quark is
replaced by another heavy quark with different flavour or spin, but
with the same velocity. Both heavy quarks lead to the same static
colour field. For $N_h$ heavy-quark flavours, there is thus an SU$(2
N_h)$ spin-flavour symmetry group, under which the effective strong
interactions are invariant. These symmetries are in close
correspondence to familiar properties of atoms. The flavour symmetry
is analogous to the fact that different isotopes have the same
chemistry, since to good approximation the wave function of the
electrons is independent of the mass of the nucleus. The electrons
only see the total nuclear charge. The spin symmetry is analogous to
the fact that the hyperfine levels in atoms are nearly degenerate.
The nuclear spin decouples in the limit $m_e/m_N\to 0$.

Heavy-quark symmetry is an approximate symmetry, and corrections
arise since the quark masses are not infinite. In many respects, it
is complementary to chiral symmetry, which arises in the opposite
limit of small quark masses. There is an important distinction,
however. Whereas chiral symmetry is a symmetry of the QCD Lagrangian
in the limit of vanishing quark masses, heavy-quark symmetry is not a
symmetry of the Lagrangian (not even an approximate one), but rather
a symmetry of an effective theory, which is a good approximation of
QCD in a certain kinematic region. It is realized only in systems in
which a heavy quark interacts predominantly by the exchange of soft
gluons. In such systems the heavy quark is almost on-shell; its
momentum fluctuates around the mass shell by an amount of order
$\Lambda_{\rm QCD}$. The corresponding fluctuations in the velocity
of the heavy quark vanish as $\Lambda_{\rm QCD}/m_Q\to 0$. The
velocity becomes a conserved quantity and is no longer a dynamical
degree of freedom\cite{Geor}. Nevertheless, results derived on the
basis of heavy-quark symmetry are model-independent consequences of
QCD in a well-defined limit. The symmetry-breaking corrections can,
at least in principle, be studied in a systematic way. To this end,
it is however necessary to recast the QCD Lagrangian for a heavy
quark,
\begin{equation}\label{QCDLag}
   {\cal L}_Q = \bar Q\,(i\,\rlap{\,/}D - m_Q)\,Q \,,
\end{equation}
into a form suitable for taking the limit $m_Q\to\infty$.

\runninghead{Heavy-Quark Symmetry} {Heavy-Quark Effective Theory}
\subsection{Heavy-Quark Effective Theory}
\noindent
The effects of a very heavy particle often become irrelevant at low
energies. It is then useful to construct a low-energy effective
theory, in which this heavy particle no longer appears. Eventually,
this effective theory will be easier to deal with than the full
theory. A familiar example is Fermi's theory of the weak
interactions. For the description of weak decays of hadrons, the weak
interactions can be approximated by point-like four-fermion
couplings, governed by a dimensionful coupling constant $G_F$. Only
at energies much larger than the masses of hadrons can the effects
of the intermediate vector bosons, $W$ and $Z$, be resolved.

The process of removing the degrees of freedom of a heavy particle
involves the following steps\cite{SVZ1}$^-$\cite{Polc}: one first
identifies the heavy-particle fields and ``integrates them out'' in
the generating functional of the Green functions of the theory. This
is possible since at low energies the heavy particle does not appear
as an external state. However, although the action of the full theory
is usually a local one, what results after this first step is a
non-local effective action. The non-locality is related to the fact
that in the full theory the heavy particle with mass $M$ can appear
in virtual processes and propagate over a short but finite distance
$\Delta x\sim 1/M$. Thus, a second step is required to obtain a local
effective Lagrangian: the non-local effective action is rewritten as
an infinite series of local terms in an Operator Product Expansion
(OPE)\cite{Wils,Zimm}. Roughly speaking, this corresponds to an
expansion in powers of $1/M$. It is in this step that the short- and
long-distance physics is disentangled. The long-distance physics
corresponds to interactions at low energies and is the same in the
full and the effective theory. But short-distance effects arising
from quantum corrections involving large virtual momenta (of order
$M$) are not reproduced in the effective theory, once the heavy
particle has been integrated out. In a third step, they have to be
added in a perturbative way using renormalization-group techniques.
These short-distance effects lead to a renormalization of the
coefficients of the local operators in the effective Lagrangian. An
example is the effective Lagrangian for non-leptonic weak decays, in
which radiative corrections from hard gluons with virtual momenta in
the range between $m_W$ and some renormalization scale $\mu\sim
1$~GeV give rise to Wilson coefficients, which renormalize the local
four-fermion interactions\cite{AltM}$^-$\cite{Gilm}.

The heavy-quark effective theory (HQET) is constructed to provide a
simplified description of processes where a heavy quark interacts
with light degrees of freedom by the exchange of soft
gluons\cite{EiFe}$^-$\cite{Mann}. Clearly, $m_Q$ is the high-energy
scale in this case, and $\Lambda_{\rm QCD}$ is the scale of the
hadronic physics we are interested in. However, a subtlety arises
since we want to describe the properties and decays of hadrons which
contain a heavy quark. Hence, it is not possible to remove the heavy
quark completely from the effective theory. What is possible,
however, is to integrate out the ``small components'' in the full
heavy-quark spinor, which describe the fluctuations around the mass
shell.

The starting point in the construction of the low-energy effective
theory is the observation that a very heavy quark bound inside a
hadron moves more or less with the hadron's velocity $v$, and is
almost on-shell. Its momentum can be written as
\begin{equation}\label{kresdef}
   p_Q^\mu = m_Q v^\mu + k^\mu \,,
\end{equation}
where the components of the so-called residual momentum $k$ are much
smaller than $m_Q$. Note that $v$ is a four-velocity, so that
$v^2=1$. Interactions of the heavy quark with light degrees of
freedom change the residual momentum by an amount of order $\Delta
k\sim\Lambda_{\rm QCD}$, but the corresponding changes in the
heavy-quark velocity vanish as $\Lambda_{\rm QCD}/m_Q\to 0$. In this
situation, it is appropriate to introduce large- and small-component
fields $h_v$ and $H_v$ by
\begin{equation}\label{hvHvdef}
   h_v(x) = e^{i m_Q v\cdot x}\,P_+\,Q(x) \,, \qquad
   H_v(x) = e^{i m_Q v\cdot x}\,P_-\,Q(x) \,,
\end{equation}
where $P_+$ and $P_-$ are projection operators defined as
\begin{equation}
   P_\pm = {1\pm\rlap/v\over 2} \,.
\end{equation}
It follows that
\begin{equation}\label{redef}
   Q(x) = e^{-i m_Q v\cdot x}\,[ h_v(x) + H_v(x) ] \,.
\end{equation}
Because of the projection operators, the new fields satisfy
$\rlap/v\,h_v=h_v$ and $\rlap/v\,H_v=-H_v$. In the rest frame, i.e.\
for $v^\mu=(1,0,0,0)$, $h_v$ corresponds to the upper two components
of $Q$, while $H_v$ corresponds to the lower ones. Whereas $h_v$
annihilates a heavy quark with velocity $v$, $H_v$ creates a heavy
antiquark with velocity $v$.

\begin{figure}[htb]
   \epsfxsize=7cm
   \centerline{\epsffile{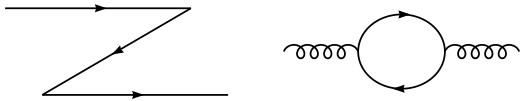}}
   \centerline{\parbox{11.5cm}{\caption{\label{fig:3.1}
Virtual fluctuations involving pair creation of heavy quarks. Time
flows to the right.
   }}}
\end{figure}

In terms of the new fields, the QCD Lagrangian (\ref{QCDLag}) for a
heavy quark takes the form
\begin{equation}\label{Lhchi}
   {\cal L}_Q = \bar h_v\,i v\!\cdot\!D\,h_v
   - \bar H_v\,(i v\!\cdot\!D + 2 m_Q)\,H_v
   + \bar h_v\,i\,\rlap{\,/}D_\perp H_v
   + \bar H_v\,i\,\rlap{\,/}D_\perp h_v \,,
\end{equation}
where $D_\perp^\mu = D^\mu - v^\mu\,v\cdot D$ is orthogonal to the
heavy-quark velocity: $v\cdot D_\perp=0$. In the rest frame,
$D_\perp^\mu=(0,\vec D\,)$ contains the spatial components of the
covariant derivative. From (\ref{Lhchi}), it is apparent that $h_v$
describes massless degrees of freedom, whereas $H_v$ corresponds to
fluctuations with twice the heavy-quark mass. These are the heavy
degrees of freedom that will be eliminated in the construction of the
effective theory. The fields are mixed by the presence of the third
and fourth terms, which describe pair creation or annihilation of
heavy quarks and antiquarks. As shown in the first diagram in
Fig.~\ref{fig:3.1}, in a virtual process, a heavy quark propagating
forward in time can turn into an antiquark propagating backward in
time, and then turn back into a quark. The energy of the intermediate
quantum state $h h\bar H$ is larger than the energy of the incoming
heavy quark by at least $2 m_Q$. Because of this large energy gap,
the virtual quantum fluctuation can only propagate over a short
distance $\Delta x\sim 1/m_Q$. On hadronic scales set by $R_{\rm
had}=1/\Lambda_{\rm QCD}$, the process essentially looks like a local
interaction of the form
\begin{equation}
   \bar h_v\,i\,\rlap{\,/}D_\perp\,{1\over 2 m_Q}\,
   i\,\rlap{\,/}D_\perp h_v \,,
\end{equation}
where we have simply replaced the propagator for $H_v$ by $1/2 m_Q$.
A more correct treatment is to integrate out the small-component
field $H_v$, thereby deriving a non-local effective action for the
large-component field $h_v$, which can then be expanded in terms of
local operators. Before doing this, let us mention a second type of
virtual corrections involving pair creation, namely heavy-quark
loops. An example is shown in the second diagram in
Fig.~\ref{fig:3.1}. Heavy-quark loops cannot be described in terms of
the effective fields $h_v$ and $H_v$, since the quark velocities
inside a loop are not conserved and are in no way related to hadron
velocities. However, such short-distance processes are proportional
to the small coupling constant $\alpha_s(m_Q)$ and can be calculated
in perturbation theory. They lead to corrections that are added onto
the low-energy effective theory in the renormalization procedure.

On a classical level, the heavy degrees of freedom represented by
$H_v$ can be eliminated using the equation of motion
$(i\,\rlap{\,/}D-m_Q)\,Q=0$. With (\ref{redef}), this gives
\begin{equation}
   i\,\rlap{\,/}D\,h_v + (i\,\rlap{\,/}D - 2 m_Q)\,H_v =0 \,,
\end{equation}
and multiplying by $P_\pm$ one derives the two equations
\begin{equation}
   -i v\!\cdot\!D\,h_v = i\,\rlap{\,/}D_\perp H_v \,, \qquad
   (i v\!\cdot\!D + 2 m_Q)\,H_v = i\,\rlap{\,/}D_\perp h_v \,.
\end{equation}
The second one can be solved to give
\begin{equation}\label{Hfield}
   H_v = {1\over 2 m_Q + i v\!\cdot\!D}\,i\,\rlap{\,/}D_\perp h_v \,,
\end{equation}
which shows that the small-component field $H_v$ is indeed of order
$1/m_Q$. We can now insert this solution into the first equation to
obtain the equation of motion for $h_v$. It is easy to see that this
equation follows from the non-local effective Lagrangian
\begin{equation}\label{Lnonloc}
   {\cal L}_{\rm eff} = \bar h_v\,i v\!\cdot\!D\,h_v
   + \bar h_v\,i\,\rlap{\,/}D_\perp\,{1\over 2 m_Q+i v\!\cdot\!D}\,
   i\,\rlap{\,/}D_\perp h_v \,.
\end{equation}
Clearly, the second term corresponds to the first class of virtual
processes shown in Fig.~\ref{fig:3.1}.

\begin{figure}[htb]
   \epsfysize=3cm
   \centerline{\epsffile{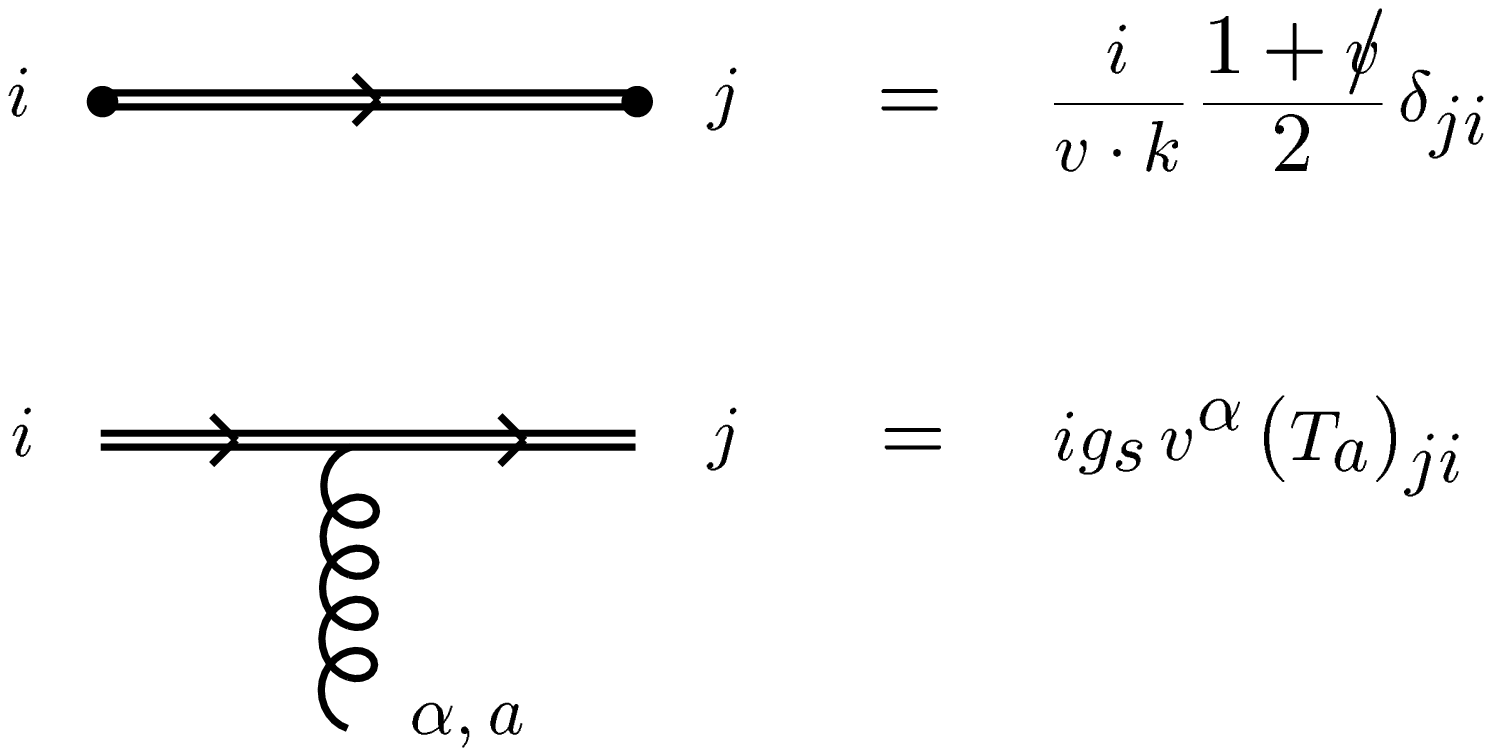}}
   \centerline{\parbox{11.5cm}{\caption{\label{fig:3.2}
Feynman rules of the HQET ($i,j$ and $a$ are colour indices). A heavy
quark is represented by a double line labelled by the velocity $v$.
The residual momentum $k$ is defined in (\protect\ref{kresdef}).
   }}}
\end{figure}

Because of the phase factor in (\ref{redef}), the $x$ dependence of
the effective heavy-quark field $h_v$ is weak. In momentum space,
derivatives acting on $h_v$ produce powers of the residual momentum
$k$, which is much smaller than $m_Q$. Hence, the non-local effective
Lagrangian (\ref{Lnonloc}) allows for a derivative expansion in
powers of $iD/m_Q$:
\begin{equation}
   {\cal L}_{\rm eff} = \bar h_v\,i v\!\cdot\!D\,h_v
   + {1\over 2 m_Q}\,\sum_{n=0}^\infty\,
   \bar h_v\,i\,\rlap{\,/}D_\perp\,\bigg( -{i v\cdot D\over 2 m_Q}
   \bigg)^n\,i\,\rlap{\,/}D_\perp h_v \,.
\end{equation}
Taking into account that $h_v$ contains a $P_+$ projection operator,
and using the identity
\begin{equation}\label{pplusid}
   P_+\,i\,\rlap{\,/}D_\perp\,i\,\rlap{\,/}D_\perp P_+
   = P_+\,\bigg[ (i D_\perp)^2 + {g_s\over 2}\,
   \sigma_{\mu\nu }\,G^{\mu\nu } \bigg]\,P_+ \,,
\end{equation}
where $[i D^\mu,i D^\nu]=i g_s\,G^{\mu\nu}$ is the gluon
field-strength tensor, one finds that\cite{EiH1,FGL}
\begin{equation}\label{Lsubl}
   {\cal L}_{\rm eff} = \bar h_v\,i v\!\cdot\!D\,h_v
   + {1\over 2 m_Q}\,\bar h_v\,(i D_\perp)^2\,h_v
   + {g_s\over 4 m_Q}\,\bar h_v\,\sigma_{\mu\nu}\,
   G^{\mu\nu}\,h_v + O(1/m_Q^2) \,.
\end{equation}
In the limit $m_Q\to\infty$, only the first terms remains:
\begin{equation}\label{Leff}
   {\cal L}_\infty = \bar h_v\,i v\!\cdot\!D\,h_v \,.
\end{equation}
This is the effective Lagrangian of the HQET. It gives rise to the
Feynman rules depicted in Fig.~\ref{fig:3.2}.

Let us take a moment to study the symmetries of this
Lagrangian\cite{Geor}. Since there appear no Dirac matrices,
interactions of the heavy quark with gluons leave its spin unchanged.
Associated with this is an SU(2) symmetry group, under which
${\cal L}_\infty$ is invariant. The action of this symmetry on the
heavy-quark fields becomes most transparent in the rest frame, where
the generators $S^i$ of SU(2) can be chosen as
\begin{equation}\label{Si}
   S^i = {1\over 2} \left( \begin{array}{cc}
                           \sigma^i ~&~ 0 \\
                           0 ~&~ \sigma^i \end{array} \right) \,;
   \qquad [S^i,S^j] = i \epsilon^{ijk} S^k \,.
\end{equation}
Here $\sigma^i$ are the Pauli matrices. An infinitesimal SU(2)
transformation $h_v\to (1 + i\vec\epsilon \cdot\vec S\,)\,h_v$ leaves
the Lagrangian invariant:
\begin{equation}\label{SU2tr}
   \delta{\cal L}_\infty = \bar h_v\,
   [i v\!\cdot\! D,i \vec\epsilon\cdot\vec S\,]\,h_v = 0 \,.
\end{equation}
Another symmetry of the HQET arises since the mass of the heavy quark
does not appear in the effective Lagrangian. For $N_h$ heavy quarks
moving at the same velocity, eq.~(\ref{Leff}) can be extended by
writing
\begin{equation}\label{Leff2}
   {\cal L}_\infty
   = \sum_{i=1}^{N_h} \bar h_v^i\,i v\!\cdot\! D\,h_v^i \,.
\end{equation}
This is invariant under rotations in flavour space. When combined
with the spin symmetry, the symmetry group is promoted to SU$(2
N_h)$. This is the heavy-quark spin-flavour symmetry\cite{Isgu,Geor}.
Its physical content is that, in the limit $m_Q\to\infty$, the strong
interactions of a heavy quark become independent of its mass and
spin.

Consider now the operators appearing at order $1/m_Q$ in the
effective Lagrangian (\ref{Lsubl}). They are easiest to identify in
the rest frame. The first operator,
\begin{equation}\label{Okin}
   {\cal O}_{\rm kin} = {1\over 2 m_Q}\,\bar h_v\,(i D_\perp)^2\,
   h_v \to - {1\over 2 m_Q}\,\bar h_v\,(i \vec D\,)^2\,h_v \,,
\end{equation}
is the gauge-covariant extension of the kinetic energy arising from
the off-shell residual motion of the heavy quark. The second operator
is the non-Abelian analogue of the Pauli interaction, which describes
the colour-magnetic coupling of the heavy-quark spin to the gluon
field:
\begin{equation}\label{Omag}
   {\cal O}_{\rm mag} = {g_s\over 4 m_Q}\,\bar h_v\,
   \sigma_{\mu\nu}\,G^{\mu\nu}\,h_v \to
   - {g_s\over m_Q}\,\bar h_v\,\vec S\!\cdot\!\vec B_c\,h_v \,.
\end{equation}
Here $\vec S$ is the spin operator defined in (\ref{Si}), and $B_c^i
= -\frac{1}{2}\epsilon^{ijk} G^{jk}$ are the components of the
colour-magnetic field. The chromo-magnetic interaction is a
relativistic effect, which scales like $1/m_Q$. This is the origin of
the heavy-quark spin symmetry.

\runninghead{Heavy-Quark Symmetry} {Spectroscopic Implications}
\subsection{Spectroscopic Implications}
\noindent
The spin-flavour symmetry leads to many interesting relations between
the properties of hadrons containing a heavy quark. The most direct
consequences concern the spectroscopy of such states\cite{IsWi}. In
the limit $m_Q\to\infty$, the spin of the heavy quark and the total
angular momentum $j$ of the light degrees of freedom are separately
conserved by the strong interactions. Because of heavy-quark
symmetry, the dynamics is independent of the spin and mass of the
heavy quark. Hadronic states can thus be classified by the quantum
numbers (flavour, spin, parity, etc.) of the light degrees of
freedom\cite{AFal}. The spin symmetry predicts that, for fixed $j\neq
0$, there is a doublet of degenerate states with total spin
$J=j\pm\frac{1}{2}$. The flavour symmetry relates the properties of
states with different heavy-quark flavour.

In general, the mass of a hadron $H_Q$ containing a heavy quark $Q$
obeys an expansion of the form
\begin{equation}\label{massexp}
   m_H = m_Q + \bar\Lambda + {\Delta m^2\over 2 m_Q} + O(1/m_Q^2) \,.
\end{equation}
The parameter $\bar\Lambda$ represents contributions arising from
terms in the Lagrangian that are independent of the heavy-quark
mass\cite{FNL}, whereas the quantity $\Delta m^2$ originates from the
terms of order $1/m_Q$ in the effective Lagrangian of the HQET. For
the ground-state pseudoscalar and vector mesons, one can parametrize
the contributions from the kinetic energy and the chromo-magnetic
interaction in terms of two quantities $\lambda_1$ and $\lambda_2$,
in such a way that\cite{FaNe}
\begin{equation}\label{FNrela}
   \Delta m^2 = -\lambda_1 + 2 \Big[ J(J+1) - \textstyle{3\over 2}
   \Big]\,\lambda_2 \,.
\end{equation}
The hadronic parameters $\bar\Lambda$, $\lambda_1$ and $\lambda_2$
are independent of $m_Q$. They characterize the properties of the
light constituents.

Consider, as a first example, the SU(3) mass splittings for heavy
mesons. The heavy-quark expansion predicts that
\begin{eqnarray}
   m_{B_S} - m_{B_d} &=& \bar\Lambda_s - \bar\Lambda_d
    + O(1/m_b) \,, \nonumber\\
   m_{D_S} - m_{D_d} &=& \bar\Lambda_s - \bar\Lambda_d
    + O(1/m_c) \,,
\end{eqnarray}
where we have indicated that the value of the parameter $\bar\Lambda$
depends on the flavour of the light quark. Thus, to the extent that
the charm and bottom quarks can both be considered sufficiently
heavy, the mass splittings should be similar in the two systems. This
prediction is confirmed experimentally, since\cite{Joe}
\begin{eqnarray}
   m_{B_S} - m_{B_d} &=& (90\pm 3)~\mbox{MeV} \,, \nonumber\\
   m_{D_S} - m_{D_d} &=& (99\pm 1)~\mbox{MeV} \,.
\end{eqnarray}

As a second example, consider the spin splittings between the
ground-state pseudoscalar ($J=0$) and vector ($J=1$) mesons, which
are the members of the spin-doublet with $j=\frac{1}{2}$. The theory
predicts that
\begin{eqnarray}
   m_{B^*}^2 - m_B^2 &=& 4\lambda_2 + O(1/m_b) \,, \nonumber\\
   m_{D^*}^2 - m_D^2 &=& 4\lambda_2 + O(1/m_c) \,.
\end{eqnarray}
The data are compatible with this:
\begin{eqnarray}\label{VPexp}
   m_{B^*}^2 - m_B^2 &\simeq& 0.49~{\rm GeV}^2 \,, \nonumber\\
   m_{D^*}^2 - m_D^2 &\simeq& 0.55~{\rm GeV}^2 \,.
\end{eqnarray}
Assuming that the $B$ system is close to the heavy-quark limit, we
obtain the value
\begin{equation}
   \lambda_2\simeq 0.12~\mbox{GeV}^2
\end{equation}
for one of the hadronic parameters in (\ref{FNrela}). This quantity
plays an important role in the phenomenology of inclusive decays of
heavy hadrons.

A third example is provided by the mass splittings between the
ground-state mesons and baryons containing a heavy quark. The HQET
predicts that
\begin{eqnarray}\label{barmes}
   m_{\Lambda_b} - m_B &=& \bar\Lambda_{\rm baryon}
    - \bar\Lambda_{\rm meson} + O(1/m_b) \,, \nonumber\\
   m_{\Lambda_c} - m_D &=& \bar\Lambda_{\rm baryon}
    - \bar\Lambda_{\rm meson} + O(1/m_c) \,.
\end{eqnarray}
This is again consistent with the experimental results
\begin{eqnarray}
   m_{\Lambda_b} - m_B &=& (346\pm 6)~\mbox{MeV} \,, \nonumber\\
   m_{\Lambda_c} - m_D &=& (416\pm 1)~\mbox{MeV} \,,
\end{eqnarray}
although in this case the data indicate sizeable symmetry-breaking
corrections. For the mass of the $\Lambda_b$ baryon, we have used the
value
\begin{equation}\label{Lbmass}
   m_{\Lambda_b} = (5625\pm 6)~\mbox{MeV} \,,
\end{equation}
which is obtained by averaging the result $m_{\Lambda_b}= (5639\pm
15)$~MeV quoted in Ref.\cite{Joe} with the preliminary value
$m_{\Lambda_b}=(5623\pm 5\pm 4)$~MeV reported by the CDF
Collaboration\cite{CDFmass}. The dominant correction to the
relations (\ref{barmes}) comes from the contribution of the
chromo-magnetic interaction to the masses of the heavy
mesons\footnote{Because of the spin symmetry, there is no such
contribution to the masses of the $\Lambda_Q$ baryons.}, which adds
a term $3\lambda_2/2 m_Q$ on the right-hand side. Including this
term, we obtain the refined prediction that the two quantities
\begin{eqnarray}
   m_{\Lambda_b} - m_B - {3\lambda_2\over 2 m_B}
   &=& (312\pm 6)~\mbox{MeV} \,, \nonumber\\
   m_{\Lambda_c} - m_D - {3\lambda_2\over 2 m_D}
   &=& (320\pm 1)~\mbox{MeV}
\end{eqnarray}
should be close to each other. This is clearly satisfied by the data.

The mass formula (\ref{massexp}) can also be used to derive
information on the heavy-quark masses from the observed hadron
masses. Introducing the ``spin-averaged'' meson masses
$\overline{m}_B=\frac{1}{4}\,(m_B+3 m_{B^*})\simeq 5.31$~GeV and
$\overline{m}_D=\frac{1}{4}\,(m_D+3 m_{D^*})\simeq 1.97$~GeV, we
find that
\begin{equation}\label{mbmc}
   m_b-m_c = (\overline{m}_B-\overline{m}_D)\,\bigg\{
   1 - {\lambda_1\over 2\overline{m}_B\overline{m}_D}
   + O(1/m_Q^3) \bigg\} \,.
\end{equation}
Using theoretical estimates for the parameter $\lambda_1$, which lie
in the range\cite{lam1}$^-$\cite{virial}
\begin{equation}\label{lam1}
   \lambda_1 = -(0.4\pm 0.2)~\mbox{GeV}^2 \,,
\end{equation}
this relation leads to
\begin{equation}\label{mbmcval}
   m_b - m_c = (3.40\pm 0.03\pm 0.03)~\mbox{GeV} \,,
\end{equation}
where the first error reflects the uncertainty in the value of
$\lambda_1$, and the second one takes into account unknown
higher-order corrections. As will be discussed in
section~\ref{sec:4}, the fact that the difference $(m_b-m_c)$ is
determined rather precisely becomes important in the analysis of
inclusive decays of heavy hadrons.

For completeness, we note that for the pole masses of the heavy
quarks we shall adopt the values
\begin{equation}
   m_b = 4.8~\mbox{GeV} \,,\qquad m_c = 1.4~\mbox{GeV} \,.
\end{equation}
The concept of the pole mass of a heavy quark has been the subject of
much discussion recently. It has been found that there is an
unavoidable ambiguity of order $\Lambda_{\rm QCD}$ in any definition
of the pole mass extending beyond perturbation
theory\cite{BeBr,Bigs}. Formally, this ambiguity appears as a
divergence of the perturbation series, which relates the pole mass to
a renormalized mass defined at short distances, such as the ``running
mass'' $\overline{m}_Q(m_Q)$. As long as we work to a finite order in
perturbation theory, however, this subtlety can be ignored. The
values given above will be used in connection with one-loop
calculations and thus refer to the one-loop pole masses of the heavy
quarks.


\runninghead{Exclusive Semileptonic Decays} {Weak Decay Form Factors}
\section{Exclusive Semileptonic Decays}
\label{sec:3}
\begin{figure}[htb]
   \epsfxsize=6.5cm
   \centerline{\epsffile{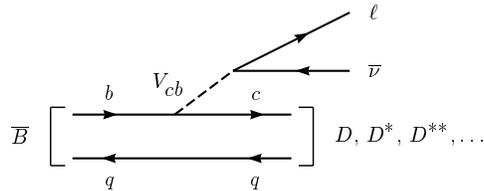}}
   \centerline{\parbox{11.5cm}{\caption{\label{fig:1}
Semileptonic decays of a $B$ meson.}}}
\end{figure}

\noindent
Semileptonic decays of $B$ mesons have received a lot of attention in
recent years. The decay channel $\bar B\to D^*\ell\,\bar\nu$ has the
largest branching fraction of all $B$-meson decay modes. From a
theoretical point of view, semileptonic decays are simple enough to
allow for a reliable, quantitative description. The analysis of these
decays provides much information about the strong forces that bind
the quarks and gluons into hadrons. Schematically, a semileptonic
decay process is shown in Fig.~\ref{fig:1}. The strength of the $b\to
c$ transition vertex is governed by the element $V_{cb}$ of the CKM
matrix. The parameters of this matrix are fundamental parameters of
the Standard Model. A primary goal of the study of semileptonic
decays of $B$ mesons is to extract with high precision the values of
$|V_{cb}|$ and $|V_{ub}|$. In this lecture, we will discuss the
theoretical basis of such analyses.

\subsection{Weak Decay Form Factors}
\noindent
Heavy-quark symmetry implies relations between the weak decay form
factors of heavy mesons, which are of particular interest. These
relations have been derived by Isgur and Wise\cite{Isgu},
generalizing ideas developed by Nussinov and Wetzel\cite{Nuss}, and
by Voloshin and Shifman\cite{Vol1,Vol2}.

\begin{figure}[htb]
   \epsfxsize=7cm
   \centerline{\epsffile{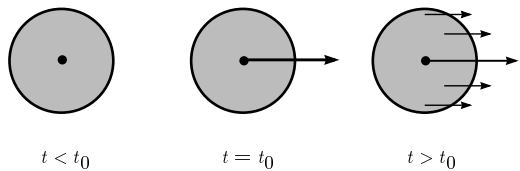}}
   \centerline{\parbox{11.5cm}{\caption{\label{fig:3.3}
Elastic transition induced by an external heavy-quark current.
   }}}
\end{figure}

Consider the elastic scattering of a $B$ meson, $\bar B(v)\to\bar
B(v')$, induced by a vector current coupled to the $b$ quark. Before
the action of the current, the light degrees of freedom in the $B$
meson orbit around the heavy quark, which acts as a static source of
colour. On average, the $b$ quark and the $B$ meson have the same
velocity $v$. The action of the current is to replace instantaneously
(at $t=t_0$) the colour source by one moving at a velocity $v'$, as
indicated in Fig.~\ref{fig:3.3}. If $v=v'$, nothing happens; the
light degrees of freedom do not realize that there was a current
acting on the heavy quark. If the velocities are different, however,
the light constituents suddenly find themselves interacting with a
moving colour source. Soft gluons have to be exchanged to rearrange
them so as to form a $B$ meson moving at velocity $v'$. This
rearrangement leads to a form-factor suppression, which reflects the
fact that as the velocities become more and more different, the
probability for an elastic transition decreases. The important
observation is that, in the limit $m_b\to\infty$, the form factor can
only depend on the Lorentz boost $\gamma = v\cdot v'$ that connects
the rest frames of the initial- and final-state mesons. Thus, in this
limit a dimensionless probability function $\xi(v\cdot v')$ describes
the transition. It is called the Isgur--Wise function\cite{Isgu}. In
the HQET, which provides the appropriate framework for taking the
limit $m_b\to\infty$, the hadronic matrix element describing the
scattering process can thus be written as
\begin{equation}\label{elast}
   {1\over m_B}\,\langle\bar B(v')|\,\bar b_{v'}\gamma^\mu b_v\,
   |\bar B(v)\rangle = \xi(v\cdot v')\,(v+v')^\mu \,.
\end{equation}
Here, $b_v$ and $b_{v'}$ are the velocity-dependent heavy-quark
fields defined in (\ref{hvHvdef}). It is important that the function
$\xi(v\cdot v')$ does not depend on $m_b$. The factor $1/m_B$ on the
left-hand side compensates for a trivial dependence on the
heavy-meson mass caused by the relativistic normalization of meson
states, which is conventionally taken to be
\begin{equation}\label{nonrelnorm}
   \langle\bar B(p')|\bar B(p)\rangle = 2 m_B v^0\,(2\pi)^3\,
   \delta^3(\vec p-\vec p\,') \,.
\end{equation}
Note that there is no term proportional to $(v-v')^\mu$ in
(\ref{elast}). This can be seen by contracting the matrix element
with $(v-v')_\mu$, which must give zero since $\rlap/v b_v = b_v$ and
$\bar b_{v'}\rlap/v' = \bar b_{v'}$.

It is more conventional to write the above matrix element in terms of
an elastic form factor $F_{\rm el}(q^2)$ depending on the momentum
transfer $q^2=(p-p')^2$:
\begin{equation}
   \langle\bar B(v')|\,\bar b\,\gamma^\mu b\,|\bar B(v)\rangle
   = F_{\rm el}(q^2)\,(p+p')^\mu \,,
\end{equation}
where $p^(\phantom{}'\phantom{}^)=m_B v^(\phantom{}'\phantom{}^)$.
Comparing this with (\ref{elast}), we find that
\begin{equation}
   F_{\rm el}(q^2) = \xi(v\cdot v') \,, \qquad
   q^2 = -2 m_B^2 (v\cdot v'-1) \,.
\end{equation}
Because of current conservation, the elastic form factor is
normalized to unity at $q^2=0$. This condition implies the
normalization of the Isgur--Wise function at the kinematic point
$v\cdot v'=1$, i.e.\ for $v=v'$:
\begin{equation}\label{Jcons2}
   \xi(1) = 1 \,.
\end{equation}
It is in accordance with the intuitive argument that the probability
for an elastic transition is unity if there is no velocity change.
Since for $v=v'$ the daughter meson is at rest in the rest frame of
the parent meson, the point $v\cdot v'=1$ is referred to as the
zero-recoil limit.

We can now use the flavour symmetry to replace the $b$ quark in the
final-state meson by a $c$ quark, thereby turning the $B$ meson into
a $D$ meson. Then the scattering process turns into a weak decay
process. In the infinite mass limit, the replacement $b_{v'}\to
c_{v'}$ is a symmetry transformation, under which the effective
Lagrangian is invariant. Hence, the matrix element
\begin{equation}
   {1\over\sqrt{m_B m_D}}\,\langle D(v')|\,\bar c_{v'}\gamma^\mu
   b_v\,|\bar B(v)\rangle = \xi(v\cdot v')\,(v+v')^\mu
\end{equation}
is still determined by the same function $\xi(v\cdot v')$. This is
interesting, since in general the matrix element of a
flavour-changing current between two pseudoscalar mesons is described
by two form factors:
\begin{equation}
   \langle D(v')|\,\bar c\gamma^\mu b\,|\bar B(v)\rangle
   = f_+(q^2)\,(p+p')^\mu - f_-(q^2)\,(p-p')^\mu \,.
\end{equation}
Comparing the above two equations, we find that
\begin{eqnarray}\label{inelast}
   f_\pm(q^2) &=& {m_B\pm m_D\over 2\sqrt{m_B m_D}}\,\xi(v\cdot v')
    \,, \nonumber\\
   q^2 &=& m_B^2 + m_D^2 - 2 m_B m_D (v\cdot v'-1) \,.
\end{eqnarray}
Thus, the heavy-quark flavour symmetry relates two {\it a priori\/}
independent form factors to one and the same function. Moreover, the
normalization of the Isgur--Wise function at $v\cdot v'=1$ now
implies a non-trivial normalization of the form factors $f_\pm(q^2)$
at the point of maximum momentum transfer, $q_{\rm max}^2=
(m_B-m_D)^2$:
\begin{equation}
   f_\pm(q_{\rm max}^2) = {m_B\pm m_D\over 2\sqrt{m_B m_D}} \,.
\end{equation}

The heavy-quark spin symmetry leads to additional relations among
weak decay form factors. It can be used to relate matrix elements
involving vector mesons to those involving pseudoscalar mesons. A
vector meson with longitudinal polarization is related to a
pseudoscalar meson by a rotation of the heavy-quark spin. Hence, the
spin-symmetry transformation $c_{v'}^\Uparrow\to c_{v'}^\Downarrow$
relates the transition matrix element for $\bar B\to D$ to that for
$\bar B\to D^*$. The result of this transformation is\cite{Isgu}:
\begin{eqnarray}
   {1\over\sqrt{m_B m_{D^*}}}\,
   \langle D^*(v',\epsilon)|\,\bar c_{v'}\gamma^\mu b_v\,
   |\bar B(v)\rangle &=& i\epsilon^{\mu\nu\alpha\beta}\,
    \epsilon_\nu^*\,v'_\alpha v_\beta\,\,\xi(v\cdot v') \,,
    \nonumber\\
   {1\over\sqrt{m_B m_{D^*}}}\,
   \langle D^*(v',\epsilon)|\,\bar c_{v'}\gamma^\mu\gamma_5\,
   b_v\,|\bar B(v)\rangle &=& \Big[ \epsilon^{*\mu}\,(v\cdot v'+1)
    - v'^\mu\,\epsilon^*\!\cdot v \Big]\,\xi(v\cdot v') \,,
    \nonumber\\
\end{eqnarray}
where $\epsilon$ denotes the polarization vector of the $D^*$ meson.
Once again, the matrix elements are completely described in terms of
the Isgur--Wise function. Now this is even more remarkable, since in
general four form factors, $V(q^2)$ for the vector current, and
$A_i(q^2)$, $i=0,1,2$, for the axial vector current, are required to
parametrize these matrix elements. In the heavy-quark limit, they
obey the relations\cite{Neu1}
\begin{eqnarray}\label{PVff}
   \xi(v\cdot v') &=& V(q^2) = A_0(q^2) = A_1(q^2)
    = \bigg[ 1 - {q^2\over(m_B+m_D)^2} \bigg]^{-1}\,A_1(q^2) \,,
    \nonumber\\
   q^2 &=& m_B^2 + m_{D^*}^2 - 2 m_B m_{D^*} (v\cdot v'-1) \,.
\end{eqnarray}

Equations (\ref{inelast}) and (\ref{PVff}) summarize the relations
imposed by heavy-quark symmetry on the weak decay form factors
describing the semileptonic decay processes $\bar B\to
D\,\ell\,\bar\nu$ and $\bar B\to D^*\ell\,\bar\nu$. These relations
are model-independent consequences of QCD in the limit where $m_b,
m_c\gg\Lambda_{\rm QCD}$. They play a crucial role in the
determination of $|V_{cb}|$. In terms of the recoil variable
$w=v\cdot v'$, the differential semileptonic decay rates in the
heavy-quark limit become:
\begin{eqnarray}\label{rates}
   {{\rm d}\Gamma(\bar B\to D\,\ell\,\bar\nu)\over{\rm d}w}
   &=& {G_F^2\over 48\pi^3}\,|V_{cb}|^2\,(m_B+m_D)^2\,m_D^3\,
    (w^2-1)^{3/2}\,\xi^2(w) \,, \nonumber\\
   {{\rm d}\Gamma(\bar B\to D^*\ell\,\bar\nu)\over{\rm d}w}
   &=& {G_F^2\over 48\pi^3}\,|V_{cb}|^2\,(m_B-m_{D^*})^2\,
    m_{D^*}^3\,\sqrt{w^2-1}\,(w+1)^2 \nonumber\\
   &&\times \Bigg[ 1 + {4w\over w+1}\,
    {m_B^2 - 2 w\,m_B m_{D^*} + m_{D^*}^2\over(m_B-m_{D^*})^2}
    \Bigg]\,\xi^2(w) \,.
\end{eqnarray}
These expressions receive symmetry-breaking corrections, since the
masses of the heavy quarks are not infinitely large. Perturbative
corrections of order $\alpha_s^n(m_Q)$, with $Q=b$ or $c$, can be
calculated order by order in perturbation theory. A more difficult
task is to control non-perturbative power corrections of order
$(\Lambda_{\rm QCD}/m_Q)^n$. The HQET provides a systematic framework
for analysing these corrections. As an example, we have discussed in
section~\ref{sec:2} the $1/m_Q$ corrections to the effective
Lagrangian. For the more complicated case of weak-decay form factors,
the analysis of the $1/m_Q$ corrections was performed by
Luke\cite{Luke}. Later, Falk and the present author have also
analysed the structure of $1/m_Q^2$ corrections for both meson and
baryon weak decay form factors\cite{FaNe}. We shall not discuss these
rather technical issues in detail, but only mention the most
important result of Luke's analysis. It concerns the zero-recoil
limit, where an analogue of the Ademollo--Gatto theorem\cite{AGTh}
can be proved. This is Luke's theorem\cite{Luke}, which states that
the matrix elements describing the leading $1/m_Q$ corrections to
weak decay amplitudes vanish at zero recoil. This theorem is valid to
all orders in perturbation theory\cite{FaNe,Neu7,ChGr}. Most
importantly, it protects the $\bar B\to D^*\ell\,\bar\nu$ decay rate
from receiving first-order $1/m_Q$ corrections at zero
recoil\cite{Vcb}. A similar statement is not true for the decay $\bar
B\to D\,\ell\,\bar\nu$, however. The reason is simple but somewhat
subtle. Luke's theorem protects only those form factors not
multiplied by kinematic factors that vanish for $v=v'$. By angular
momentum conservation, the two pseudoscalar mesons in the decay $\bar
B\to D\,\ell\,\bar\nu$ must be in a relative $p$-wave, and hence the
amplitude is proportional to the velocity $|\vec v_D|$ of the $D$
meson in the $B$-meson rest frame. This leads to a factor $(w^2-1)$
in the decay rate. In such a situation, form factors that are
kinematically suppressed can contribute\cite{Neu1}.

\runninghead{Exclusive Semileptonic Decays}
 {Short-Distance Corrections}
\subsection{Short-Distance Corrections}
\noindent
In section~\ref{sec:2}, we have discussed the first two steps in the
construction of the HQET. Integrating out the small components in the
heavy-quark fields, a non-local effective action was derived, which
was then expanded in a series of local operators. The effective
Lagrangian derived that way correctly reproduces the long-distance
physics of the full theory. It does not contain the short-distance
physics correctly, however. The reason is obvious: a heavy quark
participates in strong interactions through its coupling to gluons.
These gluons can be soft or hard, i.e.\ their virtual momenta can be
small, of the order of the confinement scale, or large, of the order
of the heavy-quark mass. But hard gluons can resolve the spin and
flavour quantum numbers of a heavy quark. Their effects lead to a
renormalization of the coefficients of the operators in the HQET. A
new feature of such short-distance corrections is that through the
running coupling constant $\alpha_s(m_Q)$ they induce a logarithmic
dependence on the heavy-quark mass\cite{Vol1}. Fortunately, since
$\alpha_s(m_Q)$ is small, these effects can be calculated in
perturbation theory.

Consider, as an example, matrix elements of the vector current
$V=\bar q\,\gamma^\mu Q$. In QCD this current is partially conserved
and needs no renormalization\cite{Prep}. Its matrix elements are
free of ultraviolet divergences. Still, these matrix elements have a
logarithmic dependence on $m_Q$ from the exchange of hard gluons with
virtual momenta of the order of the heavy-quark mass. If one goes
over to the effective theory by taking the limit $m_Q\to\infty$,
these logarithms diverge. Consequently, the vector current in the
effective theory does require a renormalization\cite{PoWi}. Its
matrix elements depend on an arbitrary renormalization scale $\mu$,
which separates the regions of short- and long-distance physics. If
$\mu$ is chosen such that $\Lambda_{\rm QCD}\ll\mu\ll m_Q$, the
effective coupling constant in the region between $\mu$ and $m_Q$ is
small, and perturbation theory can be used to compute the
short-distance corrections. These corrections have to be added to the
matrix elements of the effective theory, which contain the
long-distance physics below the scale $\mu$. Schematically, then, the
relation between matrix elements in the full and in the effective
theory is
\begin{equation}\label{OPEex}
   \langle\,V(m_Q)\,\rangle_{\rm QCD}
   = C_0(m_Q,\mu)\,\langle V_0(\mu)\rangle_{\rm HQET}
   + {C_1(m_Q,\mu)\over m_Q}\,\langle V_1(\mu)\rangle_{\rm HQET}
   + \ldots \,,
\end{equation}
where we have indicated that matrix elements in the full theory
depend on $m_Q$, whereas matrix elements in the effective theory are
mass-independent, but do depend on the renormalization scale. The
Wilson coefficients $C_i(m_Q,\mu)$ are defined by this relation.
Order by order in perturbation theory, they can be computed from a
comparison of the matrix elements in the two theories. Since the
effective theory is constructed to reproduce correctly the low-energy
behaviour of the full theory, this ``matching'' procedure is
independent of any long-distance physics, such as infrared
singularities, non-perturbative effects, the nature of the external
states used in the matrix elements, etc.

The calculation of the coefficient functions in perturbation theory
uses the powerful methods of the renormalization group. It is in
principle straightforward, yet in practice rather tedious. A
comprehensive discussion of most of the existing calculations of
short-distance corrections in the HQET can be found in
Ref.\cite{review}.

\runninghead{Exclusive Semileptonic Decays}
 {Model-Independent Determination of $|V_{cb}|$}
\subsection{Model-Independent Determination of $|V_{cb}|$}
\noindent
We will now discuss some of the most important applications and tests
of the above formalism in the context of semileptonic decays of $B$
mesons. A model-independent determination of the CKM matrix element
$|V_{cb}|$ based on heavy-quark symmetry can be obtained by measuring
the recoil spectrum of $D^*$ mesons produced in $\bar B\to
D^*\ell\,\bar\nu$ decays\cite{Vcb}. In the heavy-quark limit, the
differential decay rate for this process has been given in
(\ref{rates}). In order to allow for corrections to that limit, we
write
\begin{eqnarray}
   {{\rm d}\Gamma(\bar B\to D^*\ell\,\bar\nu)\over{\rm d}w}
   &=& {G_F^2\over 48\pi^3}\,(m_B-m_{D^*})^2\,m_{D^*}^3
    \sqrt{w^2-1}\,(w+1)^2 \nonumber\\
   &&\mbox{}\times \Bigg[ 1 + {4w\over w+1}\,
    {m_B^2-2w\,m_B m_{D^*} + m_{D^*}^2\over(m_B - m_{D^*})^2}
    \Bigg]\,|V_{cb}|^2\,{\cal{F}}^2(w) \,, \nonumber\\
\end{eqnarray}
where the hadronic form factor ${\cal F}(w)$ coincides with the
Isgur--Wise function up to symmetry-breaking corrections of order
$\alpha_s(m_Q)$ and $\Lambda_{\rm QCD}/m_Q$. The idea is to measure
the product $|V_{cb}|\,{\cal F}(w)$ as a function of $w$, and to
extract $|V_{cb}|$ from an extrapolation of the data to the
zero-recoil point $w=1$, where the $B$ and the $D^*$ mesons have a
common rest frame. At this kinematic point, heavy-quark symmetry
helps to calculate the normalization ${\cal F}(1)$ with small and
controlled theoretical errors. Since the range of $w$ values
accessible in this decay is rather small ($1<w<1.5$), the
extrapolation can be done using an expansion around $w=1$:
\begin{equation}\label{Fexp}
   {\cal F}(w) = {\cal F}(1)\,\Big[ 1 - \widehat\varrho^2\,(w-1)
   + \dots \Big] \,.
\end{equation}
The slope $\widehat\varrho^2$ is treated as a fit parameter.

\begin{figure}[htb]
   \epsfxsize=8cm
   \vspace{0.3cm}
   \centerline{\epsffile{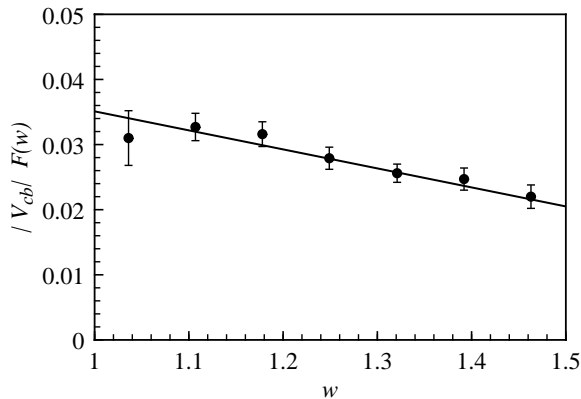}}
   \vspace{-0.3cm}
   \centerline{\parbox{11.5cm}{\caption{\label{fig:CLVcb}
CLEO data for the product $|V_{cb}|\,{\cal F}(w)$, as extracted from
the recoil spectrum in $\bar B\to D^*\ell\,\bar\nu$ decays
\protect\cite{CLEOVcb}. The line shows a linear fit to the data.
   }}}
\end{figure}

Measurements of the recoil spectrum have been performed first by the
AR\-GUS\cite{ARGVcb} and CLEO\cite{CLEOVcb} Collaborations in
experiments operating at the $\Upsilon(4s)$ resonance, and more
recently by the ALEPH\cite{ALEVcb} and DELPHI\cite{DELVcb}
Collaborations at LEP. As an example, Fig.~\ref{fig:CLVcb} shows the
data reported by the CLEO Collaboration. The results obtained by the
various experimental groups from a linear fit to their data are
summarized in Table~\ref{tab:Vcb}. The weighted average of these
results is
\begin{eqnarray}\label{VcbFraw}
   |V_{cb}|\,{\cal F}(1) &=& (34.6\pm 1.7)\times 10^{-3} \,,
    \nonumber\\
   \widehat\varrho^2 &=& 0.82\pm 0.09 \,.
\end{eqnarray}
The effect of a positive curvature of the form factor has been
investigated by Stone\cite{Stone}, who finds that the value of
$|V_{cb}|\,{\cal F}(1)$ may change by up to $+4\%$. We thus increase
the above value by $(2\pm 2)\%$ and quote the final result as
\begin{equation}\label{VcbF}
   |V_{cb}|\,{\cal F}(1) = (35.3\pm 1.8)\times 10^{-3} \,.
\end{equation}
In future analyses, the extrapolation to zero recoil should be
performed including higher-order terms in the expansion (\ref{Fexp}).
It can be shown in a model-independent way that the shape of the form
factor is highly constrained by analyticity and unitarity
requirements\cite{Boyd2,Capr}. In particular, the curvature at $w=1$
is strongly correlated with the slope of the form factor. For the
value of $\widehat\varrho^2$ given in (\ref{VcbFraw}), one obtains a
small positive curvature\cite{Capr}, in agreement with the assumption
made in Ref.\cite{Stone}.

\begin{table}[htb]
\centerline{\parbox{11.5cm}{\caption{\label{tab:Vcb}
Values for $|V_{cb}|\,{\cal F}(1)$ (in units of $10^{-3}$) and
$\widehat\varrho^2$ extracted from measurements of the recoil
spectrum in $\bar B\to D^*\ell\,\bar\nu$ decays}}}
\vspace{0.5cm}
\centerline{\begin{tabular}{|l|c|c|}\hline\hline
\rule[-0.2cm]{0cm}{0.65cm} & $|V_{cb}|\,{\cal F}(1)~(10^{-3})$ &
 $\widehat\varrho^2$ \\
\hline
ARGUS  & $38.8\pm 4.3\pm 2.5$ & $1.17\pm 0.22\pm 0.06$ \\
CLEO   & $35.1\pm 1.9\pm 2.0$ & $0.84\pm 0.12\pm 0.08$ \\
ALEPH  & $31.4\pm 2.3\pm 2.5$ & $0.39\pm 0.21\pm 0.12$ \\
DELPHI & $35.0\pm 1.9\pm 2.3$ & $0.81\pm 0.16\pm 0.10$ \\
\hline\hline
\end{tabular}}
\end{table}

Heavy-quark symmetry implies that the general structure of the
symmetry-breaking corrections to the form factor at zero recoil
is\cite{Vcb}
\begin{equation}
   {\cal F}(1) = \eta_A\,\bigg( 1 + 0\cdot
   {\Lambda_{\rm QCD}\over m_Q}
   + c_2\,{\Lambda_{\rm QCD}^2\over m_Q^2} + \dots \bigg)
   \equiv \eta_A\,(1+\delta_{1/m^2}) \,,
\end{equation}
where $\eta_A$ is a short-distance correction arising from a finite
renormalization of the flavour-changing axial current at zero recoil,
and $\delta_{1/m^2}$ parametrizes second-order (and higher) power
corrections. The absence of first-order power corrections at zero
recoil is a consequence of Luke's theorem\cite{Luke}. The one-loop
expression for $\eta_A$ has been known for a long
time\cite{Pasc,Vol2,QCD1}:
\begin{equation}\label{etaA1}
   \eta_A = 1 + {\alpha_s(M)\over\pi}\,\bigg(
   {m_b+m_c\over m_b-m_c}\,\ln{m_b\over m_c} - {8\over 3} \bigg)
   \simeq 0.96 \,.
\end{equation}
The scale $M$ in the running coupling constant can be fixed by
adopting the prescription of Brodsky, Lepage and Mackenzie
(BLM)\cite{BLM}, where it is identified with the average virtuality
of the gluon in the one-loop diagrams that contribute to $\eta_A$. If
$\alpha_s(M)$ is defined in the $\overline{\mbox{\sc ms}}$ scheme,
the result is\cite{etaVA} $M\simeq 0.51\sqrt{m_c m_b}$. Several
estimates of higher-order corrections to $\eta_A$ have been
discussed. A renormalization-group resummation of logarithms of the
type $(\alpha_s\ln m_b/m_c)^n$, $\alpha_s(\alpha_s\ln m_b/m_c)^n$ and
$m_c/m_b(\alpha_s\ln m_b/m_c)^n$ leads
to\cite{PoWi,JiMu}$^-$\cite{QCD2} $\eta_A\simeq 0.985$. On the other
hand, a resummation of ``renormalon-chain'' contributions of the form
$\beta_0^{n-1}\alpha_s^n$, where $\beta_0=11-\frac{2}{3}n_f$ is the
first coefficient of the QCD $\beta$-function, gives\cite{flow}
$\eta_A\simeq 0.945$. Using these partial resummations to estimate
the uncertainty gives $\eta_A = 0.965\pm 0.020$. Recently, Czarnecki
has improved this estimate by calculating $\eta_A$ at two-loop
order\cite{Czar}. His result,
\begin{equation}
   \eta_A = 0.960\pm 0.007 \,,
\end{equation}
is in excellent agreement with the BLM-improved one-loop expression
(\ref{etaA1}). Here the error is taken to be the size of the two-loop
correction.

The analysis of the power corrections $\delta_{1/m^2}$ is more
difficult, since it cannot rely on perturbation theory. Three
approaches have been discussed: in the ``exclusive approach'', all
$1/m_Q^2$ operators in the HQET are classified and their matrix
elements estimated, leading to\cite{FaNe,TMann}
$\delta_{1/m^2}=-(3\pm 2)\%$; the ``inclusive approach'' has been
used to derive the bound $\delta_{1/m^2}<-3\%$, and to estimate
that\cite{Shif} $\delta_{1/m^2}=-(7\pm 3)\%$; the ``hybrid approach''
combines the virtues of the former two to obtain a more restrictive
lower bound on $\delta_{1/m^2}$. This leads to\cite{Vcbnew}
\begin{equation}
   \delta_{1/m^2} = - 0.055\pm 0.025 \,.
\end{equation}

Combining the above results, adding the theoretical errors linearly
to be conservative, gives
\begin{equation}\label{F1}
   {\cal F}(1) = 0.91\pm 0.03
\end{equation}
for the normalization of the hadronic form factor at zero recoil.
Thus, the corrections to the heavy-quark limit amount to a moderate
decrease of the form factor of about 10\%. This can be used to
extract from the experimental result (\ref{VcbF}) the
model-independent value
\begin{equation}\label{Vcbexc}
   |V_{cb}| = (38.8\pm 2.0_{\rm exp}\pm 1.2_{\rm th})
   \times 10^{-3} \,.
\end{equation}

\runninghead{Exclusive Semileptonic Decays}
 {Bounds and Predictions for $\widehat\varrho^2$}
\subsection{Bounds and Predictions for $\widehat\varrho^2$}
\noindent
The slope parameter $\widehat\varrho^2$ in the expansion of the
physical form factor in (\ref{Fexp}) differs from the slope parameter
$\varrho^2$ of the Isgur--Wise function by corrections that violate
the heavy-quark symmetry. The short-distance corrections have been
calculated, with the result\cite{Vcbnew}
\begin{equation}\label{rhorel}
   \widehat\varrho^2 = \varrho^2 + (0.16\pm 0.02) + O(1/m_Q) \,.
\end{equation}
Bjorken has shown that the slope of the Isgur--Wise function is
related to the form factors of transitions of a ground-state heavy
meson into excited states, in which the light degrees of freedom
carry quantum numbers $j^P=\frac{1}{2}^+$ or $\frac{3}{2}^+$, by a
sum rule which is an expression of quark--hadron duality: in the
heavy-quark limit, the inclusive sum of the probabilities for
decays into hadronic states is equal to the probability for the free
quark transition. If one normalizes the latter probability to unity,
the sum rule takes the form\cite{Bjor}$^-$\cite{BjDT}
\begin{eqnarray}\label{inclsum}
   1 &=& {w+1\over 2}\,\bigg\{ |\xi(w)|^2 + \sum_l |\xi^{(l)}(w)|^2
    \bigg\} \nonumber\\
   &&\mbox{}+ (w-1)\,\bigg\{ 2\sum_m |\tau_{1/2}^{(m)}(w)|^2
    + (w+1)^2 \sum_n |\tau_{3/2}^{(n)}(w)|^2 \bigg\}
    + O\big[(w-1)^2\big] \,, \nonumber\\
\end{eqnarray}
where $l,m,n$ label the radial excitations of states with the same
spin-parity quantum numbers. The terms in the first line on the
right-hand side of the sum rule correspond to transitions into states
containing light constituents with quantum numbers
$j^P=\frac{1}{2}^-$. The ground state gives a contribution
proportional to the Isgur--Wise function, and excited states
contribute proportionally to analogous functions $\xi^{(l)}(w)$.
Because at zero recoil these states must be orthogonal to the ground
state, it follows that $\xi^{(l)}(1)=0$, and the corresponding
contributions to (\ref{inclsum}) are actually of order $(w-1)^2$. The
contributions in the second line correspond to transitions into
states with $j^P=\frac{1}{2}^+$ or $\frac{3}{2}^+$. Because of the
change in parity, these are $p$-wave transitions. The amplitudes are
proportional to the velocity $|\vec v_f|= (w^2-1)^{1/2}$ of the final
state in the rest frame of the initial state, which explains the
suppression factor $(w-1)$ in the decay probabilities. The functions
$\tau_j(w)$ are the analogues of the Isgur--Wise function for these
transitions\cite{IsgW}. Transitions into excited states with quantum
numbers other than the above proceed via higher partial waves and are
suppressed by at least a factor $(w-1)^2$.

For $w=1$, eq.~(\ref{inclsum}) reduces to the normalization condition
for the Isgur--Wise function. The Bjorken sum rule is obtained by
expanding in powers of $(w-1)$ and keeping terms of first order.
Taking into account the definition of the slope parameter,
$\xi'(1)=-\varrho^2$, one finds that\cite{Bjor,IsgW}
\begin{equation}\label{Bjsr}
   \varrho^2 = {1\over 4} + \sum_m |\tau_{1/2}^{(m)}(1)|^2
   + 2 \sum_n |\tau_{3/2}^{(n)}(1)|^2 > {1\over 4} \,.
\end{equation}
Notice that the lower bound is due to the prefactor
$\frac{1}{2}(w+1)$ of the first term in (\ref{inclsum}) and is of
purely kinematic origin. In the analogous sum rule for
$\Lambda_Q$ baryons, this factor is absent, and consequently the
slope parameter of the baryon Isgur--Wise function is only subject to
the trivial constraint\cite{Neu1,IWYo} $\varrho^2>0$.

Voloshin has derived another sum rule involving the form factors for
transitions into excited states, which is the analogue of the
``optical sum rule'' for the dipole scattering of light in atomic
physics. It reads\cite{Volsum}
\begin{equation}
   {m_M-m_Q\over 2} = \sum_m E_{1/2}^{(m)}\,|\tau_{1/2}^{(m)}(1)|^2
   + 2 \sum_n  E_{3/2}^{(n)}\,|\tau_{3/2}^{(n)}(1)|^2 \,,
\end{equation}
where $E_j$ are the excitation energies relative to the mass $m_M$ of
the ground-state heavy meson. The important point is that this
relation can be combined with the Bjorken sum rule to obtain an upper
bound for the slope parameter $\varrho^2$:
\begin{equation}\label{Volsr}
   \varrho^2 < {1\over 4} + {m_M-m_Q\over 2 E_{\rm min}} \,,
\end{equation}
where $E_{\rm min}$ denotes the minimum excitation energy. In the
quark model, one expects\footnote{Strictly speaking, the lowest
excited ``state'' contributing to the sum rule is $D+\pi$, which has
an excitation-energy spectrum with a threshold at $m_\pi$. However,
this spectrum is broad, so that this contribution will not invalidate
the upper bound for $\varrho^2$ derived here.} that $E_{\rm
min}\simeq m_M-m_Q$, and one may use this as an estimate to obtain
$\varrho^2<0.75$.

\begin{figure}[htb]
   \epsfxsize=8cm
   \centerline{\epsffile{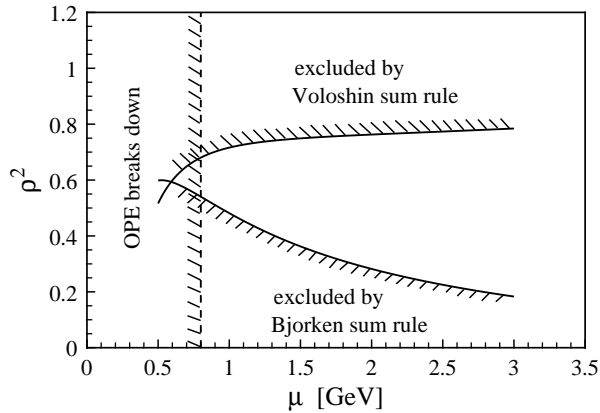}}
   \centerline{\parbox{11.5cm}{\caption{\label{fig:2}
Bounds for the slope parameter $\varrho^2$ following from
the Bjorken and Voloshin sum rules.
   }}}
\end{figure}

The above discussion of the sum rules ignores renormalization
effects. Both perturbative and non-perturbative corrections to
(\ref{Bjsr}) and (\ref{Volsr}) can be incorporated using the OPE,
where one introduces a momentum scale $\mu\sim\mbox{few}\times
\Lambda_{\rm QCD}$ large enough for $\alpha_s(\mu)$ and power
corrections of order $(\Lambda_{\rm QCD}/\mu)^n$ to be small, but
otherwise as small as possible to suppress contributions from excited
states\cite{GrKo}. The result is\cite{KoNe} $\varrho_{\rm min}^2(\mu)
< \varrho^2 < \varrho_{\rm max}^2(\mu)$, where the boundary values
are shown in Fig.~\ref{fig:2} as a function of the scale $\mu$.
Assuming that the OPE works down to values $\mu\simeq 0.8$~GeV, one
obtains rather tight bounds for the slope parameters:
\begin{equation}\label{rhobounds}
   0.5 < \varrho^2 < 0.8 \,, \qquad
   0.5 < \widehat\varrho^2 < 1.1 \,.
\end{equation}
The allowed region for $\widehat\varrho^2$ has been increased in
order to account for the unknown $1/m_Q$ corrections in the relation
(\ref{rhorel}). The experimental result given in (\ref{VcbFraw})
falls inside this region.

These bounds compare well with theoretical predictions for the slope
parameters. QCD sum rules have been used to calculate the slope of
the Isgur--Wise function. The results obtained by various authors are
$\varrho^2 = 0.84\pm 0.02$ (Bagan et al.\cite{Baga}), $0.7\pm 0.1$
(Neubert\cite{twoloop}), $0.70\pm 0.25$ (Blok and
Shifman\cite{BlSh}), and $1.00\pm 0.02$ (Narison\cite{Nari}). The
UKQCD Collaboration has presented a lattice calculation of the slope
of the form factor ${\cal F}(w)$, yielding\cite{Lattrho}
$\widehat\varrho^2 = 0.9_{-0.3-0.2}^{+0.2+0.4}$. We stress that the
sum-rule bounds in (\ref{rhobounds}) are largely model independent;
model calculations in strong disagreement with these bounds should be
discarded.

\runninghead{Exclusive Semileptonic Decays}
 {Measurement of $\bar B\to D^*\ell\,\bar\nu$ Form Factors}
\subsection{Measurement of $\bar B\to D^*\ell\,\bar\nu$ Form Factors}
\noindent
If the lepton mass is neglected, the differential decay distributions
in $\bar B\to D^*\ell\,\bar\nu$ decays can be parametrized by three
helicity amplitudes, or equivalently by three independent
combinations of form factors. It has been suggested that a good
choice for three such quantities should be inspired by the
heavy-quark limit\cite{review,subl}. One thus defines a form factor
$h_{A1}(w)$, which up to symmetry-breaking corrections coincides with
the Isgur--Wise function, and two form-factor ratios
\begin{eqnarray}
   R_1(w) &=& \bigg[ 1 - {q^2\over(m_B+m_{D^*})^2} \bigg]\,
    {V(q^2)\over A_1(q^2)} \,, \nonumber\\
   R_2(w) &=& \bigg[ 1 - {q^2\over(m_B+m_{D^*})^2} \bigg]\,
    {A_2(q^2)\over A_1(q^2)} \,.
\end{eqnarray}
The relation between $w$ and $q^2$ has been given in (\ref{PVff}).
This definition is such that in the heavy-quark limit
$R_1(w)=R_2(w)=1$ independently of $w$.

To extract the functions $h_{A1}(w)$, $R_1(w)$ and $R_2(w)$ from
experimental data is a complicated task. However, HQET-based
calculations suggest that the $w$ dependence of the form-factor
ratios, which is induced by symmetry-breaking effects, is rather
mild\cite{subl}. Moreover, the form factor $h_{A1}(w)$ is expected to
have a nearly linear shape over the accessible $w$ range. This
motivates to introduce three parameters $\varrho_{A1}^2$, $R_1$ and
$R_2$ by
\begin{eqnarray}
   h_{A1}(w) &\simeq& {\cal F}(1)\,\Big[ 1 - \varrho_{A1}^2 (w-1)
    \Big] \,, \nonumber\\
   R_1(w) &\simeq& R_1 \,, \nonumber\\
   R_2(w) &\simeq& R_2 \,,
\end{eqnarray}
where ${\cal F}(1)=0.91\pm 0.03$ from (\ref{F1}). The CLEO
Collaboration has extracted these three parameters from an analysis
of the angular distributions in $\bar B\to D^*\ell\,\bar\nu$
decays\cite{CLEff}. The result is:
\begin{eqnarray}
   \varrho_{A1}^2 &=& 0.91\pm 0.15\pm 0.06 \,, \nonumber\\
   R_1 &=& 1.18\pm 0.30\pm 0.12 \,, \nonumber\\
   R_2 &=& 0.71\pm 0.22\pm 0.07 \,.
\end{eqnarray}
Using the HQET, one obtains an essentially model-independent
prediction for the symme\-try-breaking corrections to $R_1$, whereas
the corrections to $R_2$ are somewhat model dependent. To good
approximation\cite{review}
\begin{eqnarray}
   R_1 &\simeq& 1 + {4\alpha_s(m_c)\over 3\pi}
    + {\bar\Lambda\over 2 m_c}\simeq 1.3\pm 0.1 \,, \nonumber\\
   R_2 &\simeq& 1 - \kappa\,{\bar\Lambda\over 2 m_c}
    \simeq 0.8\pm 0.2 \,,
\end{eqnarray}
with $\kappa\simeq 1$ from QCD sum rules\cite{subl}. Here
$\bar\Lambda$ is the ``binding energy'' as defined in the mass
formula (\ref{massexp}). Theoretical calculations\cite{Lamsr1,Lamsr2}
as well as phenomenological analyses\cite{FLSa,Grem} suggest that
$\bar\Lambda\simeq 0.45$--0.65~GeV is the appropriate value to be
used in one-loop calculations. A quark-model calculation of $R_1$ and
$R_2$ gives results similar to the HQET predictions\cite{ClWa}:
$R_1\simeq 1.15$ and $R_2\simeq 0.91$. The experimental data confirm
the theoretical prediction that $R_1>1$ and $R_2<1$, although the
errors are still large.

There is a model-independent relation between the three parameters
determined from the analysis of angular distributions and the slope
parameter $\widehat\varrho^2$ extracted from the semileptonic
spectrum. It reads\cite{Vcbnew}
\begin{equation}
   \varrho_{A1}^2 - \widehat\varrho^2 = {1\over 6}\,(R_1^2-1)
   + {m_B\over 3(m_B-m_{D^*})}\,(1-R_2) \,.
\end{equation}
The CLEO data give $0.07\pm 0.20$ for the difference of the slope
parameters on the left-hand side, and $0.22\pm 0.18$ for the
right-hand side. Both values are compatible within errors.

The results of this analysis are very encouraging. Within errors, the
experiment confirms the HQET predictions, starting to test them at
the level of symmetry-breaking corrections.

\runninghead{Exclusive Semileptonic Decays}
 {Decays to Charmless Final States}
\subsection{Decays to Charmless Final States}
\noindent
For completeness, we will discuss briefly semileptonic $B$-meson
decays into charmless final states, although heavy-quark symmetry
does not help much in the analysis of these processes. Recently, the
CLEO Collaboration has reported a first signal for the exclusive
semileptonic decay modes $\bar B\to\pi\,\ell\,\bar\nu$ and $\bar
B\to\rho\,\ell\,\bar\nu$. The underlying quark process for these
transitions is $b\to u\,\ell\,\bar\nu$. Thus, these decays provide
information on the strength of the CKM matrix element $V_{ub}$. The
observed branching fractions are\cite{Bpirho}:
\begin{eqnarray}\label{CLEOVub}
   \mbox{B}(\bar B\to\pi\,\ell\,\bar\nu) &=& \cases{
    (1.34\pm 0.45)\times 10^{-4} ;&ISGW, \cr
    (1.63\pm 0.57)\times 10^{-4} ;&BSW, \cr} \nonumber\\
   \mbox{B}(\bar B\to\rho\,\ell\,\bar\nu) &=& \cases{
    (2.28_{-0.83}^{+0.69})\times 10^{-4} ;&ISGW, \cr
    (3.88_{-1.39}^{+1.15})\times 10^{-4} ;&BSW. \cr}
\end{eqnarray}
There is a significant model dependence in the simulation of the
reconstruction efficiencies, for which the models of Isgur et
al.\ (ISGW)\cite{ISGW} and Bauer et al.\ (BSW)\cite{BSW}
have been used.

\begin{table}[htb]
\centerline{\parbox{11.5cm}{\caption{\label{tab:Vub}
Values for $|V_{ub}/V_{cb}|$ extracted from the CLEO measurement of
exclusive semileptonic $B$ decays into charmless final states,
taking $|V_{cb}|=0.040$. An average over the experimental results in
(\protect\ref{CLEOVub}) is used for all except the ISGW and BSW
models, where the numbers corresponding to these models are used. The
first error quoted is experimental, the second (when available) is
theoretical.}}}
\vspace{0.5cm}
\centerline{\begin{tabular}{|c|c|c|c|}\hline\hline
\rule[-0.2cm]{0cm}{0.65cm} Method & Reference
 & $\bar B\to\pi\,\ell\,\bar\nu$ & $\bar B\to\rho\,\ell\,\bar\nu$ \\
\hline
\rule[-0.2cm]{0cm}{0.65cm} QCD sum rules & Narison\cite{SNar}
 & $0.159\pm 0.019\pm 0.001$ & $0.066_{-0.009}^{+0.007}\pm 0.003$ \\
\rule[-0.2cm]{0cm}{0.65cm} & Ball\cite{PBal}
 & $0.105\pm 0.013\pm 0.011$ & $0.094_{-0.012}^{+0.010}\pm 0.016$ \\
\rule[-0.2cm]{0cm}{0.65cm} & Khodj.\ \& R\"uckl\cite{Khod}
 & $0.085\pm 0.010$ & --- \\
\hline
\rule[-0.2cm]{0cm}{0.65cm} Lattice QCD & UKQCD\cite{UKQCDVub}
 & $0.103\pm 0.012_{-0.010}^{+0.012}$ & --- \\
\rule[-0.2cm]{0cm}{0.65cm} & APE\cite{APEVub}
 & $0.084\pm 0.010\pm 0.021$ & --- \\
\hline
\rule[-0.2cm]{0cm}{0.65cm} pQCD & Li \& Yu\cite{LiYu}
 & $0.054\pm 0.006$ & --- \\
\hline
\rule[-0.2cm]{0cm}{0.65cm} Quark models & BSW\cite{BSW}
 & $0.093\pm 0.016$ & $0.076_{-0.014}^{+0.011}$ \\
\rule[-0.2cm]{0cm}{0.65cm} & KS\cite{KS} & $0.088\pm 0.011$
 & $0.056_{-0.007}^{+0.006}$ \\
\rule[-0.2cm]{0cm}{0.65cm} & ISGW2\cite{ISGW2} & $0.074\pm 0.012$
 & $0.079_{-0.014}^{+0.012}$ \\
\hline\hline
\end{tabular}}
\end{table}

The theoretical description of these heavy-to-light ($b\to u$) decays
is more model dependent than that for heavy-to-heavy ($b\to c$)
transitions, because heavy-quark symmetry does not help to constrain
the relevant hadronic form factors. A variety of calculations for
such form factors exists, based on QCD sum rules, lattice gauge
theory, perturbative QCD, or quark models. Table~\ref{tab:Vub}
contains a summary of values extracted for the ratio
$|V_{ub}/V_{cb}|$ from a selection of such calculations. Some
approaches are more consistent than others in that the extracted
values are compatible for the two decay modes. With few exceptions,
the results lie in the range
\begin{equation}
   \left| {V_{ub}\over V_{cb}} \right|_{\rm excl}
   = 0.06\mbox{--}0.11 \,,
\end{equation}
which is in good agreement with the measurement of $|V_{ub}|$
obtained from the endpoint region of the lepton spectrum in inclusive
semileptonic decays\cite{btou1,btou2}:
\begin{equation}\label{Vubval}
   \left| {V_{ub}\over V_{cb}} \right|_{\rm incl}
   = 0.08\pm 0.01_{\rm exp}\pm 0.02_{\rm th} \,.
\end{equation}

Clearly, this is only the first step towards a more reliable
determination of $|V_{ub}|$; yet, with the discovery of exclusive
$b\to u$ transitions an important milestone has been met. Efforts
must now concentrate on more reliable methods to determine the form
factors for heavy-to-light transitions. Some new ideas in this
direction have been discussed recently. They are based on lattice
calculations\cite{Flynn}, analyticity constraints\cite{Boyd1,Lell},
or variants of the form-factor relations for heavy-to-heavy
transitions\cite{Stech}.


\runninghead{Inclusive Decay Rates and Lifetimes}
 {Inclusive Decay Rates and Lifetimes}
\section{Inclusive Decay Rates and Lifetimes}
\label{sec:4}
\noindent
Inclusive decay rates determine the probability of the decay of a
particle into the sum of all possible final states with a given set
of quantum numbers. An example is provided by the inclusive
semileptonic decay rate of the $B$ meson, $\Gamma(\bar B\to
X_c\,\ell\,\bar\nu)$, where the final state consists of a
lepton--neutrino pair accompanied by any number of hadrons with total
charm-quark number $n_c=1$. Here we shall discuss the
theoretical description of inclusive decays of hadrons containing a
heavy quark\cite{Chay}$^-$\cite{Fermi}. From the theoretical point of
view, such decays have two advantages: first, bound-state effects
related to the initial state (such as the ``Fermi motion'' of the
heavy quark inside the hadron) can be accounted for in a systematic
way using the heavy-quark expansion, in much the same way as
explained in the previous sections; secondly, the fact that the final
state consists of a sum over many hadronic channels eliminates
bound-state effects related to the properties of individual hadrons.
This second feature is based on a hypothesis known as quark--hadron
duality, which is an important concept in QCD phenomenology. The
assumption of duality is that cross sections and decay rates, which
are defined in the physical region (i.e.\ the region of time-like
momenta), are calculable in QCD after a ``smearing'' or ``averaging''
procedure has been applied\cite{PQW}. In semileptonic decays, it is
the integration over the lepton and neutrino phase space that
provides a ``smearing'' over the invariant hadronic mass of the final
state (so-called ``global'' duality). For non-leptonic decays, on the
other hand, the total hadronic mass is fixed, and it is only the fact
that one sums over many hadronic states that provides an
``averaging'' (so-called ``local'' duality). Clearly, local duality
is a stronger assumption than global duality. It is important to
stress that quark--hadron duality cannot yet be derived from first
principles, although it is a necessary assumption for many
applications of QCD. The validity of global duality has been tested
experimentally using data on hadronic $\tau$ decays\cite{Maria}. A
more formal attempt to address the problem of quark--hadron duality
can be found in Ref.\cite{Shifm}.

Using the optical theorem, the inclusive decay width of a hadron
$H_b$ containing a $b$ quark can be written in the form
\begin{equation}\label{ImT}
   \Gamma(H_b\to X) = {1\over 2 m_{H_b}}\,2\,\mbox{Im}\,
   \langle H_b|\,{\bf T}\,|H_b\rangle \,,
\end{equation}
where the transition operator ${\bf T}$ is given by the time-ordered
product of two effective Lagrangians:
\begin{equation}
   {\bf T} = i\!\int{\rm d}^4x\,T\{\,
   {\cal L}_{\rm eff}(x),{\cal L}_{\rm eff}(0)\,\} \,.
\end{equation}
In fact, inserting a complete set of states inside the time-ordered
product, we find the standard expression
\begin{equation}
   \Gamma(H_b\to X) = {1\over 2 m_{H_b}}\,\sum_X\,
   (2\pi)^4\,\delta^4(p_H-p_X)\,|\langle X|\,{\cal L}_{\rm eff}\,
   |H_b\rangle|^2
\end{equation}
for the decay rate. For the case of semileptonic and non-leptonic
decays, ${\cal L}_{\rm eff}$ is the effective weak Lagrangian given
in (\ref{LFermi}), which in practice is corrected for short-distance
effects\cite{cpcm1}$^-$\cite{cpcm5} arising from the exchange of
gluons with virtualities between $m_W$ and $m_b$. If some quantum
numbers of the final states $X$ are specified, the sum over
intermediate states is restricted appropriately. In the case of the
inclusive semileptonic decay rate, for instance, the sum would
include only those states $X$ containing a lepton--neutrino pair.

\begin{figure}[htb]
   \epsfxsize=7cm
   \centerline{\epsffile{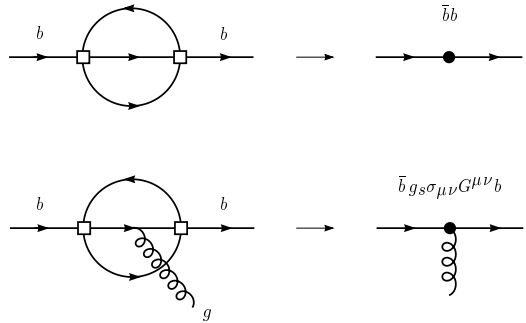}}
   \centerline{\parbox{11.5cm}{\caption{\label{fig:Toper}
Perturbative contributions to the transition operator ${\bf T}$
(left), and the corresponding operators in the OPE (right). The open
squares represent a four-fermion interaction of the effective
Lagrangian ${\cal L}_{\rm eff}$, while the black circles represent
local operators in the OPE.
   }}}
\end{figure}

In perturbation theory, some contributions to the transition operator
are given by the two-loop diagrams shown on the left-hand side in
Fig.~\ref{fig:Toper}. Because of the large mass of the $b$ quark, the
momenta flowing through the internal lines in these diagrams are
large. It is thus possible to construct an OPE for the transition
operator, in which ${\bf T}$ is represented as a series of local
operators containing the heavy-quark fields. The operator with the
lowest dimension, $d=3$, is $\bar b b$. It arises from integrating
over the internal lines in the first diagram shown in the figure. The
only gauge-invariant operator with dimension $d=4$ is $\bar
b\,i\rlap{\,/}D\,b$; however, the equation of motion implies that
between physical states this operator can be replaced by $m_b\bar b
b$. The first operator that is different from $\bar b b$ has
dimension $d=5$ and contains the gluon field. It is given by $\bar
b\,g_s\sigma_{\mu\nu} G^{\mu\nu} b$. This operator arises from
diagrams in which a gluon is emitted from one of the internal lines,
such as the second diagram shown in the figure.

For dimensional reasons, the matrix elements of such
higher-dimensional operators are suppressed by inverse powers of the
heavy-quark mass. Thus, any inclusive decay rate of a hadron $H_b$
can be written in the form\cite{Bigi}$^-$\cite{MaWe}:
\begin{equation}\label{gener}
   \Gamma(H_b\to X_f) = {G_F^2 m_b^5\over 192\pi^3}\,
   \bigg\{ c_3^f\,\langle\bar b b\rangle_H
   + c_5^f\,{\langle\bar b\,g_s\sigma_{\mu\nu} G^{\mu\nu} b
   \rangle_H\over m_b^2} + \dots \bigg\} \,,
\end{equation}
where the prefactor arises naturally from the loop integrations,
$c_n^f$ are calculable coefficient functions (which also contain the
relevant CKM matrix elements) depending on the quantum numbers $f$ of
the final state, and $\langle O\rangle_H$ are the (normalized)
forward matrix elements of local operators, for which we use the
short-hand notation
\begin{equation}
   \langle O\rangle_H = {1\over 2 m_{H_b}}\,\langle H_b|\,
   O\,|H_b\rangle \,.
\end{equation}

In the next step, these matrix elements are systematically expanded
in powers of $1/m_b$, using the technology of the HQET. Introducing
the velocity-dependent fields $b_v$ of the HQET, where $v$ denotes
the velocity of the hadron $H_b$, one finds\cite{FaNe,MaWe,Adam}
\begin{eqnarray}
   \langle\bar b b\rangle_H &=& 1
    - {\mu_\pi^2(H_b)-\mu_G^2(H_b)\over 2 m_b^2} + O(1/m_b^3) \,,
    \nonumber\\
   \langle\bar b\,g_s\sigma_{\mu\nu} G^{\mu\nu} b\rangle_H
   &=& 2\mu_G^2(H_b) + O(1/m_b) \,,
\end{eqnarray}
where we have defined the HQET matrix elements
\begin{eqnarray}
   \mu_\pi^2(H_b) &=& {1\over 2 m_{H_b}}\,
    \langle H_b(v)|\,\bar b_v\,(i\vec D)^2\,b_v\,|H_b(v)\rangle \,,
    \nonumber\\
   \mu_G^2(H_b) &=& {1\over 2 m_{H_b}}\,
    \langle H_b(v)|\,\bar b_v {g_s\over 2}\sigma_{\mu\nu}
     G^{\mu\nu} b_v\,|H_b(v)\rangle \,.
\end{eqnarray}
Here $(i\vec D)^2=(i v\cdot D)^2-(i D)^2$; in the rest frame, this is
the square of the operator for the
spatial momentum of the heavy quark. Inserting these results into
(\ref{gener}), we obtain
\begin{equation}\label{generic}
   \Gamma(H_b\to X_f) = {G_F^2 m_b^5\over 192\pi^3}\,
   \bigg\{ c_3^f\,\bigg( 1 - {\mu_\pi^2(H_b)\over 2 m_b^2} \bigg)
   + (4 c_5^f + c_3^f)\,{\mu_G^2(H_b)\over 2 m_b^2}
   + \dots \bigg\} \,.
\end{equation}
It is instructive to understand the appearance of the ``kinetic
energy'' contribution $\mu_\pi^2$, which is the gauge-covariant
extension of the square of the $b$-quark momentum inside the heavy
hadron. This contribution is the field-theory analogue of the Lorentz
factor $(1-\vec v_b^{\,2})^{1/2}\simeq 1-\vec p_b^{\,2}/2 m_b^2$, in
accordance with the fact that the lifetime, $\tau=1/\Gamma$, for a
moving particle increases due to time dilation.

The main result of the heavy-quark expansion for inclusive decay
rates is that the free quark decay (i.e.\ the parton model) provides
the first term in a systematic $1/m_b$ expansion, i.e.\
\begin{equation}
   \Gamma(H_b\to X_f) = {G_F^2 m_b^5\over 192\pi^3}\,c_3^f\,
   \Big\{ 1 + O(1/m_b^2) \Big\} \,.
\end{equation}
For dimensional reasons, the free-quark decay rate is proportional to
the fifth power of the $b$-quark mass. The non-perturbative
corrections to this picture, which arise from bound-state effects
inside the hadron $H_b$, are suppressed by (at least) two powers of
the heavy-quark mass, i.e.\ they are of relative order $(\Lambda_{\rm
QCD}/m_b)^2$. Note that the absence of first-order power corrections
is a simple consequence of the equation of motion, as there is no
independent gauge-invariant operator of dimension $d=4$ that could
appear in the OPE. The fact that bound-state effects in inclusive
decays are strongly suppressed explains {\it a posteriori\/} the
success of the parton model in describing such processes.

The hadronic matrix elements appearing in the heavy-quark expansion
(\ref{generic}) can be determined to some extent from the known
masses of heavy hadron states. For the $B$ meson, one finds
that\cite{FaNe}
\begin{eqnarray}\label{mupimuG}
   \mu_\pi^2(B) &=& - \lambda_1 = (0.4\pm 0.2)~\mbox{GeV}^2 \,,
    \nonumber\\
   \mu_G^2(B) &=& 3\lambda_2 = {3\over 4}\,(m_{B^*}^2 - m_B^2)
    \simeq 0.36~\mbox{GeV}^2 \,,
\end{eqnarray}
where $\lambda_1$ and $\lambda_2$ are the parameters appearing in the
mass formula (\ref{FNrela}). For the ground-state baryon $\Lambda_b$,
in which the light constituents have total spin zero, it follows that
\begin{equation}
   \mu_G^2(\Lambda_b) = 0 \,,
\end{equation}
while the matrix element $\mu_\pi^2(\Lambda_b)$ obeys the relation
\begin{equation}
   (m_{\Lambda_b}-m_{\Lambda_c}) - (\overline{m}_B-\overline{m}_D)
   = \Big[ \mu_\pi^2(B)-\mu_\pi^2(\Lambda_b) \Big]\,\bigg(
   {1\over 2 m_c} - {1\over 2 m_b} \bigg) + O(1/m_Q^2) \,,
\end{equation}
where $\overline{m}_B$ and $\overline{m}_D$ denote the spin-averaged
masses introduced in connection with (\ref{mbmc}). With the value of
$m_{\Lambda_b}$ given in (\ref{Lbmass}), this leads to
\begin{equation}\label{mupidif}
   \mu_\pi^2(B) - \mu_\pi^2(\Lambda_b) = (0.01\pm 0.03)~\mbox{GeV}^2
   \,.
\end{equation}
What remains to be calculated, then, is the coefficient functions
$c_n^f$ for a given inclusive decay channel. We shall now discuss the
most important applications of this general formalism.

\runninghead{Inclusive Decay Rates and Lifetimes}
 {Determination of $|V_{cb}|$ from Inclusive Semileptonic Decays}
\subsection{Determination of $|V_{cb}|$ from Inclusive Semileptonic
Decays}
\noindent
The extraction of $|V_{cb}|$ from the inclusive semileptonic decay
rate of the $B$ meson is based on the
expression\cite{Bigi}$^-$\cite{MaWe}
\begin{eqnarray}\label{Gamsl}
   \Gamma(\bar B\to X_c\,\ell\,\bar\nu)
   &=& {G_F^2 m_b^5\over 192\pi^3}\,|V_{cb}|^2\,\Bigg\{ \bigg(
    1 + {\lambda_1+3\lambda_2\over 2 m_b^2} \bigg)\,\bigg[
    f\bigg( {m_c\over m_b} \bigg)
    + {\alpha_s(M)\over\pi}\,g\bigg( {m_c\over m_b} \bigg) \bigg]
    \nonumber\\
   &&\phantom{ {G_F^2 m_b^5\over 192\pi^3}\,|V_{cb}|^2\,\Bigg\{ }
    - {6\lambda_2\over m_b^2}\,\bigg( 1 - {m_c^2\over m_b^2}
    \bigg)^4  + \dots \Bigg\} \,,
\end{eqnarray}
where $m_b$ and $m_c$ are the poles mass of the $b$ and $c$ quarks
(defined to a given order in perturbation theory), and $f(x)$ and
$g(x)$ are phase-space functions:
\begin{equation}
   f(x) = 1 - 8 x^2 + 8 x^6 - x^8 - 12 x^4\ln x^2 \,,
\end{equation}
and $g(x)$ is given elsewhere\cite{fgrefs}. The theoretical
uncertainties in this determination of $|V_{cb}|$ are quite
different from those entering the analysis of exclusive decays. In
particular, in inclusive decays there appear the quark masses rather
than the meson masses. Moreover, the theoretical description relies
on the assumption of global quark--hadron duality, which is not
necessary for exclusive decays. One should distinguish three sources
of theoretical uncertainties:

\paragraph{Hadronic parameters:}
The non-perturbative corrections are very small; with
$-\lambda_1=(0.4\pm 0.2)$~GeV$^2$ and $\lambda_2=0.12$~GeV$^2$, one
finds a reduction of the free-quark decay rate by $-(4.2\pm 0.5)\%$.
The uncertainty in this number is below 1\% and thus completely
negligible.

\begin{figure}[htb]
   \epsfysize=5cm
   \centerline{\epsffile{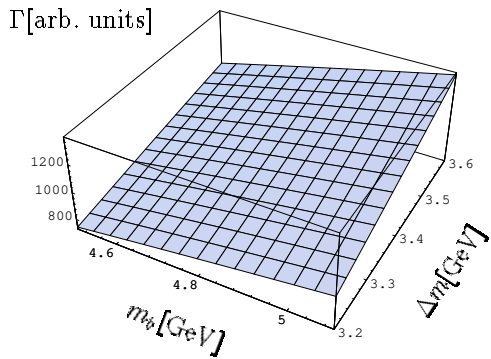}}
   \centerline{\parbox{11.5cm}{\caption{\label{fig:masses}
Dependence of the inclusive semileptonic decay rate on
$m_b$ and $\Delta m=m_b-m_c$.
   }}}
\end{figure}

\paragraph{Quark-mass dependence:}
The fact that $\Gamma\sim m_b^5\,f(m_c/m_b)$ suggests a strong
dependence of the decay rate on the value of the $b$-quark mass.
However, this dependence becomes milder if one chooses $m_b$ and
$\Delta m=m_b-m_c$ as independent variables. This is apparent from
Fig.~\ref{fig:masses}, which shows that $\Gamma\sim m_b^{2.3}\,\Delta
m^{2.7}$. This choice of variables is also preferred from a
conceptual point of view, since it leads to essentially uncorrelated
theoretical uncertainties: whereas $m_b=m_B-\bar\Lambda+\dots$ is
mainly determined by the $\bar\Lambda$ parameter of the HQET, the
mass difference $\Delta m$ obeys the expansion shown in (\ref{mbmc}),
i.e.\ it is sensitive to the kinetic-energy parameter $\lambda_1$.
Theoretical uncertainties of 60~MeV on $\Delta m$ and 200~MeV on
$m_b$ are reasonable; values much smaller than this are probably too
optimistic. This leads to
\begin{equation}
   \bigg( {\delta\Gamma\over\Gamma} \bigg)_{\rm masses}
   = \sqrt{ \bigg( 0.10\,{\delta m_b\over 200~\mbox{MeV}} \bigg)^2
   + \bigg( 0.05\,{\delta\Delta m\over 60~\mbox{MeV}} \bigg)^2 }
   \simeq 11\% \,.
\end{equation}

\paragraph{Perturbative corrections:}
The perturbative corrections are the most subtle part of the
analysis. The semileptonic rate is known exactly to order $\alpha_s$
only\cite{fgrefs}, although a partial calculation of the coefficient
of order $\alpha_s^2$ exists\cite{LSW}. The result is
\begin{equation}
   {\Gamma\over\Gamma_{\rm tree}} = 1 - 1.67\,{\alpha_s(m_b)\over\pi}
   - (1.68\beta_0+\dots)\,\bigg( {\alpha_s(m_b)\over\pi} \bigg)^2
   + \dots \,.
\end{equation}
The one-loop correction is moderate; it amounts to about $-11\%$. Of
the two-loop coefficient, only the part proportional to the
$\beta$-function coefficient $\beta_0$ is known. For $n_f=3$ light
quark flavours, this term gives $1.68\beta_0\simeq 15.1$,
corresponding to a rather large correction of about $-6\%$. One may
take this as an estimate of the perturbative uncertainty. The
dependence of the result on the choice of the renormalization scale
and scheme has been investigated, too, and found to be of
order\cite{BaNi} 6\%. Yet, the actual perturbative uncertainty may be
larger than that. A subset of higher-order corrections, the so-called
``renormalon-chain'' contributions of the form
$\beta_0^{n-1}\alpha_s^n$, can be summed to all orders in
perturbation theory, leading to\cite{BaBB} $\Gamma/\Gamma_{\rm
tree}=0.77\pm 0.05$, which is equivalent to choosing the rather low
scale $M\simeq 1$~GeV in (\ref{Gamsl}). This estimate is 12\% lower
than the one-loop result. These considerations show that there are
substantial perturbative uncertainties in the calculation of the
semileptonic decay rate. They could only be reduced with a complete
two-loop calculation, which is however quite a formidable task. At
present, we consider $(\delta\Gamma/\Gamma)_{\rm pert}\simeq 10\%$ a
reasonable estimate.

Adding, as previously, the theoretical errors linearly, and taking
the
square root, leads to
\begin{equation}
   {\delta|V_{cb}|\over|V_{cb}|} \simeq 10\%
\end{equation}
for the theoretical uncertainty in the determination of $|V_{cb}|$
from inclusive decays, keeping in mind that this method relies in
addition on the assumption of global quark--hadron duality. Taking
the result of Ball et al.\cite{BaBB} for the central value, we quote
\begin{equation}
   |V_{cb}| = (0.0400\pm 0.0040)\,\bigg( {B_{\rm SL}\over 10.9\%}
   \bigg)^{1/2}\,\bigg( {\tau_B\over 1.6~\mbox{ps}}
   \bigg)^{-1/2} \,.
\end{equation}
With the new world averages for the semileptonic branching ratio,
$B_{\rm SL}=(10.90\pm 0.46)\%$ (see below), and for the average
$B$-meson lifetime\cite{Joe}, $\tau_B=(1.60\pm 0.03)$~ps, we
obtain
\begin{equation}
   |V_{cb}| = (40.0\pm 0.9_{\rm exp}\pm 4.0_{\rm th})
   \times 10^{-3} \,.
\end{equation}
This is in excellent agreement with the value in (\ref{Vcbexc}),
which has been extracted from the analysis of the exclusive decay
$\bar B\to D^*\ell\,\bar\nu$. This agreement is gratifying given the
differences of the methods used, and it provides an indirect test of
global quark--hadron duality. Combining the two measurements gives
the final result
\begin{equation}
   |V_{cb}| = 0.039\pm 0.002 \,.
\end{equation}
After $V_{ud}$ and $V_{us}$, this is now the third-best known entry
in the CKM matrix.

\runninghead{Inclusive Decay Rates and Lifetimes}
 {Semileptonic Branching Ratio and Charm Counting}
\subsection{Semileptonic Branching Ratio and Charm Counting}
\noindent
The semileptonic branching ratio of $B$ mesons is defined as
\begin{equation}
   B_{\rm SL} = {\Gamma(\bar B\to X\,e\,\bar\nu)\over
   \sum_\ell \Gamma(\bar B\to X\,\ell\,\bar\nu) + \Gamma_{\rm NL}
   + \Gamma_{\rm rare}} \,,
\end{equation}
where $\Gamma_{\rm NL}$ and $\Gamma_{\rm rare}$ are the inclusive
rates for non-leptonic and rare decays, respectively. The main
difficulty in calculating $B_{\rm SL}$ is not in the semileptonic
width, but in the non-leptonic one. As mentioned above, the
calculation of non-leptonic decays in the heavy-quark expansion
relies on the strong assumption of local quark--hadron duality.

Measurements of the semileptonic branching ratio have been performed
by various experimental groups, using both model-dependent and
model-in\-de\-pen\-dent analyses. The status of the results is
controversial, as there is a discrepancy between low-energy
measurements performed at the $\Upsilon(4s)$ resonance and
high-energy measurements performed at the $Z^0$ resonance. The
average value at low energies is\cite{Tomasz} $B_{\rm SL}=(10.37\pm
0.30)\%$. High-energy measurements performed at LEP, on the other
hand, give\cite{Pascal} $B_{\rm SL}^{(b)}=(11.11\pm 0.23)\%$. The
superscript $(b)$ indicates that this value refers not to the $B$
meson, but to a mixture of $b$ hadrons (approximately 40\% $B^-$,
40\% $\bar B^0$, 12\% $B_s$, and 8\% $\Lambda_b$). Assuming that the
corresponding semileptonic width $\Gamma_{\rm SL}^{(b)}$ is close to
that of the $B$ meson\footnote{Theoretically, this is expected to be
a very good approximation.}, we can correct for this and find $B_{\rm
SL}=(\tau(B)/\tau(b))\, B_{\rm SL}^{(b)}=(11.30\pm 0.26)\%$, where
$\tau(b)=(1.57\pm 0.03)$~ps is the average lifetime corresponding to
the above mixture of $b$ hadrons\cite{Joe}. The discrepancy between
the low-energy and high-energy measurements of the semileptonic
branching ratio is therefore larger than 3 standard deviations. If we
take the average and inflate the error to account for this disturbing
fact, we obtain
\begin{equation}\label{Bslval}
   B_{\rm SL} = (10.90\pm 0.46)\% \,.
\end{equation}
In understanding this result, an important aspect is charm counting,
i.e.\ the measurement of the average number $n_c$ of charm hadrons
produced per $B$ decay. Recently, two new (preliminary) measurements
of this quantity have been performed. The CLEO Collaboration has
presented the value\cite{Tomasz,ncnew} $n_c=1.16\pm 0.05$, and the
ALEPH Collaboration has reported the result\cite{ALEPHnc}
$n_c=1.20\pm 0.08$. The average is
\begin{equation}\label{ncval}
   n_c = 1.17\pm 0.04 \,.
\end{equation}

In the parton model, one finds\cite{Alta} $B_{\rm SL}\simeq 13\%$ and
$n_c\simeq 1.15$. Whereas $n_c$ is in agreement with experiment, the
semileptonic branching ratio is predicted to be too large. With the
establishment of the $1/m_Q$ expansion the non-perturbative
corrections to the parton model could be computed, and they turned
out to be too small to improve the prediction. This led Bigi et al.\
to conclude that values $B_{\rm SL}<12.5\%$ cannot be accommodated by
theory, thus giving rise to a puzzle referred to as the ``baffling
semileptonic branching ratio''\cite{baff}. The situation has changed
recently, however, when it was shown that higher-order perturbative
corrections lower the value of $B_{\rm SL}$
significantly\cite{BSLnew1}. The exact order-$\alpha_s$ corrections
to the non-leptonic width have been computed for $m_c\ne 0$, and an
analysis of the renormalization scale and scheme dependence has been
performed. In particular, it turns out that radiative corrections
increase the partial width $\Gamma(\bar B\to X_{c\bar c s})$ by a
large amount. This has two effects: it lowers the semileptonic
branching ratio, but at the price of a higher value of $n_c$.

\begin{figure}[htb]
   \epsfxsize=6.5cm
   \centerline{\epsffile{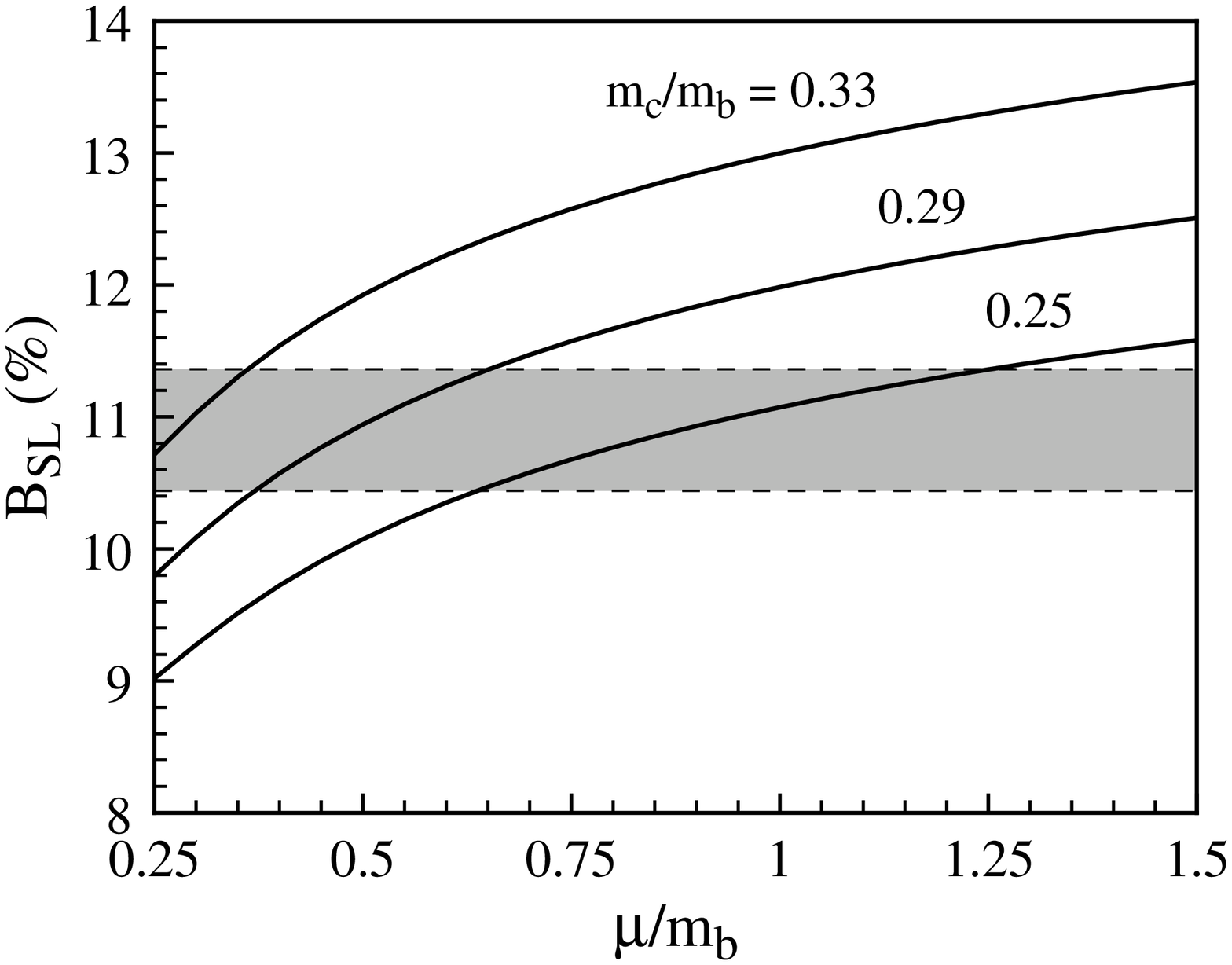}}
   \epsfxsize=6.5cm
   \centerline{\epsffile{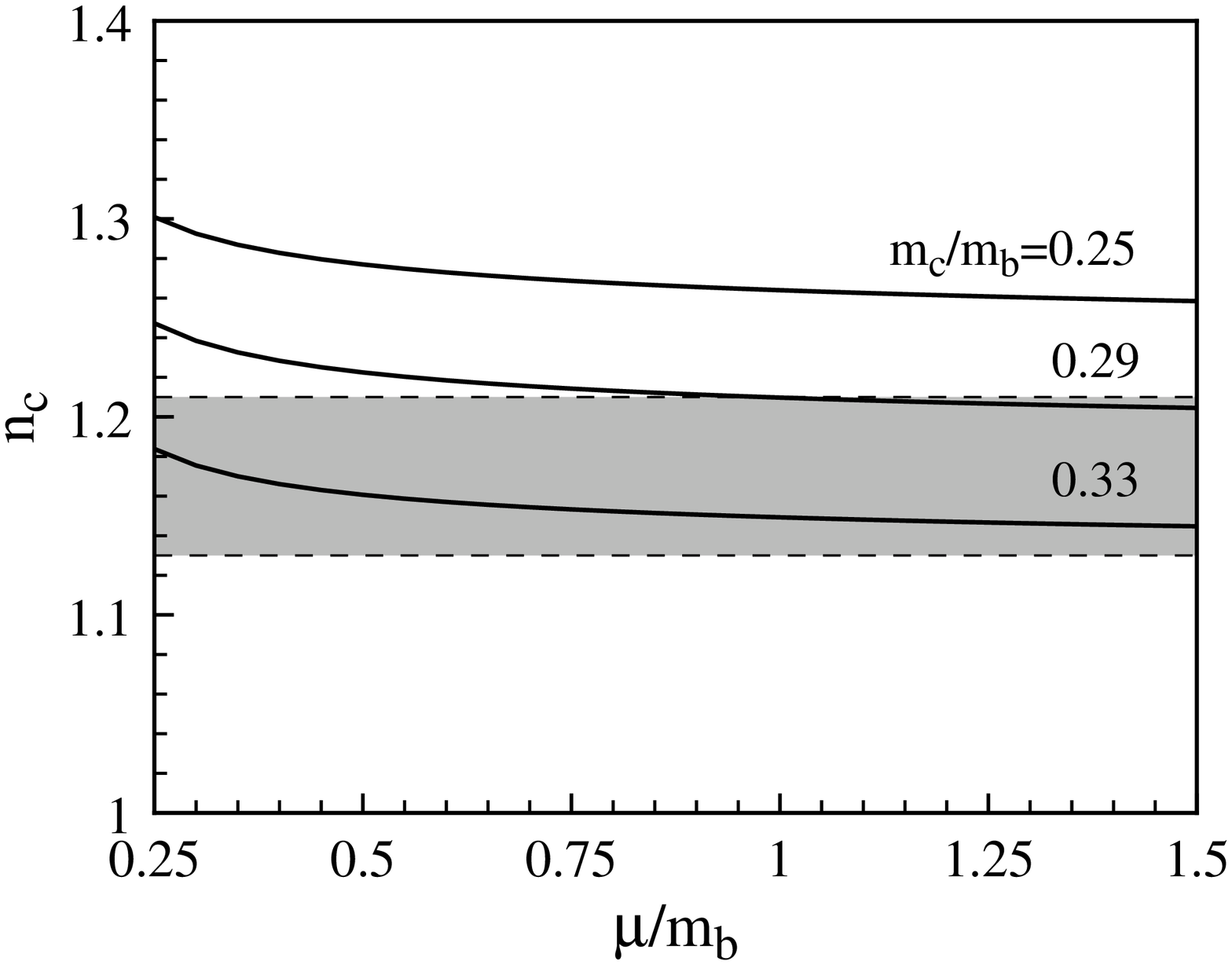}}
   \centerline{\parbox{14cm}{\caption{\label{fig:mudep}
Scale dependence of the theoretical predictions for the semileptonic
branching ratio and $n_c$. The bands show the average experimental
values given in (\protect\ref{Bslval}) and (\protect\ref{ncval}).
   }}}
\end{figure}

The original analysis of Bagan et al.\ has recently been corrected in
an erratum\cite{BSLnew1}. Here we shall present the results of an
independent numerical analysis using the same theoretical input (for
a detailed discussion, see Ref.\cite{Chris}). The semileptonic
branching ratio and $n_c$ depend on the quark-mass ratio $m_c/m_b$
and on the ratio $\mu/m_b$, where $\mu$ is the scale used to
renormalize the coupling constant $\alpha_s(\mu)$ and the Wilson
coefficients appearing in the non-leptonic decay rate. The freedom in
choosing the scale $\mu$ reflects our ignorance of higher-order
corrections, which are neglected when the perturbative expansion is
truncated at order $\alpha_s$. Below we shall consider several
choices for the renormalization scale. We allow the pole masses of
the heavy quarks to vary in the range [see (\ref{mbmcval})]
\begin{equation}
   m_b = (4.8\pm 0.2)~\mbox{GeV} \,, \qquad
   m_b - m_c = (3.40\pm 0.06)~\mbox{GeV} \,,
\end{equation}
corresponding to $0.25<m_c/m_b<0.33$. Non-perturbative effects
appearing at order $1/m_b^2$ in the heavy-quark expansion are
described by the single parameter $\lambda_2\simeq 0.12~\mbox{GeV}^2$
defined in (\ref{mupimuG}); the dependence on the parameter
$\lambda_1$ is the same for all inclusive decay rates and cancels out
in $B_{\rm SL}$ and $n_c$. For the two choices $\mu=m_b$ and
$\mu=m_b/2$, we obtain
\begin{eqnarray}
   B_{\rm SL} &=& \cases{
    12.0\pm 1.0 \% ;& $\mu=m_b$, \cr
    10.9\pm 0.9 \% ;& $\mu=m_b/2$, \cr} \nonumber\\
   \phantom{ \bigg[ }
   n_c &=& \cases{
    1.21\mp 0.06 ;& $\mu=m_b$, \cr
    1.22\mp 0.06 ;& $\mu=m_b/2$. \cr}
\end{eqnarray}
The uncertainties in the two quantities, which result from the
variation of $m_c/m_b$ in the range given above, are anticorrelated.
Notice that the semileptonic branching ratio has a stronger scale
dependence than $n_c$. This is illustrated in Fig.~\ref{fig:mudep},
which shows the two quantities as a function of $\mu$. By choosing a
low renormalization scale, values $B_{\rm SL}<12\%$ can easily be
accommodated. The experimental data prefer a scale $\mu/m_b\sim 0.5$,
which is indeed not unnatural. Using the BLM scale setting
method\cite{BLM}, Luke et al.\ have estimated that $\mu\gsim 0.32
m_b$ is an appropriate scale in this case\cite{LSW}.

\begin{figure}[htb]
   \epsfxsize=7cm
   \centerline{\epsffile{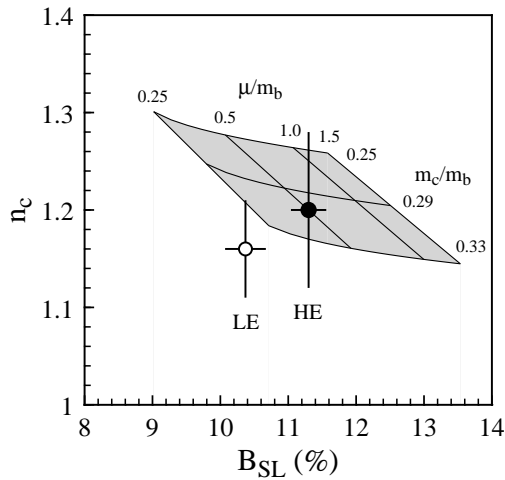}}
   \centerline{\parbox{14cm}{\caption{\label{fig:BSL}
Combined theoretical predictions for the semileptonic branching ratio
and charm counting as a function of the quark-mass ratio $m_c/m_b$
and the renormalization scale $\mu$. The data points show the average
experimental values for $B_{\rm SL}$ and $n_c$ obtained in low-energy
(LE) and high-energy (HE) measurements, as discussed in the text.
   }}}
\end{figure}

The combined theoretical predictions for the semileptonic branching
ratio and charm counting are shown in Fig.~\ref{fig:BSL}. They are
compared with the experimental results obtained from low- and
high-energy measurements. It was argued that the combination of a low
semileptonic branching ratio and a low value of $n_c$ would
constitute a potential problem for the Standard Model\cite{Buch}.
However, with the new experimental and theoretical numbers, only for
the low-energy measurements a small discrepancy remains between
theory and experiment.

Previous attempts to resolve the ``problem of the semileptonic
branching ratio'' have focused on four possibilities:

\begin{itemize}
\item
It has been argued that the experimental value of $n_c$ may depend on
model assumptions about the production of charm hadrons, which are
sometimes questionable\cite{Buch,FDW}.
\item
It has been pointed out that the assumption of local quark--hadron
duality could fail in non-leptonic $B$
decays\cite{PaSt}$^-$\cite{BGM}. If so, this will most likely happen
in the channel $b\to c\bar c s$, where the energy release, $E=m_B -
m_{X(c\bar cs)}$, is of order 1.5~GeV or less. However, if one
assumes that sizeable duality violations occur only in this channel,
it is impossible to improve the agreement between theory and
experiment\cite{Beijing}.
\item
Another possibility is that higher-order corrections in the $1/m_b$
expansion, which were previously thought to be negligible, give a
sizeable contribution. As will be discussed in more detail below,
certain corrections involving the participation of a spectator quark
are enhanced by phase space, so that they lead to effects of relative
size\cite{Beijing} $16\pi^2 (\Lambda_{\rm QCD}/m_b)^3$ rather than
$(\Lambda_{\rm QCD}/m_b)^3$. They could lower the semileptonic
branching ratio by up to 1\%, depending on the size of some hadronic
matrix elements\cite{Chris}. Lattice calculations could help to
confirm or rule out this possibility.
\item
Finally, there is also the possibility to invoke New
Physics\cite{Grza}$^-$\cite{CGG}. One may, for instance, consider
extensions of the Standard Model with enhanced flavour-changing
neutral currents such as $b\to s\,g$. The effect of such a
contribution would be that both $B_{\rm SL}$ and $n_c$ are reduced by
a factor $(1+\eta B_{\rm SL}^{\rm SM})^{-1}$, where
$\eta=(\Gamma_{\rm rare}-\Gamma_{\rm rare}^{\rm SM})/ \Gamma_{\rm
SL}$. To obtain a sizeable decrease requires values $\eta\sim
0.5$, which are large (in the Standard Model, $\Gamma_{\rm
rare}^{\rm SM}/ \Gamma_{\rm SL}\sim 0.1$), but not excluded by
current experiments.
\end{itemize}

For completeness, we briefly discuss the semileptonic branching ratio
for $B$ decays into a $\tau$ lepton, which is suppressed by phase
space. The ratio of the semileptonic rates for decays into $\tau$
leptons and into electrons can be calculated reliably. The result
is\cite{incltau}
\begin{equation}
   {B(\bar B\to X\,\tau\,\bar\nu_\tau)\over
    B(\bar B\to X\,e\,\bar\nu_\tau)} = 0.22\pm 0.02 \,.
\end{equation}
This ratio has been measured at LEP and is found to be\cite{Tomasz}
\begin{equation}
   {B(\bar B\to X\,\tau\,\bar\nu_\tau)\over
    B(\bar B\to X\,e\,\bar\nu_\tau)} = 0.234\pm 0.029 \,,
\end{equation}
in good agreement with the theoretical prediction.

\runninghead{Inclusive Decay Rates and Lifetimes}
 {Lifetime Ratios of $b$ Hadrons}
\subsection{Lifetime Ratios of $b$ Hadrons}
\noindent
The heavy-quark expansion shows that the lifetimes of all hadrons
containing a $b$ quark agree up to non-perturbative corrections
suppressed by at least two powers of $1/m_b$. In particular, it
predicts that
\begin{eqnarray}\label{taucrude}
   {\tau(B^-)\over\tau(B^0)} &=& 1 + O(1/m_b^3) \,,
    \nonumber\\
   {\tau(B_s)\over\tau(B_d)} &=& (1.00\pm 0.01) + O(1/m_b^3) \,,
    \nonumber\\
   {\tau(\Lambda_b)\over\tau(B^0)} &=& 1
    + {\mu_\pi^2(\Lambda_b)-\mu_\pi^2(B)\over 2 m_b^2}
    - c_G\,{\mu_G^2(B)\over m_b^2} + O(1/m_b^3) \nonumber\\
   &\simeq& 0.98 + O(1/m_b^3) \,,
\end{eqnarray}
where $c_G\simeq 1.1$, and we have used (\ref{mupimuG}) and
(\ref{mupidif}). The uncertainty in the value of the ratio
$\tau(B_s)/\tau(B_d)$ arises from unknown SU(3)-violating effects in
the matrix elements of $B_s$ mesons. The above theoretical
predictions may be compared with the average experimental values for
the lifetime ratios, which are\cite{Joe,CDFtb}:
\begin{eqnarray}\label{taudata}
   {\tau(B^-)\over\tau(B^0)} &=& 1.02\pm 0.04 \,, \nonumber\\
   {\tau(B_s)\over\tau(B_d)} &=& 1.01 + 0.07 \,,
    \nonumber\\
   {\tau(\Lambda_b)\over\tau(B^0)} &=& 0.78\pm 0.05 \,.
\end{eqnarray}
Whereas the lifetime ratios of the different $B$ mesons are in good
agreement with the theoretical prediction, the low value of the
lifetime of the $\Lambda_b$ baryon is surprising.

\begin{figure}[htb]
   \epsfxsize=7cm
   \centerline{\epsffile{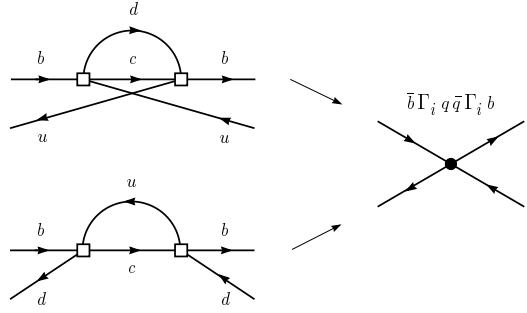}}
   \centerline{\parbox{11.5cm}{\caption{\label{fig:Tspec}
Spectator contributions to the transition operator ${\bf T}$ (left),
and the corresponding operators in the OPE (right). Here $\Gamma_i$
denotes some combination of Dirac and colour matrices.
   }}}
\end{figure}

To understand the structure of the lifetime differences requires to
go further in the $1/m_b$ expansion\cite{liferef}. Although at first
sight it appears that higher-order corrections could be safely
neglected given the smallness of the $1/m_b^2$ corrections, this
impression is erroneous for two reasons: first, at order $1/m_b^3$
in the heavy-quark expansion for non-leptonic decay rates there
appear four-quark operators, whose matrix elements explicitly depend
on the flavour of the spectator quark(s) in the hadron $H_b$, and
hence are responsible for lifetime differences between hadrons with
different light constituents; secondly, these spectator effects
receive a phase-space enhancement factor of $16\pi^2$ with respect to
the leading terms in the OPE\cite{Beijing}. This can be seen from
Fig.~\ref{fig:Tspec}, which shows that the corresponding
contributions to the transition operator ${\bf T}$ arise from
one-loop rather than two-loop diagrams. The presence of this
phase-space enhancement factor leads to a peculiar structure of the
heavy-quark expansion for non-leptonic rates, which may be displayed
as follows:
\begin{eqnarray}
   \Gamma &=& \Gamma_0\,\Bigg\{ 1
    + x_2 \bigg( {\Lambda_{\rm QCD}\over m_b} \bigg)^2
    + x_3 \bigg( {\Lambda_{\rm QCD}\over m_b} \bigg)^3 + \dots
    \nonumber\\
   &&\mbox{}+ 16\pi^2\,\bigg[
    y_3 \bigg( {\Lambda_{\rm QCD}\over m_b} \bigg)^3
    + y_4 \bigg( {\Lambda_{\rm QCD}\over m_b} \bigg)^4 + \dots
    \bigg] \Bigg\} \,.
\end{eqnarray}
Here $x_n$ and $y_n$ are coefficients of order unity. It is
conceivable that the terms of order $16\pi^2\,(\Lambda_{\rm
QCD}/m_b)^3$ could be larger than the ones of order $(\Lambda_{\rm
QCD}/m_b)^2$. It is thus important to include this type of
corrections to all predictions for non-leptonic rates. Moreover,
there is a challenge to calculate the hadronic matrix elements of the
corresponding four-quark operators with high accuracy. Lattice
calculations could help to improve the existing estimates of such
matrix elements.

In total, a set of four four-quark operators is induced by spectator
effects. They are:
\begin{eqnarray}
   O_{\rm V-A}^q &=& \bar b\,\gamma_\mu (1-\gamma_5)\,q\,
    \bar q\,\gamma^\mu (1-\gamma_5)\,b \,, \nonumber\\
   O_{\rm S-P}^q &=& \bar b\,(1-\gamma_5)\,q\,
    \bar q\,(1+\gamma_5)\,b \,, \nonumber\\
   T_{\rm V-A}^q &=& \bar b\,\gamma_\mu (1-\gamma_5)\,t_a\,q\,
    \bar q\,\gamma^\mu (1-\gamma_5)\,t_a\,b \,, \nonumber\\
   T_{\rm S-P}^q &=& \bar b\,(1-\gamma_5)\,t_a\,q\,
    \bar q\,(1+\gamma_5)\,t_a\,b \,,
\end{eqnarray}
where $q$ is a light quark, and $t_a$ are the generators of colour
SU(3). In most previous analyses of spectator effects the hadronic
matrix elements of these operators have been estimated making
simplifying assumptions\cite{liferef}$^-$\cite{ShiV}. For the matrix
elements between $B$-meson states the vacuum saturation
approximation\cite{SVZ} was assumed, i.e.\ the matrix elements of the
four-quark operators have been evaluated by inserting the vacuum
inside the current products. This leads to
\begin{eqnarray}\label{fact}
   \langle\bar B_q|\,O_{\rm V-A}^q\,|\bar B_q\rangle
   &=& \langle\bar B_q|\,O_{\rm S-P}^q\,|\bar B_q\rangle
    = f_{B_q}^2 m_{B_q}^2 \,, \nonumber\\
   \langle\bar B_q|\,T_{\rm V-A}^q\,|\bar B_q\rangle
   &=& \langle\bar B_q|\,T_{\rm S-P}^q\,|\bar B_q\rangle = 0 \,,
\end{eqnarray}
where $f_{B_q}$ is the decay constant of the $B_q$ meson, which is
defined as
\begin{equation}
   \langle 0\,|\,\bar q\,\gamma^\mu\gamma_5\,b\,
   |\bar B_q(v)\rangle = i f_{B_q} m_{B_q} v^\mu \,.
\end{equation}
This approach has been criticized by Chernyak\cite{Chern}, who
estimates that corrections to the vacuum saturation approximation can
be as large as 50\%.

An unbiased analysis of spectator effects, which avoids assumptions
about had\-ro\-nic matrix elements, can be performed if instead of
(\ref{fact}) one defines\cite{Chris}
\begin{eqnarray}\label{Biepsi}
   \langle\bar B_q|\,O_{\rm V-A}^q\,|\bar B_q\rangle
   &=& B_1\,f_{B_q}^2 m_{B_q}^2 \,, \nonumber\\
   \langle\bar B_q|\,O_{\rm S-P}^q\,|\bar B_q\rangle
   &=& B_1\,f_{B_q}^2 m_{B_q}^2 \,, \nonumber\\
   \langle\bar B_q|\,T_{\rm V-A}^q\,|\bar B_q\rangle
   &=& \varepsilon_1\,f_{B_q}^2 m_{B_q}^2 \,, \nonumber\\
   \langle\bar B_q|\,T_{\rm S-P}^q\,|\bar B_q\rangle
   &=& \varepsilon_2\,f_{B_q}^2 m_{B_q}^2 \,.
\end{eqnarray}
The values of the dimensionless hadronic parameters $B_i$ and
$\varepsilon_i$ are currently not known; ultimately, they may be
calculated using some field-theoretic approach such as lattice gauge
theory or QCD sum rules. The vacuum saturation approximation
corresponds to setting $B_i=1$ and $\varepsilon_i=0$ (at some scale
$\mu$, where the approximation is believed to be valid). For real
QCD, however, it is known that
\begin{equation}
   B_i = O(1) \,,\qquad \varepsilon_i=O(1/N_c) \,,
\end{equation}
where $N_c$ is the number of colours. Below, we shall treat $B_i$ and
$\varepsilon_i$ (renormalized at the scale $m_b$) as unknown
parameters. Similarly, the relevant hadronic matrix elements of the
four-quark operators between $\Lambda_b$-baryon states can be
parametrized by two parameters, $\widetilde B$ and $r$, where
$\widetilde B=1$ in the valence-quark approximation, in which the
colour of the quark fields in the operators is identified with the
colour of the quarks inside the baryon\cite{Chris}.

\subsubsection{Lifetime ratio for $B^-$ and $B^0$}
\begin{figure}[htb]
   \epsfxsize=9cm
   \centerline{\epsffile{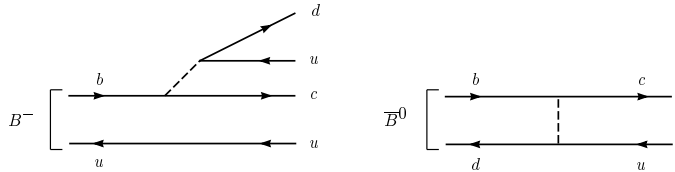}}
   \centerline{\parbox{11.5cm}{\caption{\label{fig:Wint}
Pauli interference and $W$ exchange contributions to the lifetimes of
the $B^-$ and the $\bar B^0$ mesons. The spectator effect in the
first diagram arises from the interference due to the presence of two
identical $\bar u$ quarks in the final state.
   }}}
\end{figure}

\noindent
The lifetimes of the charged and neutral $B$ mesons differ because of
two types of spectator effects illustrated in Fig.~\ref{fig:Wint}.
They are referred to as Pauli interference and $W$
exchange\cite{Gube}$^-$\cite{ShiV}. In the operator language, these
effects are represented by the hadronic matrix elements of the local
four-quark operators given in (\ref{Biepsi}). In fact, the diagrams
in Fig.~\ref{fig:Wint} can be obtained from those in
Fig.~\ref{fig:Tspec} by cutting the internal lines, which corresponds
to taking the imaginary part in (\ref{ImT}).

The explicit calculation of these spectator effects leads
to\cite{Chris}
\begin{equation}\label{DeltaGam}
   \Delta\Gamma_{\rm spec}(B_q) = {G_F^2 m_b^5\over 192\pi^3}\,
   |V_{cb}|^2\,16\pi^2\,{f_B^2\,m_B\over m_b^3}\,\zeta_{B_q} \,,
\end{equation}
where
\begin{equation}
   \zeta_{B^-}\simeq -0.4 B_1 + 6.6\varepsilon_1 \,, \qquad
   \zeta_{B^0}\simeq -2.2\varepsilon_1 + 2.4\varepsilon_2 \,.
\end{equation}
Note the factor of $16\pi^2$ in (\ref{DeltaGam}), which arises from
the phase-space enhancement of spectator effects. Given that the
parton-model result for the total decay width is
\begin{equation}
   \Gamma_{\rm tot}(B)\simeq 3.7\times
   {G_F^2 m_b^5\over 192\pi^3}\,|V_{cb}|^2 \,,
\end{equation}
we see that the characteristic scale of spectator contributions is
\begin{equation}
   4\pi^2\,{f_B^2\,m_B\over m_b^3}\simeq
   \bigg( {2\pi f_B\over m_b} \bigg)^2 \simeq 5\% \,.
\end{equation}
Thus, it is natural that the lifetimes of different $b$ hadrons
differ by a few per cent.

The precise value of the lifetime ratio depends crucially on the size
of the hadronic matrix elements. Taking $f_B=200$~MeV for the decay
constant of the $B$ meson (see Ref.\cite{review} and references
therein), i.e.\ absorbing the uncertainty in this parameter into the
definition of $B_i$ and $\varepsilon_i$, leads to\cite{Chris}
\begin{equation}
   {\tau(B^-)\over\tau(B^0)} \simeq 1 + 0.03 B_1
   - 0.71\varepsilon_1 + 0.20\varepsilon_2 \,.
\end{equation}
The most striking feature of this result is that the coefficients of
the colour-octet operators $T_{\rm V-A}$ and $T_{\rm S-P}$ are orders
of magnitude larger than those of the colour-singlet operator $O_{\rm
V-A}$. As a consequence, the vacuum insertion approximation, which
was adopted in Ref.\cite{liferef} to predict that
$\tau(B^-)/\tau(B_d)$ is larger than unity by an amount of order 5\%,
cannot be trusted. With $\varepsilon_i$ of order $1/N_c$, it is
conceivable that the non-factorizable contributions actually dominate
the result. Thus, without a detailed calculation of the parameters
$\varepsilon_i$ no reliable prediction can be obtained. Given our
present ignorance about the true values of the hadronic matrix
elements, we must conclude that even the sign of the sum of the
spectator contributions cannot be predicted. A lifetime ratio in the
range $0.8<\tau(B^-)/\tau(B^0)<1.2$ could be easily accommodated by
theory.

In view of these considerations, the experimental fact that the
lifetime ratio turns out to be very close to unity is somewhat of a
surprise. It implies a constraint on a certain combination of the
colour-octet matrix elements, which reads
\begin{equation}\label{epsbound}
   \varepsilon_1 - 0.3\varepsilon_2 = \mbox{few~\%}.
\end{equation}

\subsubsection{Lifetime ratio for $B_s$ and $B_d$}
\noindent
The lifetimes of the two neutral mesons $B_s$ and $B_d$ differ
because spectator effects depend on the flavour of the light quark,
and moreover because the hadronic matrix elements in the two cases
differ by SU(3) symmetry-breaking corrections. It is difficult to
predict the sign of the net effect, but the magnitude cannot be
larger than one or two per cent\cite{Chris,liferef}. Hence
\begin{equation}
   {\tau(B_s)\over\tau(B_d)} = 1\pm O(1\%) \,,
\end{equation}
which is consistent with the experimental value in (\ref{taudata}).
Note that $\tau(B_s)$ denotes the average lifetime of the two $B_s$
states.

\subsubsection{Lifetime ratio for $\Lambda_b$ and $B^0$}
\noindent
Although, as shown in (\ref{taucrude}), lifetime differences between
heavy mesons and baryons start at order $1/m_b^2$, the main effects
are expected to appear at order $1/m_b^3$ in the heavy-quark
expansion. However, here one encounters the problem that the matrix
elements of four-quark operators are needed between baryon states.
Very little is known about such matrix elements. Bigi et al.\ have
adopted a simple non-relativistic quark model and conclude
that\cite{liferef}
\begin{equation}
   {\tau(\Lambda_b)\over\tau(B^0)} = 0.90\mbox{--}0.95 \,.
\end{equation}
An even smaller lifetime difference has been obtained by
Rosner\cite{Rosner}.

An unbiased analysis gives, in the present case\cite{Chris}:
\begin{equation}
   {\tau(\Lambda_b)\over\tau(\bar B^0)}\simeq 0.98
   - 0.17\varepsilon_1 + 0.20\varepsilon_2
   - (0.013 + 0.022\widetilde B) r \,,
\end{equation}
where $\widetilde B$ and $r$ are expected to be positive and of order
unity. Given the structure of this result, it seems difficult to
explain the experimental value $\tau(\Lambda_b)/\tau(B^0) = 0.78\pm
0.05$ without violating the bound (\ref{epsbound})\footnote{Another
constraint arises if one does not want to spoil the theoretical
prediction for the semileptonic branching
ratio\protect\cite{Chris}.}. Essentially the only possibility is to
have $r$ of order 2--4 or so, as there are good theoretical arguments
why $\widetilde B$ cannot be much larger than unity. On the other
hand, in a constituent quark picture, $r$ is the ratio of the wave
functions determining the probability to find a light quark at the
location of the $b$ quark inside the $\Lambda_b$ baryon and the $B$
meson, i.e.\
\begin{equation}
   r = {|\psi_{bq}^{\Lambda_b}(0)|^2
        \over |\psi_{b\bar q}^{B_q}(0)|^2} \,,
\end{equation}
and it is hard to see how this ratio could be much different from
unity.

In view of the above discussion, the problem of the short $\Lambda_b$
lifetime appears as a puzzle, whose explanation may lie beyond the
heavy-quark expansion. If the current experimental value persists,
one may have to question the validity of local quark--hadron duality,
which is assumed in the theoretical calculation of lifetimes and
(non-leptonic) inclusive decay rates.


\runninghead{CP Violation} {CP Violation}
\section{CP Violation}
\label{sec:5}
\noindent
The violation of CP symmetry is one of the most intriguing aspects of
high-energy physics. Experimentally, it is one of the least tested
properties of the Standard Model. To date, there is only a single
unambiguous measurement of a CP-violating parameter: the measurement
of $\epsilon_K$ in $K$ decays\cite{CCFT}. The Standard Model
description of CP violation is very predictive, on the other hand;
all CP-violating effects are related to the phase $\delta$ of the CKM
matrix. Yet, this description has two major difficulties: first, CP
violation is a necessary prerequisite for baryogenesis\cite{Sakh},
but CP violation in the Standard Model is believed to be too small to
account for the observed baryon asymmetry in the Universe; secondly,
there is the so-called strong CP problem. The symmetries of the
strong interactions allow a term in the QCD Lagrangian that violates
CP:
\begin{equation}\label{strongCP}
   \theta\,{\alpha_s\over 4\pi}\,\mbox{Tr}\,G_{\mu\nu}
   \widetilde{G}^{\mu\nu} \,.
\end{equation}
The problem is that such a term would contribute to the electric
dipole moment of the neutron\cite{Balu,Crew}. In general, electric
dipole moments of elementary particles are sensitive probes of
CP-violating effects. Since the only vector that characterizes an
elementary particle is its spin\footnote{In this sense, the term
``elementary particle'' applies to the neutron, too.}, we must have
$\vec D=d\,\vec J$. However, $\vec J$ and $\vec D$ have different
transformation properties under parity and time-reversal
transformations. Consequently, if either P or T is a good symmetry,
we must have $d=0$. According to the CPT theorem, a violation of
time-reversal symmetry implies CP violation. This is why measurements
of electric dipole moments can be used to constrain CP-violating
parameters. The current experimental upper bound on the electric
dipole moment of the neutron
\begin{equation}
   |d_n| < 1.1\times 10^{-25} e\,\mbox{cm}\quad (95\%~\mbox{CL})
\end{equation}
implies $|\theta|<10^{-9}$, corresponding to an extreme fine-tuning
of a parameter in the QCD Lagrangian. The above two problems call for
extensions of the Standard Model, such as the Peccei--Quinn
symmetry\cite{PeQu}. The prospects are thus good that detailed
studies of CP-violating phenomena at future $B$ factories will
provide hints to physics beyond the Standard Model.

Besides measurements of the electric dipole moments of the electron
and the neutron, the most interesting observables for CP violation
are the weak decays of $K$ and $B$ mesons. In this section, we will
present first a general, model-independent discussion and
classification of CP-violating effects in meson decays. We will
distinguish three types of CP violation: direct CP violation in weak
decays, indirect CP violation in the mixing of neutral meson states,
and CP violation in the interference of mixing and decay. We will
then focus on an analysis of these effects in the Standard Model. A
more detailed discussion of CP violation in and beyond the Standard
Model can be found in the comprehensive review articles by
Nir\cite{Yosi}, and by Buchalla et al.\cite{BBL}.

\subsection{P, C, and CP Transformations}
\noindent
We start with a discussion of the parity, charge-conjugation and CP
transformations acting on meson states. The parity transformation is
a space-time transformation, under which $t\to t,\,\vec x\to-\vec x$.
It changes the sign of momenta, $\vec p\to-\vec p$, leaving spins
unchanged. For pseudoscalar mesons $P$ and $\bar P$, the parity
transformation implies (adopting the common phase conventions)
\begin{equation}
   \mbox{\bf P}\,|P(\vec p\,)\rangle = -|P(-\vec p\,)\rangle \,,
   \qquad \mbox{\bf P}\,|\bar P(\vec p\,)\rangle
   = -|\bar P(-\vec p\,)\rangle \,.
\end{equation}

Charge conjugation is a transformation that relates particles and
antiparticles, leaving all space-time coordinates unchanged, i.e.
\begin{equation}
   \mbox{\bf C}\,|P(\vec p\,)\rangle = |\bar P(\vec p\,)\rangle \,,
   \qquad \mbox{\bf C}\,|\bar P(\vec p\,)\rangle
   = |P(\vec p\,)\rangle \,.
\end{equation}

The combined transformation, CP, acts on the pseudoscalar meson
states as follows:
\begin{equation}
   \mbox{\bf CP}\,|P(\vec p\,)\rangle = -|\bar P(-\vec p\,)\rangle
   \,, \qquad \mbox{\bf CP}\,|\bar P(\vec p\,)\rangle
   = -|P(-\vec p\,)\rangle \,.
\end{equation}
For neutral mesons, $P^0$ and $\bar P^0$, one can construct the CP
eigenstates
\begin{equation}
   |P_1^0\rangle = {1\over\sqrt{2}}\,\Big( |P^0\rangle
    - |\bar P^0\rangle \Big) \,, \qquad
   |P_2^0\rangle = {1\over\sqrt{2}}\,\Big( |P^0\rangle
    + |\bar P^0\rangle \Big) \,,
\end{equation}
which obey
\begin{equation}
   \mbox{\bf CP}\,|P_1^0\rangle = |P_1^0\rangle \,, \qquad
   \mbox{\bf CP}\,|P_2^0\rangle = -|P_2^0\rangle \,.
\end{equation}

\runninghead{CP Violation} {Direct CP Violation in Weak Decays}
\subsection{Direct CP Violation in Weak Decays}
\noindent
Consider two decay processes related to each other by a CP
transformation. Let $P$ and $\bar P$ be CP-conjugated pseudoscalar
meson states, and $f$ and $\bar f$ some CP-conjugated final states:
\begin{equation}
   \mbox{\bf CP}\,|P\rangle = e^{i\varphi_P}\,|\bar P\rangle \,,
   \qquad
   \mbox{\bf CP}\,|f\rangle = e^{i\varphi_f}\,|\bar f\rangle \,.
\end{equation}
The phases $\varphi_P$ and $\varphi_f$ are arbitrary. The
CP-conjugated decay amplitudes, $A$ and $\bar A$, can then be written
as
\begin{eqnarray}
   A &= \langle f|\,{\cal H}\,|P\rangle
   &= \sum_i A_i\,e^{i\delta_i}\,e^{i\phi_i} \,, \nonumber\\
   \bar A &= \langle\bar f|\,{\cal H}\,|\bar P\rangle
   &= e^{i(\varphi_P-\varphi_f)} \sum_i A_i\,e^{i\delta_i}\,
    e^{-i\phi_i} \,,
\end{eqnarray}
where ${\cal H}$ is the effective Hamiltonian for weak decays, and
$A_i$ are real partial amplitudes. Two types of phases may appear in
the decay amplitudes: the weak phases $\phi_i$ are parameters of the
Lagrangian that violate CP. They usually appear in the electroweak
sector of the theory and enter $A$ and $\bar A$ with opposite signs.
The strong phases $\delta_i$ appear in scattering amplitudes even if
the Lagrangian is CP invariant. They usually arise from rescattering
effects due to the strong interactions and enter $A$ and $\bar A$
with the same sign.

Although the definition of strong and weak phases is to a large
degree convention dependent, one can show that the ratio
\begin{equation}
   \left|\,{\bar A\over A}\,\right| = \left|\,
   {\sum_i A_i\,e^{i\delta_i}\,e^{i\phi_i} \over
    \sum_i A_i\,e^{i\delta_i}\,e^{-i\phi_i}}\,\right|
\end{equation}
is independent of phase conventions and therefore physically
meaningful. The condition
\begin{equation}
   \left|\,{\bar A\over A}\,\right|\ne 1 \quad\Rightarrow\quad
   \mbox{direct CP violation}
\end{equation}
implies CP violation, which results from the interference of decay
amplitudes leading to the same final state. Note that this requires
at least two partial amplitudes that differ in both their weak and
strong phases.

\begin{figure}[htb]
   \epsfxsize=10cm
   \centerline{\epsffile{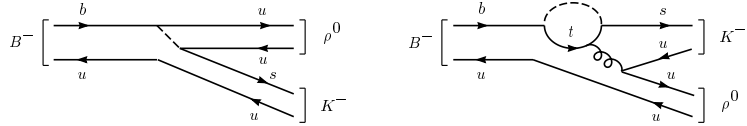}}
   \centerline{\parbox{11.5cm}{\caption{\label{fig:BrhoK}
Tree and penguin diagrams for the decay $B^\pm\to K^\pm\rho^0$.
   }}}
\end{figure}

\begin{figure}[htb]
   \epsfxsize=10.5cm
   \centerline{\epsffile{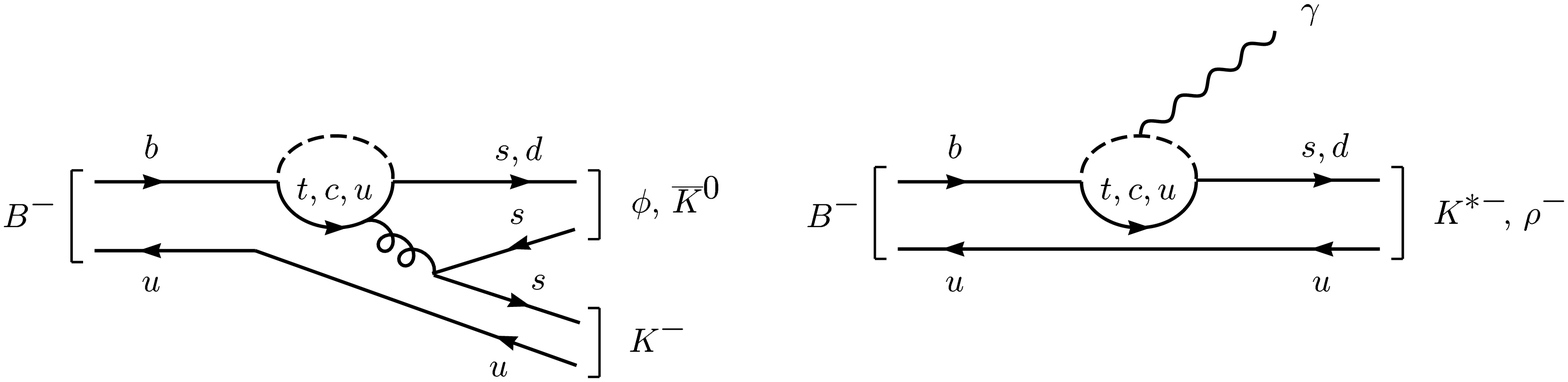}}
   \centerline{\parbox{11.5cm}{\caption{\label{fig:BKK}
Penguin diagrams for some tree-forbidden $B$ decays.
   }}}
\end{figure}

\paragraph{Experimental observation of direct CP violation:}
Since mixing is unavoidable in neutral meson decays, it is best to
observe direct CP violation in the decays of charged mesons. One
defines the CP asymmetry:
\begin{equation}
   a_f = {\Gamma(P^+\to f) - \Gamma(P^-\to\bar f) \over
          \Gamma(P^+\to f) + \Gamma(P^-\to\bar f)}
   = {1-|\bar A/A|^2\over 1+|\bar A/A|^2} \,.
\end{equation}
The requirement of at least two partial amplitudes with different
phases forces us to consider non-leptonic decays, since leptonic and
semileptonic decays are usually dominated by a single diagram.
Non-leptonic decays, on the other hand, can receive so-called
``tree'' and ``penguin''\footnote{It is a challenge to draw a penguin
diagram in such a way that it would actually deserve its name. Among
the various rumours about the origin of the name ``penguin'', the
author tends to believe the one of a well-known CERN theorist, who
had a bet that he could introduce any name he wanted into high-energy
physics.} contributions\cite{SVZ1}. Penguin diagrams contain a
$W$-boson--quark loop and typically involve other weak phases than
tree diagrams. In order to get large interference effects, one needs
partial amplitudes with similar magnitude\cite{BaSS}. A possibility
is to consider decays in which the tree contribution is suppressed,
with respect to the penguin contribution, by small CKM parameters.
This compensates for the loop suppression of penguin diagrams. In the
Standard Model, an example of this type is the decay $B^\pm\to
K^\pm\rho^0$ shown in Fig.~\ref{fig:BrhoK}, for which the tree
diagram is proportional to the small CKM parameters $|V_{ub}
V_{us}^*|\sim 10^{-3}$, whereas the penguin diagram is proportional
to $(\alpha_s/12\pi) \ln(m_t^2/m_b^2)\,|V_{tb} V_{ts}^*|\simeq
0.02\times 0.04\sim 10^{-3}$. Another possibility is to consider
tree-forbidden decays, which can only proceed through penguin
diagrams. In this case, it is the possibility to have different
quarks in the loop ($t,c,u$) that leads to the interference. Examples
are $B^\pm\to K^\pm K$ and $B^\pm\to K^\pm\phi$, as well as the
radiative decays $B^\pm\to K^{*\pm}\gamma$ and
$B^\pm\to\rho^\pm\gamma$, see Fig.~\ref{fig:BKK}. Unfortunately, the
decays $K^\pm\to\pi^\pm\pi^0$ and $B^\pm\to\pi^\pm\pi^0$ are pure
$\Delta I=\frac{3}{2}$ transitions and are thus governed by a single
strong phase $\delta_2$ corresponding to a $\pi\pi$ final state with
isospin $I=2$; isospin $I=1$ is not allowed because of Bose symmetry.
It follows that $a_{\pi\pi}=0$. There is no unambiguous experimental
evidence for direct CP violation yet.

\paragraph{Hadronic uncertainties:}
Calculations of direct CP-violating asymmetries have large
theoretical uncertainties. They are limited by our incapacity to
calculate hadronic matrix elements of quark operators with high
accuracy. Moreover, direct CP violation requires non-trivial strong
phase shifts, which are notoriously hard to calculate. In some cases,
however, part of the uncertainty can be eliminated using isospin
analysis\cite{iso1}$^-$\cite{iso5}.

\runninghead{CP Violation}
 {Indirect CP Violation in the Mixing of Neutral Mesons}
\subsection{Indirect CP Violation in the Mixing of Neutral Mesons}
\noindent
The neutral mesons $P^0$ and $\bar P^0$ can mix via common decay
channels:
\begin{equation}
   P^0 \leftrightarrow X \leftrightarrow \bar P^0 \,.
\end{equation}
An arbitrary neutral meson state can thus be written as a
superposition of the flavour eigenstates, $a|P^0\rangle + b|\bar
P^0\rangle$, which obeys the time-dependent Schr\"odinger equation
\begin{equation}
   i {{\rm d}\over{\rm d}t}\,
   \bigg( \begin{array}{c} a \\ b \end{array} \bigg)
   = {\bf H}\,\bigg( \begin{array}{c} a \\ b \end{array} \bigg)
   = \Big( {\bf M} - \textstyle{i\over 2}\,{\bf\Gamma} \Big)
   \bigg( \begin{array}{c} a \\ b \end{array} \bigg) \,,
\end{equation}
where ${\bf M}$ and ${\bf\Gamma}$ are Hermitian $2\times 2$ matrices,
which are called the mass and decay matrices, respectively. Since the
Hamilton operator, ${\bf H}$, is not Hermitian, its eigenvectors
\begin{equation}
   |P_{1,2}\rangle = p|P^0\rangle \pm q|\bar P^0\rangle \,;\quad
   |p^2| + |q^2| = 1
\end{equation}
are not orthogonal, and the eigenvalues
\begin{equation}
   \mu_i = M_i - {i\over 2}\,\Gamma_i \,;\quad i=1,2
\end{equation}
are complex. This reflects that the states $P_1$ and $P_2$ are
resonances, not elementary particles. $M_i$ are the masses of these
resonances, and $\Gamma_i$ are their decay widths. The states $P_i$
have a diagonal time evolution given by
\begin{equation}\label{timeevol}
   |P_i(t)\rangle = e^{-i M_i t}\,e^{-{1\over 2}\Gamma_i t}\,
   |P_i(0)\rangle \,.
\end{equation}

One can show that the ratio
\begin{equation}
   \bigg|\,{q\over p}\,\bigg|^2 = \left|\,
   {M_{12}^* - {i\over 2}\,\Gamma_{12}^* \over
    M_{12} - {i\over 2}\,\Gamma_{12}}\,\right|
\end{equation}
is independent of phase conventions and therefore physically
meaningful. The condition
\begin{equation}
   \bigg|\,{q\over p}\,\bigg|\ne 1 \quad\Rightarrow\quad
   \mbox{indirect CP violation}
\end{equation}
implies CP violation, which results from the fact that the flavour
eigenstates are different from the CP eigenstates.

Let us collect some useful equations related to the mixing of neutral
mesons. Define the mass difference $\Delta m=m_2-m_1$ and the width
difference $\Delta\Gamma=\Gamma_2-\Gamma_1$. Then the following
relations hold:
\begin{eqnarray}\label{qprela}
   (\Delta m)^2 - \textstyle{1\over 4}\,(\Delta\Gamma)^2
   &=& 4 |M_{12}|^2 - |\Gamma_{12}|^2 \,, \nonumber\\
   \phantom{ \bigg[ }
   \Delta m\cdot\Delta\Gamma &=& 4\,\mbox{Re}\,(M_{12}\Gamma_{12}^*)
    \,, \nonumber\\
   {q\over p} &=& -{1\over 2}\,
    {\Delta m - {i\over 2}\,\Delta\Gamma\over
     M_{12} - {i\over 2}\,\Gamma_{12}} = -2\,
    {M_{12}^* - {i\over 2}\,\Gamma_{12}^* \over
     \Delta m - {i\over 2}\,\Delta\Gamma} \,.
\end{eqnarray}
An alternative common notation is to define $\bar\epsilon$ such that
\begin{equation}
   p = {1+\bar\epsilon\over\sqrt{2(1+|\bar\epsilon|^2)}} \,,
   \qquad q = {1-\bar\epsilon\over\sqrt{2(1+|\bar\epsilon|^2)}} \,,
   \qquad {q\over p} = {1-\bar\epsilon\over 1+\bar\epsilon} \,.
\end{equation}

\paragraph{$q/p$ in the kaon system:}
One defines the ``short-lived'' and ``long-lived'' neutral kaon
states $K_S=K_1$ and $K_L=K_2$, which differ significantly in their
lifetimes: $\tau_S=(8.926\pm 0.012)\times 10^{-11}$~s and
$\tau_L=(5.17\pm 0.04)\times 10^{-8}$~s. Experimentally,
\begin{eqnarray}
   \Delta m_K &=& m_L - m_S
    = (3.510\pm 0.018)\times 10^{-15}~\mbox{GeV} \,, \nonumber\\
   \Delta\Gamma_K &=& \Gamma_L - \Gamma_S
    = -(7.361\pm 0.010)\times 10^{-15}~\mbox{GeV} \,,
\end{eqnarray}
so that
\begin{equation}\label{DGDM}
   \Delta\Gamma_K\simeq -2\Delta m_K \,.
\end{equation}
If we define
\begin{equation}
   {\Gamma_{12}^*\over M_{12}^*}
   = - \left|{\Gamma_{12}\over M_{12}}\right|\,
   e^{i\phi_{12}} \,,
\end{equation}
the experimental observation that there is only a small CP violation
in the kaon system is reflected in the fact that
$|\phi_{12}|=O(10^{-3})$. From (\ref{qprela}), we find to first order
in this small angle
\begin{equation}
   \bigg( {q\over p} \bigg)_K
   \simeq {\Gamma_{12}^*\over|\Gamma_{12}|}\,
   \Bigg\{1 - i\phi_{12}\,{1+i({\Delta\Gamma_K\over 2\Delta m_K})
    \over 1+({\Delta\Gamma_K\over 2\Delta m_K})^2} \Bigg\} \,,
\end{equation}
so that with (\ref{DGDM})
\begin{equation}
   \bigg|\,{q\over p}\,\bigg|_K - 1 \simeq - 2\,\mbox{Re}\,
   \bar\epsilon_K \simeq -\phi_{12} = O(10^{-3}) \,.
\end{equation}

\paragraph{$q/p$ in the $B$-meson system:}
Decay channels common to $B^0$ and $\bar B^0$, which are responsible
for the difference $\Delta\Gamma_B$, are known to have branching
fractions of order $10^{-3}$ or less. Hence, although
$\Delta\Gamma_B$ has not yet been measured directly, it follows that
$|\Delta\Gamma_B|/\Gamma_B<10^{-2}$. The observed $B^0$--$\bar B^0$
mixing rate implies\cite{DmBexp} $\Delta m_B/\Gamma_B=0.74\pm 0.04$,
on the other hand, so that model independently
\begin{equation}
   |\Delta\Gamma_B| \ll \Delta m_B \,.
\end{equation}
Thus, there is a negligible lifetime difference between the CP
eigenstates, and one therefore refers to these states as ``light''
and ``heavy'', $B_L=B_1$ and $B_H=B_2$. It follows that
$|\Gamma_{12}|\ll |M_{12}|$, and to first order in
$\Gamma_{12}/M_{12}$ we obtain from (\ref{qprela})
\begin{equation}
   \bigg( {q\over p} \bigg)_B \simeq - {M_{12}^*\over|M_{12}|}\,
   \bigg( 1 - {1\over 2}\,\mbox{Im}\,{\Gamma_{12}\over M_{12}}
   \bigg) \,.
\end{equation}
Hence
\begin{equation}
   \bigg|\,{q\over p}\,\bigg|_B - 1 \simeq - 2\,\mbox{Re}\,
   \bar\epsilon_B = O(10^{-2}) \,.
\end{equation}
As in the kaon system, CP violation in $B^0$--$\bar B^0$ mixing is a
small effect.

\paragraph{Experimental observation of indirect CP violation in the
kaon system:}
One uses the fact that the semileptonic decays of neutral mesons are
flavour-tagging, i.e.\ $P^0\rlap{\,/}{\to}\ell^-\nu$ and $\bar
P^0\rlap{\,/}{\to}\ell^+\bar\nu$, and defines
\begin{equation}
   A = \langle\ell^+\bar\nu X|\,{\cal H}\,|P^0\rangle \,, \qquad
   A^* = \langle\ell^-\nu X|\,{\cal H}\,|\bar P^0\rangle \,.
\end{equation}
Because of the large lifetime difference between the two neutral kaon
states, it is possible to prepare a beam of $K_L$ particles and to
measure the asymmetry
\begin{equation}
   a_{\rm SL}^K =
   {\Gamma(K_L\to\ell^+\bar\nu X) - \Gamma(K_L\to\ell^-\nu X)\over
    \Gamma(K_L\to\ell^+\bar\nu X) + \Gamma(K_L\to\ell^-\nu X)} \,.
\end{equation}
Using that $|K_L\rangle=p|K^0\rangle - q|\bar K^0\rangle$, and hence
\begin{equation}
   \langle\ell^+\bar\nu X|\,{\cal H}\,|K_L\rangle = p A \,, \qquad
   \langle\ell^-\nu X|\,{\cal H}\,|K_L\rangle = - q A^* \,,
\end{equation}
we find that
\begin{equation}
   a_{\rm SL}^K = {1-|q/p|^2\over 1+|q/p|^2}
   = {2\,\mbox{Re}\,\bar\epsilon_K\over 1+|\bar\epsilon_K|^2}
   \simeq 2\,\mbox{Re}\,\bar\epsilon_K \,.
\end{equation}
Experimentally, it was found that
\begin{equation}
   a_{\rm SL}^K = (3.27\pm 0.12)\times 10^{-3} \,.
\end{equation}
This was the observation of indirect CP violation in the kaon
system\cite{CCFT}.

\paragraph{Experimental observation of indirect CP violation in the
$B$-meson system:}
Since $B_L$ and $B_H$ have almost identical lifetimes, it is not
possible to produce selectively beams of $B_L$ or $B_H$ particles.
With $m_{H,L}=m_B\pm\frac{1}{2} \Delta m_B$ and $\Gamma_{H,L}\simeq
\Gamma_B$, eq.~(\ref{timeevol}) gives for the time evolution of an
initially pure $B^0$ state:
\begin{eqnarray}
   |B^0(0)\rangle &=& {1\over 2p}\,
    \Big( |B_H\rangle + |B_L\rangle \Big) \,, \nonumber\\
   |B^0(t)\rangle &=& {1\over 2p}\,e^{-i m_B t}\,
    e^{-{1\over 2}\Gamma_B t}\,\bigg\{
    e^{-{i\over 2}\Delta m_B t}\,|B_H\rangle
    + e^{{i\over 2}\Delta m_B t}\,|B_L\rangle \bigg\} \nonumber\\
   &=& e^{-i m_B t}\,e^{-{1\over 2}\Gamma_B t}\,\bigg\{
    \cos\Big( {\textstyle{1\over 2}}\,\Delta m_B t \Big)\,
    |B^0\rangle + {i q\over p}\,\sin\Big( \textstyle{1\over 2}\,
    \Delta m_B t \Big)\,|\bar B^0\rangle \bigg\} \,. \nonumber\\
\end{eqnarray}
Similarly:
\begin{equation}
   |\bar B^0(t)\rangle = e^{-i m_B t}\,e^{-{1\over 2}\Gamma_B t}\,
   \bigg\{ \cos\Big( {\textstyle{1\over 2}}\,\Delta m_B t \Big)\,
   |\bar B^0\rangle + {i p\over q}\,\sin\Big(
   \textstyle{1\over 2}\,\Delta m_B t \Big)\,|B^0\rangle \bigg\} \,.
\end{equation}
Defining the semileptonic asymmetry as
\begin{equation}
   a_{\rm SL}^B = {\Gamma(\bar B^0(t)\to\ell^+\bar\nu X)
   - \Gamma(B^0(t)\to\ell^-\nu X)\over
   \Gamma(\bar B^0(t)\to\ell^+\bar\nu X)
   + \Gamma(B^0(t)\to\ell^-\nu X)} \,,
\end{equation}
and taking into account that $B^0\rlap{\,/}{\to}\ell^-\nu$ and $\bar
B^0\rlap{\,/}{\to}\ell^+\bar\nu$, we obtain
\begin{equation}
   a_{\rm SL}^B = {1-|q/p|^4\over 1+|q/p|^4}
   \simeq 4\,\mbox{Re}\,\bar\epsilon_B = O(10^{-2}) \,.
\end{equation}
To date, there is no experimental evidence for indirect CP violation
in the $B$-meson system.

\paragraph{Hadronic uncertainties:}
The calculation of $|q/p|$ involves hadronic matrix elements of local
four-quark operators (so-called $B$ parameters). The theoretical
uncertainty in the calculation of such matrix elements is about 30\%.

\runninghead{CP Violation}
 {CP Violation in the Interference of Mixing and Decay}
\subsection{CP Violation in the Interference of Mixing and Decay}
\noindent
Consider decays of neutral mesons into CP eigenstates:
\begin{equation}
   A = \langle f_{\rm CP}|\,{\cal H}\,|P^0\rangle \,, \qquad
   A^* = \langle f_{\rm CP}|\,{\cal H}\,|\bar P^0\rangle \,.
\end{equation}
It can be shown that the product
\begin{equation}
   \lambda = {q\over p}\cdot{\bar A\over A}
\end{equation}
is independent of phase conventions and thus physically meaningful.
In other words, the convention dependence of $q/p$ cancels against
that of $\bar A/A$. The condition
\begin{equation}
   \lambda\ne 1 \quad\Rightarrow\quad \mbox{CP violation}
\end{equation}
implies CP violation. Note that direct CP violation ($|\bar A/A|\ne
1$) and indirect CP violation ($|q/p|\ne 1$) imply $|\lambda|\ne 1$,
but they are not necessary for the weaker condition $\lambda\ne 1$.
In fact, the case $|\lambda|=1$ but $\mbox{Im}\,\lambda\ne 0$ is the
theoretically favoured situation. In that case $\lambda$ is a pure
phase, which can be calculated without hadronic uncertainties.

Many decays of neutral $B$ mesons are of the kind described above. If
one defines the CP asymmetry\cite{Cart}$^-$\cite{DuRo}
\begin{equation}
   a_{f_{\rm CP}} = {\Gamma(B^0(t)\to f_{\rm CP})
    - \Gamma(\bar B^0(t)\to f_{\rm CP})\over
    \Gamma(B^0(t)\to f_{\rm CP})
    + \Gamma(\bar B^0(t)\to f_{\rm CP})}
\end{equation}
and takes into account that $|q/p|_B\simeq 1$, it follows that
\begin{eqnarray}
   a_{f_{\rm CP}} &\simeq& {(1-|\lambda|^2)\,\cos(\Delta m_B t)
    - 2\,\mbox{Im}\,\lambda\,\sin(\Delta m_B t)\over 1+|\lambda|^2}
    \nonumber\\
   &\stackrel{|\lambda|=1}{\to}& -\mbox{Im}\,\lambda\,
    \sin(\Delta m_B t) \,.
\end{eqnarray}
The ``clean modes'' with $|\lambda|\simeq 1$ are those dominated by a
single weak phase $\phi$, so that
\begin{equation}
   {\bar A\over A} \simeq e^{-2 i\phi}
\end{equation}
is close to a pure phase. Examples of such decays are discussed in
detail below. Unfortunately, this method is not useful in kaon
decays, since
\begin{equation}
   \mbox{Im}\,\lambda(K\to\pi\pi) = O(10^{-3}) \,,
\end{equation}
i.e.\ very small.

\runninghead{CP Violation} {CP Violation in the Standard Model}
\subsection{CP Violation in the Standard Model}
\noindent
We will now specify the general framework described above and discuss
CP violation in the context of the Standard Model. Below mass scales
of order $m_W\sim 80$~GeV, the Standard Model gauge group
$\mbox{SU}_{\rm C}(3)\times\mbox{SU}_{\rm L}(2)\times\mbox{U}_{\rm
Y}(1)$ is spontaneously broken to $\mbox{SU}_{\rm
C}(3)\times\mbox{U}_{\rm em}(1)$, since the scalar Higgs doublet
$\phi$ acquires a vacuum expectation value. This gives masses to the
$W$ and $Z$ bosons, as well as to the quarks and leptons. The quark
masses arise from the Yukawa couplings to the Higgs doublet, which in
the unbroken theory are assumed to be of the most general form
invariant under local gauge transformations. The Yukawa interactions
are written in terms of the weak eigenstates $q'$ of the quark
fields, which have simple transformation properties under
$\mbox{SU}_{\rm L}(2)\times \mbox{U}_{\rm Y}(1)$. After the symmetry
breaking, one redefines the quark fields so as to obtain the mass
terms in the canonical form. This has an interesting effect on the
form of the flavour-changing charged-current interactions. In the
weak basis, these interactions have the form
\begin{equation}
   {\cal L}_{\rm int} = - {g\over\sqrt{2}}\,
   (\bar u'_{\rm L}, \bar c'_{\rm L}, \bar t'_{\rm L})\,\gamma^\mu
   \left( \begin{array}{c} d'_{\rm L} \\ s'_{\rm L} \\
   b'_{\rm L} \end{array} \right) W_\mu^\dagger +
   \mbox{h.c.}
\end{equation}
In terms of the mass eigenstates $q$, however, this becomes
\begin{equation}
   {\cal L}_{\rm int} = - {g\over\sqrt{2}}\,
   (\bar u_{\rm L}, \bar c_{\rm L}, \bar t_{\rm L})\,\gamma^\mu\,
   V_{\rm CKM}
   \left( \begin{array}{c} d_{\rm L} \\ s_{\rm L} \\
   b_{\rm L} \end{array} \right) W_\mu^\dagger +
   \mbox{h.c.}
\end{equation}
The CKM mixing matrix $V_{\rm CKM}$ is a unitary matrix in flavour
space. In the general case of $n$ quark generations, $V_{\rm CKM}$
would be an $n\times n$ matrix.

For two generations, the mixing matrix can be parametrized by one
angle and three phases:
\begin{eqnarray}
   V &=& \left( \begin{array}{rl}
    \cos\theta_{\rm C}\,e^{i\alpha} & ~\sin\theta_{\rm C}
    \,e^{i\beta} \\
    -\sin\theta_{\rm C}\,e^{i\gamma} & ~\cos\theta_{\rm C}\,
    e^{i(\beta+\gamma-\alpha)}
   \end{array} \right) \nonumber\\
   && \nonumber\\
   &=& \left( \begin{array}{cc}
    e^{i\alpha} & 0 \\
    0 & e^{i\gamma}
   \end{array} \right)
   \left( \begin{array}{rc}
    \cos\theta_{\rm C} & ~\sin\theta_{\rm C} \\
    -\sin\theta_{\rm C} & ~\cos\theta_{\rm C}
   \end{array} \right)
   \left( \begin{array}{cc}
    1 & 0 \\
    0 & e^{i(\beta-\alpha)}
   \end{array} \right) \,.
\end{eqnarray}
The phases are not observable, however, as they can be absorbed into
a redefinition of the phases of the quark fields $u_{\rm L}$, $c_{\rm
L}$, $s_{\rm L}$ relative to $d_{\rm L}$. After this redefinition,
the matrix takes the standard form due to Cabibbo\cite{Cabi}:
\begin{equation}
   V_{\rm C} = \left( \begin{array}{rc}
    \cos\theta_{\rm C} & ~\sin\theta_{\rm C} \\
    -\sin\theta_{\rm C} & ~\cos\theta_{\rm C}
   \end{array} \right) \,.
\end{equation}
In the case of three generations, $V_{\rm CKM}$ can be parametrized
by three Euler angles and six phases, five of which can be removed by
adjusting the relative phases of the left-handed quark fields. Hence,
three angles $\theta_{ij}$ and one observable phase $\delta$ remain
in the quark mixing matrix, as was first pointed out by Kobayashi and
Maskawa\cite{KoMa}. For completeness, we note that in the general
case of $n$ generations, it is easy to show that there are
$\frac{1}{2} n(n-1)$ angles and $\frac{1}{2}(n-1)(n-2)$ observable
phases\cite{Jarl}. Therefore, whereas the Cabibbo matrix is real and
has only one parameter, the CKM matrix is complex and can be
parametrized by four quantities. The imaginary part of the mixing
matrix is necessary to describe CP violation in the Standard Model.
In general, CP is violated in flavour-changing decays if there is no
degeneracy of any two quark masses, and if the quantity $J_{\rm
CP}\ne 0$, where
\begin{equation}\label{JCPdef}
   J_{\rm CP} = |\,\mbox{Im}\,(V_{ij} V_{kl} V_{il}^* V_{kj}^*)\,|
   \,;\quad i\ne k \,,~j\ne l \,.
\end{equation}
It can be shown that all CP-violating amplitudes in the Standard
Model are proportional to $J_{\rm CP}$, and that this quantity is
invariant under phase redefinitions of the quark fields\cite{Jar1}.

Ignoring the strong CP problem, i.e.\ assuming that $\theta=0$ in
(\ref{strongCP}), the complex phase of the CKM matrix is the only
parameter in the Standard Model that violates CP symmetry. Hence, the
Standard Model is very predictive in describing CP-violating effects:
all CP-violating observables are in principle calculable in terms of
only one parameter. In practice, however, strong interaction effects
have to be controlled before such calculations can be performed.

Let us mention two of the most convenient parametrizations of the
CKM matrix. The ``standard parametrization''\cite{Chau} recommended
by the Particle Data Group is
\begin{equation}\label{VKMstand}
   V_{\rm CKM} = \left( \begin{array}{ccc}
    c_{12}\,c_{13} & s_{12}\,c_{13} & s_{13}\,e^{-i\delta} \\
    -s_{12}\,c_{23} - c_{12}\,s_{23}\,s_{13}\,e^{i\delta} &
    c_{12}\,c_{23} - s_{12}\,s_{23}\,s_{13}\,e^{i\delta} &
    s_{23}\,c_{13} \\
    s_{12}\,s_{23} - c_{12}\,c_{23}\,s_{13}\,e^{i\delta} &
    -c_{12}\,s_{23} - s_{12}\,c_{23}\,s_{13}\,e^{i\delta} &
    c_{23}\,c_{13}
   \end{array} \right) \,.
\end{equation}
Here, the short-hand notation $c_{ij}=\cos\theta_{ij}$ and
$s_{ij}=\sin\theta_{ij}$ is used. Some advantages of this
parametrization are the following ones:

\begin{itemize}
\item
$|V_{ub}|=s_{13}$ is given by a single angle, which experimentally
turns out to be very small.
\item
Because of this, several other entries are given by single angles to
an accuracy of better than four digits. They are: $V_{ud}\simeq
c_{12}$, $V_{us}\simeq s_{12}$, $V_{cb}\simeq s_{23}$, and
$V_{tb}\simeq c_{23}$.
\item
The CP-violating phase $\delta$ appears together with the small
parameter $s_{13}$, making it explicit that CP violation in the
Standard Model is a small effect. Indeed, one finds
\begin{equation}\label{JCPstan}
   J_{\rm CP} = |\,s_{13}\,s_{23}\,s_{12}\,s_\delta\,
   c_{13}^2\,c_{23}\,c_{12}\,| \,.
\end{equation}
\end{itemize}

For many purposes and applications, it is more convenient to use an
approximate parametrization of the CKM matrix, which makes explicit
the strong hierarchy observed experimentally. Setting $c_{13}=1$
(experimentally, it is known that $c_{13}>0.99998$) and neglecting
$s_{13}$ compared with terms of order unity, we find
\begin{equation}
   V_{\rm CKM} \simeq \left( \begin{array}{ccc}
    c_{12} & s_{12} & s_{13}\,e^{-i\delta} \\
    -s_{12}\,c_{23} & c_{12}\,c_{23} & s_{23} \\
    s_{12}\,s_{23} - c_{12}\,c_{23}\,s_{13}\,e^{i\delta} &
    ~-c_{12}\,s_{23}~ & c_{23}
   \end{array} \right) \,.
\end{equation}
Now denote $\lambda=s_{12}\simeq 0.22$. Experiments indicate that
$s_{23}=O(\lambda^2)$ and $s_{13}=O(\lambda^3)$. Hence, it is natural
to define $s_{23}=A\,\lambda^2$ and $s_{13}\,e^{-i\delta}=A\,
\lambda^3 (\rho-i\eta)$, with $A$, $\rho$ and $\eta$ of order unity.
An expansion in powers of $\lambda$ then leads to the Wolfenstein
parametrization\cite{Wolf}
\begin{equation}\label{Wolpar}
   V_{\rm CKM} \simeq \left( \begin{array}{ccc}
    1-{\lambda^2\over 2} & \lambda & A\,\lambda^3(\rho-i\eta) \\
    \phantom{ \bigg[ }
    -\lambda & 1-{\lambda^2\over 2} & A\,\lambda^2 \\
    A\,\lambda^3(1-\rho-i\eta) & ~-A\,\lambda^2~ & 1
   \end{array} \right) + O(\lambda^4) \,.
\end{equation}
It nicely exhibits the hierarchy of the mixing matrix: the entries on
the diagonal are close to unity, $V_{us}$ and $V_{cd}$ are of order
20\%, $V_{cb}$ and $V_{ts}$ are of order 4\%, and $V_{ub}$ and
$V_{td}$ are of order 1\% and thus the smallest entries in the
matrix. Some care has to be taken when one wants to calculate the
quantity $J_{\rm CP}$ in the Wolfenstein parametrization, since the
result is of order $\lambda^6$ and thus beyond the accuracy of the
approximation. However, taking $i=u$, $j=d$, $k=t$, and $l=b$ in
(\ref{JCPdef}), we obtain the correct answer
\begin{equation}
   J_{\rm CP} \simeq A^2\,\eta\,\lambda^6
   \simeq 1.1\times 10^{-4} A^2\,\eta \,,
\end{equation}
which shows that $J_{\rm CP}$ is generically of order $10^{-4}$ for
$\lambda\simeq 0.22$.

In principle, the entries in the first two rows of the mixing matrix
are accessible in so-called direct (tree-level) processes, i.e.\ in
weak decays of hadrons containing the corresponding quarks. In
practice, $|V_{ud}|$ and $|V_{us}|$ are known to an accuracy of
better than 1\%, $|V_{cb}|$ is known to 5\%, and $|V_{cd}|$ and
$|V_{cs}|$ are known to about 10--20\%. Hence, the two Wolfenstein
parameters $\lambda$ and $A$ are rather well determined
experimentally:
\begin{equation}\label{AWoval}
   \lambda = |V_{us}| = 0.2205\pm 0.0018 \,, \qquad
   A = \left| {V_{cb}\over V_{us}^2} \right| = 0.80\pm 0.04 \,.
\end{equation}
On the other hand, $|V_{ub}|$ has an uncertainty of about 30\%, and
the same is true for $|V_{td}|$, which is obtained from $B^0$--$\bar
B^0$ mixing. This implies a rather significant uncertainty in the
values of the Wolfenstein parameters $\rho$ and $\eta$. A more
precise determination of these parameters will be a challenge to
experiments and theory over the next decade.

\runninghead{CP Violation} {The Unitarity Triangle}
\subsection{The Unitarity Triangle}
\noindent
A simple but beautiful way to visualize the implications of unitarity
is provided by the so-called unitarity triangle\cite{Bjtr}, which
uses the fact that the unitarity equation
\begin{equation}
   V_{ij}\,V_{ik}^* = 0 \qquad (j\ne k)
\end{equation}
can be represented as the equation of a closed triangle in the
complex plane. There are six such triangles, all of which have the
same area\cite{JaSt}
\begin{equation}\label{AreaJcp}
   |A_\Delta| = {1\over 2}\,J_{\rm CP} \,.
\end{equation}
Under phase reparametrizations of the quark fields, the triangles
change their orientation in the complex plane, but their shape
remains unaffected.

\begin{figure}[htb]
\vspace{-1.2cm}
   \epsfysize=4.5cm
   \centerline{\epsffile{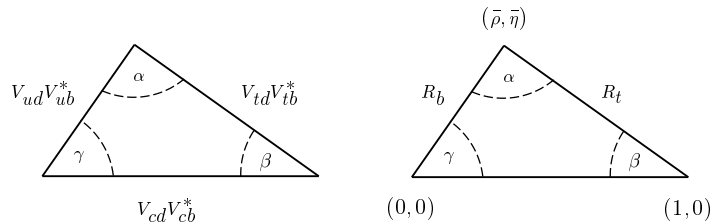}}
\vspace{-0.8cm}
   \centerline{\parbox{11.5cm}{\caption{\label{fig:2.2}
The unitarity triangle (left), and its rescaled form in the
$\bar\rho$--$\bar\eta$ plane (right). The angle $\gamma$ coincides
with the phase $\delta$ of the standard parametrization.
   }}}
\end{figure}

Most useful from the phenomenological point of view is the triangle
relation
\begin{equation}
   V_{ud} V_{ub}^* + V_{cd} V_{cb}^* + V_{td} V_{tb}^* = 0 \,,
\end{equation}
since it contains the most poorly known entries in the CKM matrix. It
has been widely discussed in the
literature\cite{Bjtr}$^-$\cite{HaRo}.
In the standard parametrization, $V_{cd} V_{cb}^*$ is real, and the
unitarity triangle has the form shown in Fig.~\ref{fig:2.2}. It is
useful to rescale the triangle by dividing all sides by $V_{cd}
V_{cb}^*$. The rescaled triangle has the coordinates $(0,0)$,
$(1,0)$, and $(\bar\rho,\bar\eta)$, where\cite{Oster}
\begin{equation}
   \bar\rho = \bigg( 1 - {\lambda^2\over 2} \bigg)\,\rho \,,\qquad
   \bar\eta = \bigg( 1 - {\lambda^2\over 2} \bigg)\,\eta
\end{equation}
are related to the Wolfenstein parameters $\rho$ and $\eta$ appearing
in (\ref{Wolpar}). Unitarity amounts to the statement that the
triangle is closed, and CP is violated when the area of the triangle
does not vanish, i.e.\ when all the angles are different from zero.

\begin{figure}[htb]
   \epsfxsize=11cm
   \centerline{\epsffile{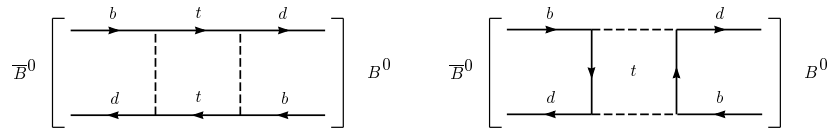}}
   \centerline{\parbox{11.5cm}{\caption{\label{fig:box}
Box diagrams for $B^0$--$\bar B^0$ mixing in the Standard Model.
   }}}
\end{figure}

To determine the shape of the triangle, one can aim for measurements
of the two sides $R_b$ and $R_t$, and of the angles $\alpha$,
$\beta$, and $\gamma$. So far, experimental information is available
only on the sides of the triangle. The current value of $|V_{ub}|$
in (\ref{Vubval}) implies
\begin{equation}\label{Rbexp}
   R_b = \sqrt{\bar\rho^2 + \bar\eta^2}
   = \bigg( 1 - {\lambda^2\over 2} \bigg)\,{1\over\lambda}\,
   \left|{V_{ub}\over V_{cb}}\right| = 0.35\pm 0.09 \,.
\end{equation}
To determine $R_t$, one needs information on $|V_{td}|$, which can be
extracted from $B^0$--$\bar B^0$ mixing. In the Standard Model, the
mass difference $\Delta m_B$ between the two neutral meson states is
calculable from the box diagrams shown in Fig.~\ref{fig:box}. The
resulting theoretical expression is
\begin{equation}\label{dmBSM}
   \Delta m_B = {G_F^2 m_W^2\over 6\pi^2}\,\eta_B\,
   B_B f_B^2\,m_B\,S(m_t/m_W)\,|V_{td} V_{tb}^*|^2 \,,
\end{equation}
where $\eta_B=0.55\pm 0.01$ accounts for the QCD
corrections\cite{BJWe}, and $S(m_t/m_W)$ is a function of the top
quark mass\cite{InLi,Bur1}. The product $B_B f_B^2$ parametrizes the
hadronic matrix element of a local four-quark operator between
$B$-meson states. There exists a vast literature on calculations of
the decay constant $f_B$ and the $B_B$ parameter. Combining the
results of some recent QCD sum-rule\cite{Lamsr2,fBsr1} and lattice
calculations\cite{fBlat1}$^-$\cite{fBlat4}, we quote the value
\begin{equation}
   f_B = 185\pm 40~\mbox{MeV} \,.
\end{equation}
Together with the prediction $B_B\simeq 1.08$ obtained from lattice
calculations\cite{Abad}, this gives
\begin{equation}
   B_B^{1/2} f_B = (200\pm 40)~\mbox{MeV} \,.
\end{equation}
Solving then (\ref{dmBSM}) for $|V_{td}|$, one obtains\cite{BuHa}
\begin{equation}
   |V_{td}| = 8.53\times 10^{-3}\,
   \bigg( {200~\mbox{MeV}\over B_B^{1/2} f_B} \bigg)\,
   \bigg( {170~\mbox{GeV}\over\bar m_t(m_t)} \bigg)^{0.76}\,
   \bigg( {\Delta m_B\over 0.465~\mbox{ps}^{-1}} \bigg)^{1/2} \,.
\end{equation}
Taking $\bar m_t(m_t)=(170\pm 15)$~GeV for the running top-quark
mass, and using the average experimental value\cite{DmBexp}
\begin{equation}
   \Delta m_B = (0.465\pm 0.024)~\mbox{ps}^{-1} \,,
\end{equation}
gives
\begin{equation}
   |V_{td}| = (8.53\pm 1.81)\times 10^{-3} \,.
\end{equation}
The corresponding range of values for $R_t$ is
\begin{equation}\label{Rtexp}
   R_t = \sqrt{(1-\bar\rho)^2 + \bar\eta^2}
   = {1\over\lambda}\,\left|{V_{td}\over V_{cb}}\right|
   = 0.99\pm 0.22 \,.
\end{equation}

Equations (\ref{Rbexp}) and (\ref{Rtexp}) yield constraints on the
Wolfenstein parameters $\bar\rho$ and $\bar\eta$, which have the form
of rings centred at $(\bar\rho,\bar\eta)=(0,0)$ and $(0,1)$. Another
constraint can be obtained from the measurement of indirect CP
violation in the kaon system. The experimental result on the
parameter $\epsilon_K$ measuring CP violation in $K^0$--$\bar K^0$
mixing implies that the unitarity triangle lies in the upper half
plane. The constraint arising in the $\bar\rho$--$\bar\eta$ plane has
the form of a hyperbola, the shape of which depends on a hadronic
parameter $B_K$. The theoretical prediction is\cite{BuHa}
\begin{equation}
   \bar\eta\,\bigg[ (1- \bar\rho)\,A^2\,\bigg( {m_t\over m_W}
   \bigg)^{1.52} + (0.69\pm 0.05) \bigg]\,A^2\,B_K \simeq 0.50 \,,
\end{equation}
where $A=0.80\pm 0.04$ according to (\ref{AWoval}). In the last few
years, theoretical calculations of the $B_K$ parameter have converged
and give results in the ball park of
\begin{equation}
   B_K = 0.75\pm 0.15 \,.
\end{equation}
In particular, the most recent lattice
calculations\cite{BKlat1,BKlat2} are in good agreement with the
results obtained using the $1/N_c$ expansion\cite{Nc1,Nc2}, and the
differences with previous, lower predictions for $B_K$ based on
duality and chiral symmetry\cite{DGHo,PdeR} have been understood.

In principle, the measurement of the ratio $\mbox{Re}\,
(\epsilon'/\epsilon)$ in the kaon system could provide a
determination of $\eta$ independent of $\rho$. In practice, however,
the experimental situation is unclear\cite{NA31,E731}, and the
theoretical calculations\cite{Burs}$^-$\cite{Marti} of this ratio are
affected by large uncertainties, so that there currently is no useful
bound to be derived.

\begin{figure}[htb]
   \epsfxsize=8cm
   \centerline{\epsffile{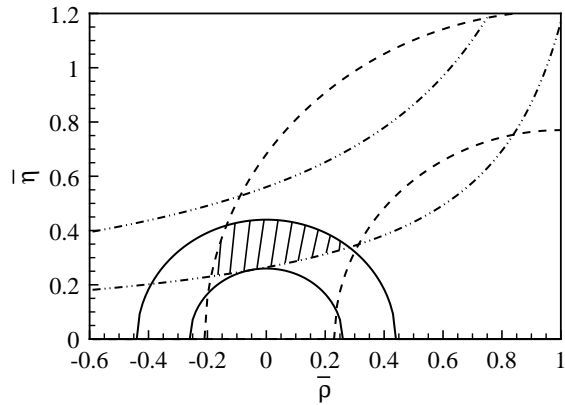}}
   \centerline{\parbox{11.5cm}{\caption{\label{fig:2.3}
Experimental constraints on the unitarity triangle in the
$\bar\rho$--$\bar\eta$ plane. The region between the solid (dashed)
circles is allowed by the measurement of $R_b$ ($R_t$) discussed
above. The dash-dotted curves show the constraint following from the
measurement of the $\epsilon_K$ parameter in the kaon system. The
shaded region shows the allowed range for the tip of the unitarity
triangle. The base of the triangle has the coordinates $(0,0)$ and
$(1,0)$.
   }}}
\end{figure}

In Fig.~\ref{fig:2.3}, we show the constraints which the measurements
of $R_b$, $R_t$, and $\epsilon_K$ imply in the $\bar\rho$--$\bar\eta$
plane. Given the present theoretical and experimental uncertainties
in the analysis of charmless $B$ decays, $B^0$--$\bar B^0$ mixing,
and CP violation in the kaon system, there is still a rather large
region allowed for the Wolfenstein parameters. This has important
implications. For instance, the allowed region for the angle $\beta$
of the unitarity triangle (see Fig.~\ref{fig:2.2}) is such that
\begin{equation}\label{sinbe}
   0.34 < \sin 2\beta < 0.75 \,.
\end{equation}
Below, we will discuss that the CP asymmetry in the decay $\bar
B\to\psi\,K_S$, which is one of the favoured modes to search for CP
violation at a future $B$ factory, is proportional to $\sin 2\beta$.
Obviously, the prospects for discovering CP violation with such a
machine depend on whether $\sin 2\beta$ is closer to the upper or
lower bound in (\ref{sinbe}). A more reliable determination of the
shape of the unitarity triangle is thus of great importance.

On the other hand, our knowledge of the unitarity triangle has
already improved a lot in the last few years, and we are now reaching
a state where the analysis described in this section becomes a
serious test of the Standard Model. If the three bands in
Fig.~\ref{fig:2.3} did not overlap, this would be an indication of
New Physics.

\runninghead{CP Violation}
 {CP Asymmetries in Neutral $B$-Meson Decays}
\subsection{CP Asymmetries in Neutral $B$-Meson Decays}
\noindent
As mentioned above, decays of neutral $B$ mesons into CP eigenstates
provide for largely model-independent determinations of the angles of
the unitarity triangle. In the $B$-meson system, up to corrections of
order $10^{-2}$, we have
\begin{equation}
   \bigg( {q\over p} \bigg)_B \simeq - {M_{12}^*\over |M_{12}|}
   = {(V_{tb}^* V_{td})^2\over |V_{tb}^* V_{td}|^2}
   = {V_{tb}^* V_{td}\over V_{tb} V_{td}^*} = e^{-2i\beta} \,.
\end{equation}
This combination of CKM parameters can be read off directly from the
vertices of the box diagrams in Fig.~\ref{fig:box}, which in the
Standard Model are responsible for the non-diagonal element
$M_{12}^*$ of the mass matrix. Notice that for the real part of the
box diagrams, which determines $M_{12}$, the contributions of $c$ and
$u$ quarks in the loops can be neglected.

To eliminate hadronic uncertainties, one has to choose decay modes
dominated by a single diagram. However, most channels receive
contributions from ``tree'' and ``penguin'' diagrams, which for a
generic $b\to q\,\bar q' q'$ decay contribute in the
ratio\cite{LoPe}$^-$\cite{Grins}
\begin{equation}
   {\mbox{penguin}\over\mbox{tree}} \sim {\alpha_s\over 12\pi}\,
   \ln{m_t^2\over m_b^2} \cdot r \cdot
   {V_{tb} V_{tq}^*\over V_{q'b} V_{q'q}^*} \,.
\end{equation}
The first factor arises from the loop suppression of the penguin
diagrams and is of order 2\%, the second factor accounts for the fact
that hadronic matrix elements of penguin operators are usually
enhanced with respect to those of the operators appearing in tree
diagrams by $r\sim 2$--5, and the last factor is the ratio of CKM
matrix elements.

It follows that there are three possibilities to obtain the dominance
of a single diagram:
\begin{itemize}
\item
If the CKM parameters of the penguin diagram are not enhanced with
respect to the tree diagram, i.e.\ if
\begin{equation}
   \left| {V_{tb} V_{tq}^*\over V_{q'b} V_{q'q}^*} \right| \le 1 \,,
\end{equation}
the tree diagram dominates over the penguin diagram. Examples of
such decays are $\bar B\to\pi\pi$, $\bar B\to D\bar D$, $B_s\to\rho
K_S$ and $B_s\to\psi K_S$.
\item
If tree diagrams are forbidden, the penguin diagram dominates.
Examples of such decays are $\bar B\to\phi K_S$, $\bar B\to K_S K_S$,
$B_s\to\eta'\eta'$ and $B_s\to\phi K_S$.
\item
If
\begin{equation}
   \mbox{arg}\bigg( {V_{tb} V_{tq}^*\over V_{q'b} V_{q'q}^*}
   \bigg) = 0 ~\mbox{or}~\pi \,,
\end{equation}
both the tree and the penguin diagram have the same weak phase. In
that case one still has $|\bar A/A|=1$, i.e.\ no hadronic
uncertainties. Examples of such decays are $\bar B\to\phi K_S$ and
$B_s\to\psi\phi$.
\end{itemize}
Let us illustrate these three classes of decays with explicit
examples.

\subsubsection{Tree-dominant decays: $\bar B\to\pi\pi$}
\begin{figure}[htb]
   \epsfxsize=9.9cm
   \centerline{\epsffile{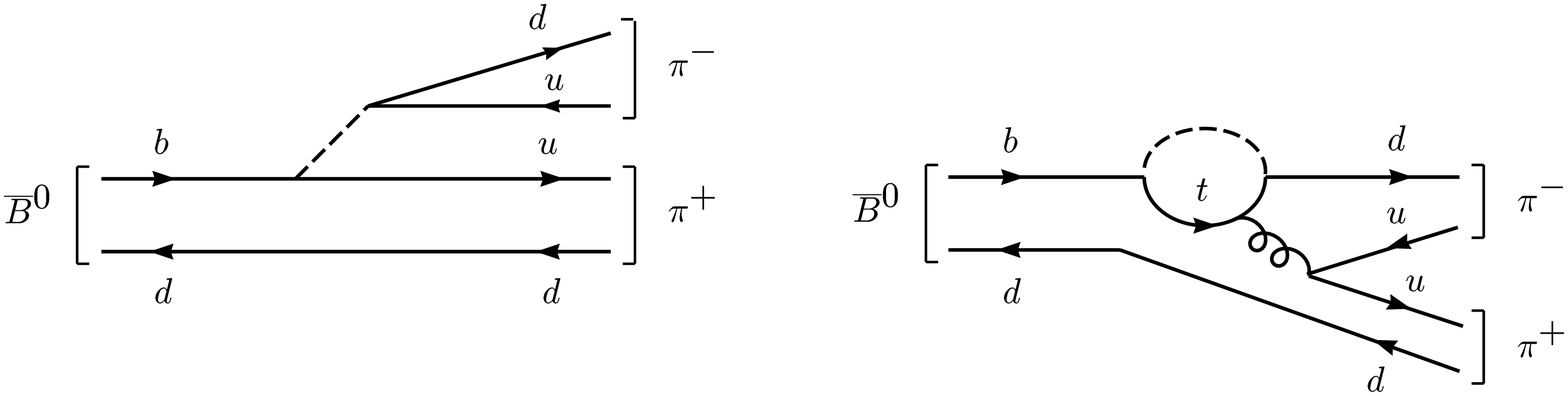}}
   \centerline{\parbox{11.5cm}{\caption{\label{fig:Bpipi}
Tree and penguin diagrams for the decay $\bar B\to\pi\pi$.
   }}}
\end{figure}

\noindent
The decay $\bar B\to\pi\pi$ proceeds through the quark decay $b\to
u\bar u d$, for which both the tree and the penguin diagram have CKM
parameters of order $\lambda^3$, as shown in Fig.~\ref{fig:Bpipi}.
Thus, the tree diagram is dominant, and to a good approximation
\begin{equation}
   \lambda_{\pi\pi} = {q\over p} \cdot {\bar A\over A}
   \simeq {V_{tb}^* V_{td}\over V_{tb} V_{td}^*} \cdot
   {V_{ub} V_{ud}^*\over V_{ub}^* V_{ud}}
   = e^{-2i\beta}\,e^{-2i\gamma} = e^{2i\alpha} \,,
\end{equation}
and
\begin{equation}
   \mbox{Im}\,\lambda_{\pi\pi} \simeq \sin 2\alpha \,.
\end{equation}
Hadronic uncertainties arise from the small admixture of penguin
contributions, which lead to $|\lambda|\ne 1$. They are
expected\cite{Yosi} to be of order 10\%, and can be reduced further
by using isospin analysis\cite{iso1}$^-$\cite{iso5}.

\subsubsection{Tree-forbidden decays: $\bar B\to\phi K_S$}
\begin{figure}[htb]
   \epsfxsize=4.7cm
   \centerline{\epsffile{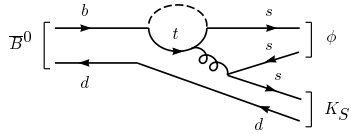}}
   \centerline{\parbox{11.5cm}{\caption{\label{fig:BphiKS}
Penguin diagram for the decay $\bar B\to\phi K_S$.
   }}}
\end{figure}

\noindent
The decay $\bar B\to\phi K_S$ proceeds through the quark transition
$b\to s\bar s s$, i.e.\ it involves a flavour-changing neutral
current, which is forbidden at the tree level in the Standard Model.
Thus, the relevant diagram is the penguin transition shown in
Fig.~\ref{fig:BphiKS}. A new ingredient is the presence of $K$--$\bar
K$ mixing, which adds a factor
\begin{equation}
   \bigg( {q\over p} \bigg)_K \simeq
   {V_{cs} V_{cd}^*\over V_{cs}^* V_{cd}}
\end{equation}
in the definition of $\lambda$. This is essential for decays with a
single $K_S$, since only $B^0\to K^0$ and $\bar B^0\to\bar K^0$
transitions are allowed, and interference between them is possible
only due to $K$--$\bar K$ mixing. It follows that
\begin{equation}
   \lambda_{\phi K_S} = \bigg( {q\over p} \bigg)_B \cdot
   \bigg( {q\over p} \bigg)_K \cdot {\bar A\over A}
   \simeq {V_{tb}^* V_{td}\over V_{tb} V_{td}^*} \cdot
   {V_{cs} V_{cd}^*\over V_{cs}^* V_{cd}} \cdot
   {V_{tb} V_{ts}^*\over V_{tb}^* V_{ts}}
   = e^{-2i\beta} \,,
\end{equation}
and therefore
\begin{equation}
   \mbox{Im}\,\lambda_{\phi K_S} \simeq - \sin 2\beta \,.
\end{equation}

\subsubsection{Decays with a single weak phase: $\bar B\to\phi K_S$}
\begin{figure}[htb]
   \epsfxsize=10cm
   \centerline{\epsffile{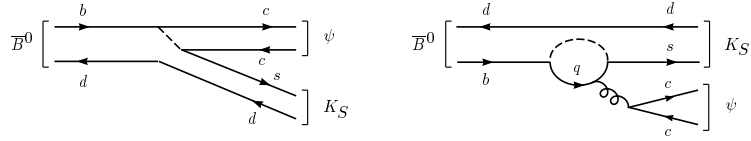}}
   \centerline{\parbox{11.5cm}{\caption{\label{fig:BpsiKS}
Tree and penguin diagrams for the decay $\bar B\to\psi K_S$.
   }}}
\end{figure}

\noindent
The decay $\bar B\to\psi K_S$ is based on the quark transition $b\to
c\bar c s$, for which the tree diagram is dominant. As shown in
Fig.~\ref{fig:BpsiKS}, the tree amplitude is proportional to $V_{cb}
V_{cs}^*\sim\lambda^2$. One finds
\begin{equation}
   \lambda_{\psi K_S} = - \bigg( {q\over p} \bigg)_B \cdot
   \bigg( {q\over p} \bigg)_K \cdot {\bar A\over A}
   \simeq {V_{tb}^* V_{td}\over V_{tb} V_{td}^*} \cdot
   {V_{cs} V_{cd}^*\over V_{cs}^* V_{cd}} \cdot
   {V_{cb} V_{cs}^*\over V_{cb}^* V_{cs}}
   = - e^{-2i\beta} \,,
\end{equation}
and therefore
\begin{equation}
   \mbox{Im}\,\lambda_{\psi K_S} \simeq \sin 2\beta \,.
\end{equation}
In the present case, the contamination from the penguin contribution
is extremely small\cite{BiSa}. Depending on the flavour $q$ of the
quark in the loop, the penguin contributions are proportional to
$V_{tb} V_{ts}^*\simeq\lambda^2$ (for $q=t$), $V_{cb}
V_{cs}^*\simeq\lambda^2$ (for $q=c$), and $V_{ub}
V_{us}^*\simeq\lambda^4$ (for $q=u$). Because of the relation $V_{tb}
V_{ts}^*=-V_{cb} V_{cs}^* + O(\lambda^4)$, it follows that up to very
small corrections the penguin contributions have the same weak phase
as the tree diagram. Hence, their presence affects neither
$|\lambda|$ nor Im$\lambda$. Detailed estimates show that the
hadronic uncertainties are only of order $10^{-3}$. This makes the
measurement of $\sin 2\beta$ in $\bar B\to\psi K_S$ the theoretically
cleanest determination of any CKM parameter. For this reason, this
decay is often considered the ``gold-plated'' mode of a $B$ factory.

The above-mentioned examples are only meant to illustrate the range
of possibilities for performing model-independent measurements of
CP-violating CKM parameters in neutral $B$-meson decays into CP
eigenstates. A summary and some more examples are given in
Table~\ref{tab:CP}. The angle $\beta'$ appearing in the CP
asymmetries for $B_s$-meson decays is the analogue of the angle
$\beta$ in the unitarity triangle defined by the relation
\begin{equation}
   V_{us} V_{ub}^* + V_{cs} V_{cb}^* + V_{ts} V_{tb}^* = 0 \,.
\end{equation}
Experimentally, $|\sin 2\beta'|<0.06$.

\begin{table}[htb]
\centerline{\parbox{11.5cm}{\caption{\label{tab:CP}
Examples of CP asymmetries for $B$ and $B_s$ decays into CP
eigenstates}}}
\vspace{0.5cm}
\centerline{\begin{tabular}{|c||c|c||c|c|}\hline\hline
\rule[-0.2cm]{0cm}{0.65cm} & \multicolumn{2}{c||}{$B$ decays}
 & \multicolumn{2}{c|}{$B_s$ decays} \\
\hline
\rule[-0.2cm]{0cm}{0.65cm} Quark decay & Final state & SM prediction
 & Final state & SM prediction \\
\hline
\rule[-0.2cm]{0cm}{0.65cm}
 $b\to c\bar c s$ & $\psi K_S$ & $-\sin 2\beta$ &
 $D_s^+ D_s^-$ & $-\sin 2\beta'$ \\
\rule[-0.2cm]{0cm}{0.65cm}
 $b\to c\bar c d$ & $D^+ D^-$ & $-\sin 2\beta$ &
 $\psi K_S$ & $-\sin 2\beta'$ \\
\rule[-0.2cm]{0cm}{0.65cm}
 $b\to u\bar u d$ & $\pi^+\pi^-$ & $\sin 2\alpha$ &
 $\rho K_S$ & $-\sin 2(\gamma+\beta')$ \\
\rule[-0.2cm]{0cm}{0.65cm}
 $b\to s\bar s s$ & $\phi K_S$ & $-\sin 2(\beta-\beta')$ &
 $\eta'\eta'$ & 0 \\
\rule[-0.2cm]{0cm}{0.65cm}
 $b\to s\bar s d$ & $K_S K_S$ & 0 &
 $\phi K_S$ & $\sin 2(\beta-\beta')$ \\
\hline\hline
\end{tabular}}
\end{table}

At the end of this section, let us stress again that the Standard
Model description of CP violation is at the same time very predictive
(since all CP violation is related to a single parameter) and most
likely wrong (because of the problems with baryogenesis and strong CP
violation). Thus, the prospects are good that once the various CP
asymmetries in $B$-meson decays can be explored at a $B$ factory,
deviations from the picture described here will arise. Those
deviations would indicate New Physics beyond the Standard Model.

\runninghead{Concluding Remarks} {Concluding Remarks}
\section{Concluding Remarks}
\noindent
We have presented a review of the theory and phenomenology of
heavy-flavour physics. The theoretical tools that allow to perform
quantitative calculations in this area are the heavy-quark symmetry,
the heavy-quark effective theory, and the $1/m_Q$ expansion. We have
discussed in detail exclusive weak decays of $B$ mesons, inclusive
decay rates and lifetimes of $b$ hadrons, and CP violation in
$B$-meson decays. Besides presenting the status of the latest
developments in these fields, our hope was to convinced the reader
that heavy-flavour physics is a rich and diverse area of research,
which is at present characterized by a fruitful interplay between
theory and experiments. This has led to many significant discoveries
and developments on both sides. Heavy-quark physics has the potential
to determine many important parameters of the electroweak theory and
to test the Standard Model at low energies. At the same time, it
provides an ideal laboratory to study the nature of non-perturbative
phenomena in QCD, still one of the least understood properties of the
Standard Model. The phenomenon of CP violation, finally, is one of
the most intriguing aspects of high-energy physics. Today, there is
only a single unambiguous measurement of a CP-violating quantity. But
already in a few years, when CP violation in the $B$-meson system can
be explored at the $B$ factories, this will very likely provide some
clues about the physics beyond the Standard Model.

Indeed, the prospects for further significant developments in the
field of heavy-flavour physics look rather promising. With the
approval of the first asymmetric $B$ factories at SLAC and KEK, with
ongoing $B$-physics programs at the existing facilities at Cornell,
Fermilab and CERN, and with plans for future $B$ physics at HERA-B
and the LHC-B, there are $B$eautiful times ahead of us!

\vspace{0.3cm}
{\it Acknowledgements:\/}
I am grateful to Irinel Caprini, Maria Girone and Chris Sachrajda for
useful discussions and collaboration on some of the subjects
presented here, and to Patricia Ball for exchanges concerning the
calculation of the semileptonic branching ratio.

\runninghead{Bibliography} {Bibliography}

\end{document}